\documentclass[aps,nofootinbib,notitlepage,superscriptaddress,twocolumn,10pt,prd]{revtex4-1}

\usepackage{graphicx}
\usepackage{amsmath,amssymb}
\usepackage{hyperref}
\usepackage{braket}
\usepackage{subfigure}
\usepackage{float}
\usepackage[dvipsnames]{xcolor}
\usepackage{soul}
\usepackage{rotating}
\usepackage{multirow}
\usepackage{mathtools}

\usepackage{makecell}

\usepackage[margin=0.75in]{geometry}

\hypersetup{
	colorlinks=true,
	linkcolor=red,
	citecolor=blue,
}

\usepackage{xcolor}
\newcommand{\joint}{J}
\newcommand{\pcopy}{C}
\newcommand{\single}{S}
\newcommand{\unsplitjoint}{{UJ}}
\newcommand{\splitjoint}{{CJ}}
\newcommand{\splitunsplitjoint}{{CUJ}}
\newcommand{\unsplitcopy}{{U}}
\newcommand{\splitcopy}{{SC}}
\newcommand{\Nsplit}{C}
\newcommand{\Nunsplit}{U}

\newcommand{\m}{{\mathfrak{m}}}
\newcommand{\M}{{\mathfrak{M}}}

\usepackage{natbib}

\begin{document}
	
\title{Quantifying concordance of correlated cosmological data sets} 

\author{Marco Raveri}
\affiliation{Center for Particle Cosmology, Department of Physics and Astronomy, University of Pennsylvania, Philadelphia, PA 19104, USA}
\author{Georgios Zacharegkas}
\affiliation{Kavli Institute for Cosmological Physics, Department of Astronomy \& Astrophysics, The University of Chicago, Chicago, IL 60637, USA}
\author{Wayne Hu}
\affiliation{Kavli Institute for Cosmological Physics, Department of Astronomy \& Astrophysics, Enrico Fermi Institute, The University of Chicago, Chicago, IL 60637, USA}

\begin{abstract}	
We develop estimators of agreement and disagreement between correlated cosmological data sets.
These account for data correlations when computing the significance of both tensions and excess confirmation while remaining statistically optimal.
We discuss and thoroughly characterize different approaches commenting on the ones that have the best behavior in practical applications.
We complement the calculation of their statistical distribution within the Gaussian model with one estimator that  takes non-Gaussianities fully into account.
To illustrate the use of our techniques, we apply these estimators to supernovae measurements of the distance-redshift relation, absolutely calibrated by the local distance ladder.
The suite of best estimators that we discuss finds results that are in excellent agreement between estimators and find no indications of significant internal inconsistencies in this data set above the $1\%$ probability threshold.
This shows the robustness of local determinations of the Hubble constant to features in the distance-redshift relation.
\end{abstract}
	
\maketitle

\section{Introduction}\label{Sec:Introduction}

The remarkable ability of the $\Lambda$CDM model to explain a wide range of observations, such as the spectrum of the fluctuations in the cosmic microwave background, observations of gravitational lensing and the clustering of galaxies, has made it widely accepted as the standard cosmological model.

However, despite its successes, as the  precision of different experiments has increased, so too has the statistical significance of discrepancies between their inferences of the $\Lambda$CDM model parameters (for a recent review see~\cite{Verde:2019ivm}).
Such discrepancies deserve close attention since they may hint at the existence of new physical phenomena or to the presence of residual systematic effects that are not yet understood.

In parallel with increased experimental precision, cosmological data sets have become increasingly complex to the point that understanding whether different probes agree or not requires the use of dedicated statistical tools~\cite{Marshall:2004zd,Feroz:2008wr,March:2011rv,Amendola:2012wc,Verde:2013wza,Bennett:2014tka,Martin:2014lra,Karpenka:2014moa,Larson:2014roa,Addison:2015wyg,Raveri:2015maa,Seehars:2014ora,Seehars:2015qza,Grandis:2016fwl,Addison:2017fdm,Nicola:2018rcd,Weiland:2018kon,Huang:2018xle,Raveri:2018wln,Motloch:2018pjy,Motloch:2018pyt,Adhikari_2019,Kerscher:2019pzk,Handley:2019wlz,Huang:2019nor,Lin:2019zdn}.

In this paper we develop concordance and discordance estimators (CDE) for data sets that are correlated.
This allows us to study both internal consistency of a data set and the mutual consistency of different correlated data sets.
In the former case splitting a data set in different parts naturally leads to correlated data pieces.
The latter case, even though present cosmological measurements are only weakly correlated, will become increasingly important in the future as correlations will become more relevant, reflecting the fact that different experiments will be measuring different properties of the same underlying sky.

We focus on extending  the CDEs introduced for uncorrelated data sets~\cite{Raveri:2018wln} to correlated data sets. 
In particular, we discuss estimators quantifying parameter shifts between two correlated data sets and 
goodness of fit loss when two data sets are joined together.
We analyze  these estimators under the Gaussian linear model (GLM), assuming Gaussianity of the data and model parameters. We comment on mitigation against non-Gaussianities that is built in some of them in practical applications.

To complement and check these results we also discuss a purely Markov chain Monte Carlo (MCMC) approach, which uses the full distribution of parameters, as inferred from MCMC samples, to compute the probability of a parameter shift.
This technique fully takes into account parameter non-Gaussianities and any non-linear aspect of the model.

When considering correlated data sets we follow two different strategies.
In the first approach we consider the two disjoint data sets and build estimators working on their separate parameter inference though properly including their correlation.
In the second one, we always consider the joint data set but fit the two parts of the data set with different cosmological parameters.
This approach was also employed in~\cite{Zhang:2003ii,Chu:2004qx,Wang:2007fsa,Abate:2008au,Ruiz:2014hma,Bernal:2015zom,Kohlinger:2018sxx,Lemos:2019txn}.
We call the first strategy the data split one while we refer to the second one as the parameter split strategy.
The data and parameter split techniques are equivalent when data sets are uncorrelated but different when they are correlated.

Overall we find that the parameter split methodology is more convenient in practice. 
All estimators that we consider, in this case, can be easily obtained from the posterior distribution while the same is not true for the data split estimators.
Moreover the different parameter split estimators should agree if non-Gaussianities are negligible, providing essential cross checks of the reported results.

As a demonstration of our methodology we apply these estimators to the Pantheon type Ia supernova (SN) sample~\cite{Scolnic:2017caz} calibrated with measurements of the Hubble constant from~\cite{Riess:2019cxk}.
We choose to split the SN data into two subsets at the redshift values of $z=0.3$ and $z=0.7$, loosely corresponding, respectively, to the time of dark energy-dark matter equality and the time at which cosmic acceleration begins.

We find that the $\Lambda$CDM model provides a good fit to these data with the exception of the $z>0.7$ SN measurements for which the agreement seems to be too good at the $94 \%$ confidence level.
We also find excellent agreement between the results of different parameter split estimators, regardless of mild non-Gaussianities in the SN posterior that are effectively mitigated.
As reported by different estimators the first SN split at $z=0.3$ is in good agreement with the $\Lambda$CDM model while the second  split, at $z=0.7$ results in parameters that are too close to each other at about $97\%$ probability.
Furthermore, we do not find significant indications of differences in the estimates of the Hubble constant between different SN redshift splits indicating that its direct measurement is robust against split in the SN catalog and to features in the SN distance-redshift relation.

This paper is organized as follows. 
In Sec.~\ref{Sec:GLMReview} we briefly review the Gaussian Linear Model. 
In Sec.~\ref{Sec:ModelDataset} we describe the SN data set that we employ in this work as an illustration of our statistical techniques.
Sec.~\ref{Sec:ImpactCorrelations} includes a quantification of the importance of SN data correlations and their impact on cosmological parameters.
The properties and differences of the two approaches that we follow, data splits and parameter split, are discussed in Sec.~\ref{Sec:DataParamSplits}, before a detailed discussion of the two methodologies separately in Sec.~\ref{Sec:CDEdatasplit} and Sec.~\ref{Sec:CDEparamsplit}, respectively. 
We summarize our conclusions in Sec.~\ref{Sec:Conclusions}. 

Details of our techniques are presented in a series of Appendices.
In App.~\ref{App:CCA} we discuss the Canonical Correlation Analysis for quantifying the impact of
correlations.
In App.~\ref{App:ExampleDataParamSplit} we provide a worked  pedagogical example that clarifies
the differences between data splits and parameter split. 
In App.~\ref{App:QDMAPdataExact} and App.~\ref{App:QDMAPcopyExact} we derive the exact distributions of the goodness of fit loss statistic in the cases of data and parameter splits, respectively. 
In App.~\ref{App:UnsplitMultisplit}, we generalize our discussion of splitting the data into an arbitrary number of subsets.

\section{The Gaussian linear model} \label{Sec:GLMReview}

In this section we gather some basic definitions that we will later use throughout the paper.
For an in depth discussion of the Gaussian Linear Model (GLM) we refer the reader to~\cite{Raveri:2018wln}.
	
We denote the multivariate Gaussian distribution in $N$ dimensions with mean $\bar{\theta}$ and covariance $\mathcal{C}$ as $\mathcal{N}_N(\theta; \bar{\theta},\mathcal{C})$.
For a given data set, $D$, described by a model $\mathcal{M}$ that depends on a set of parameters $\theta$, the posterior probability distribution of the parameters is given by:
\begin{equation}\label{Eq:PosteriorDef}
P(\theta | D, \mathcal{M}) = \frac{\mathcal{L}(\theta) \Pi (\theta)}{\mathcal{E}} \; ,
\end{equation}
where the likelihood is the probability of the data at any given choice of parameters $\mathcal{L}(\theta) = P( D | \theta, \mathcal{M})$ and any prior knowledge is encoded in $\Pi (\theta)$.
The normalization of the posterior, $\mathcal{E}\equiv P(D|\mathcal{M})$, is the evidence that provides the probability distribution of the data given the model $\mathcal{M}$.

In this section, we assume that the prior distribution is Gaussian in the model parameters, $\Pi(\theta)= \mathcal{N}_N(\theta; \theta_\Pi, \mathcal{C}_{\Pi})$, with  an unbiased mean $\theta_{\Pi}$ and covariance $\mathcal{C}_{\Pi}$.
As discussed in~\cite{Raveri:2018wln} this is a good choice to use in practice as it allows us to treat Gaussian priors on nuisance parameters exactly and models the most relevant features of informative flat priors: the scale of the prior and its central value.

We further assume that the likelihood is a Gaussian distribution in data space, $\mathcal{L}(\theta)= \mathcal{N}_d(x; m, \Sigma)$ and we denote by $d$ the number of data points $x$ and $\Sigma$ their covariance matrix.
The mean of the distribution is given by the model prediction, $m(\theta)$.

The GLM assumes that one can linearly expand the model prediction, $m(\theta)$, around a given parameter point. 
Since we are working with Gaussian priors, for simplicity, in the following we assume that the linear model expansion point is the prior center $\theta_\Pi$ and we can write
\begin{equation}\label{Eq:GLMdef}
m(\theta) \approx m_\Pi + M (\theta -\theta_\Pi) \; ,
\end{equation}
where $m_\Pi \equiv m(\theta_\Pi)$ and $M \equiv \left. (\partial m/\partial \theta) \right|_{\theta_\Pi}$ is the Jacobian of the transformation between parameter and data space.

Given the model prediction $m_\Pi$, the residual of a randomly chosen data point $x$, henceforth $X \equiv x -m_\Pi$, can be projected onto a component along the linear  model, $\mathbb{P} X$, and another component orthogonal to it, $X - \mathbb{P} X = (\mathbb{I} -\mathbb{P}) X$.  The projector can be thought of as a two-step process. The first step is to construct the linear combinations of data, namely $\tilde{M} X$, that give the parameter estimates: $\theta -\theta_\Pi = \tilde{M} X$, where   $\tilde{M} = \mathcal{C} M^T \Sigma^{-1}$,
with 
\begin{align} \label{Eq:GLMMLCovariance}
\mathcal{C} = \langle (\theta -\theta_\Pi ) (\theta -\theta_\Pi )^T \rangle = (M^T \Sigma^{-1} M)^{-1} \; ,
\end{align}
as the parameter covariance or inverse Fisher matrix.
Then, as a second step, given the parameter estimates we transform back into data space using the Jacobian, $M(\theta -\theta_\Pi) = M\tilde{M}X \equiv \mathbb{P}X$.  Thus  $\mathbb{P} \equiv M\tilde{M}$ is the full projector:
\begin{equation}\label{Eq:GLMProjector}
\mathbb{P} = M \mathcal{C} M^T \Sigma^{-1} \; ,
\end{equation}
 and $(\mathbb{I} - \mathbb{P})$ is its complement.
 
In the GLM the maximum likelihood (ML) is given by:
\begin{align} \label{Eq:MaximumLikelihood}
\ln \mathcal{L}_{\max} =& -\frac{1}{2} X^T (\mathbb{I} - \mathbb{P})^T \Sigma^{-1} (\mathbb{I} - \mathbb{P}) X \nonumber \\
& -\frac{d}{2}\ln(2\pi) -\frac{1}{2}\ln(|\Sigma|) \,,
\end{align}
where we used $|\cdot|$ to denote the determinant of a matrix. Notice that the first line of Eq. \eqref{Eq:MaximumLikelihood} contains all the pieces that depend on the data while the second one contains normalization constants that are often neglected.
The parameters corresponding to the maximum likelihood model are given by:
\begin{align} \label{Eq:GLMMaximumLikeParams}
\theta_{\rm ML} = \tilde{M}(x-m_{\Pi}+M\theta_\Pi) \,.
\end{align}
Over realizations of data, the maximum likelihood parameters are distributed as $\mathcal{N}_N(\theta; \theta_\Pi, \mathcal{C})$. 
These expressions refer to the true maximum likelihood of a model and they should be obtained without reference to the prior; $\theta_\Pi$ appears here due to the assumption that the prior mean is unbiased.

The maximum posterior (MAP) parameters combine the ML parameters with the prior:
\begin{align} \label{Eq:GLMMaximumPosterior}
\theta_{p} = \mathcal{C}_p (\mathcal{C}_\Pi^{-1} \theta_\Pi + \mathcal{C}^{-1} \theta_{\rm ML})  \,,
\end{align}
where $\mathcal{C}_p^{-1}=\mathcal{C}_\Pi^{-1}+\mathcal{C}^{-1}$.
Under the GLM the maximum posterior parameters are distributed as $\mathcal{N}(\theta; \theta_\Pi, \mathcal{C}_p)$.

Within the GLM the probability of the data, i.e.~the evidence, is Gaussian distributed for the Gaussian priors that we consider and is given by $\mathcal{E} = \mathcal{N}_d(x;m(\theta_\Pi), \Sigma + M\mathcal{C}_\Pi M^T)$.

We define all the statistics, $Q$, that we discuss in this paper to follow the convention that: if $P(Q>Q^{\rm obs})$ approaches zero then the observed value lies in the tail of the distribution that we would associate with a tension; if it approaches one the observed value would be in the tail associated with excess confirmation.

\section{SN dataset and model} \label{Sec:ModelDataset}

As an example case we study the internal consistency of the Pantheon type Ia supernovae (SN) sample~\cite{Scolnic:2017caz} to redshift splits under the $\Lambda$CDM model.

The Pantheon collaboration provides measurements of the SN magnitude, corrected for stretch, color, etc., 
relative to a fiducial absolute magnitude $\m-\M_{\rm fid}$, with its covariance $\Sigma_{\rm SN}$. $\M_{\rm fid}=-19.34$ is predetermined by a fit under the assumption of a Hubble constant of $H_{0}^{\rm fid}=70$  for definiteness (see e.g.~\cite{Guy:2005me}).\footnote{Note that {\texttt{\url{https://github.com/dscolnic/Pantheon}} provides $\m-\M_{\rm fid}$
whereas  \texttt{\url{https://github.com/cmbant/CosmoMC}}  provides   $\m$ from which $\M_{\rm fid}$ can be extracted.}} 
Here and throughout $H_0$ is quoted in units of km\,s$^{-1}$\,Mpc$^{-1}$, whereas $c=1$ in 
general formulae.  
The likelihood is then analytically marginalized over the true absolute magnitude $\M$ when considering the distance modulus $\m-\M$. Hence the reference Pantheon SN likelihood is not Gaussian in $\m-\M$.
Some of the methods that we discuss rely on Gaussianity of the likelihood in data space and in addition we wish to explore compatibility of $H_0$ determinations between subsets of the data.   To achieve this we introduce the SN absolute magnitude $\M$ as a model parameter.
We take the SN based measurements of the Hubble constant in~\cite{Riess:2019cxk}, of $H_0^{\rm SN}= 74.03 \pm 1.42$, to infer a measurement of $\M =\hat \M\pm \sigma_{\M}$ as 
\begin{align}
\hat\M & = 5 \log_{10}\frac{ H_0^{\rm SN}}{ H_0^{\rm fid} }+ \M_{\rm fid} \;, \nonumber\\
\sigma_{\M} & = \frac{5}{ \ln 10} \frac{ \sigma_{H_0^{\rm SN}}}{H_0^{\rm SN}} \;.
\end{align}
The SN data likelihood, as provided by the Pantheon collaboration, remains a Gaussian distribution
\begin{equation}
{\cal L}_{\rm SN} = {\cal N}\left(\m-\M_{\rm fid}; 5\log \frac{d_L}{10{\rm pc}}+\M-\M_{\rm fid},\Sigma_{\rm SN}\right) \;,
\end{equation} 
but the relationship to the luminosity distance $d_L$ comes through the likelihood for the absolute
magnitude data
\begin{equation}
{\cal L}_{\M}  = {\cal N}(\M; \hat \M, \sigma^2_{\M}),
\end{equation} 
such that
\begin{equation}
{\cal L} = {\cal L}_{\rm SN}  \times {\cal L}_{\M} .
\end{equation}
This obviates the need to marginalize $\M$ when considering cosmological constraints.  
Note that by inferring ${\cal L}_\M$ from the determination of $H_0$ from \cite{Riess:2019cxk} rather than directly calibrating $\M$ in the process of Pantheon data reduction, we force the Pantheon dataset as a whole to return the same mean value for $H_0$ but allow for nontrivial consistency tests with subsets of the data.
  
As a function of SN redshift, the flat $\Lambda$CDM  model for $d_L$ is
\begin{equation}
d_L(z) = \frac{ 1+z }{H_0}  \int_0^z\frac{dz'}{\sqrt{\Omega_m (1+z')^3 + 1-\Omega_m}}  .
\end{equation} 
We place flat priors on the range of the two cosmological parameters
$H_0 \in [40,100]$ and $\Omega_m \in [0,1]$.
These control respectively the amplitude and shape of $d_L(z)$.

The Pantheon SN sample covers the redshift range $z\in[0.01,2.26]$ with $1048$ SN distance measurements.
Given this many data points it is important to fix the data splits with a solid a priori criterium to avoid look-elsewhere type corrections to statistical significance that are hard to quantify.

We therefore choose  two relevant physical times in the SN redshift range with which
to cut the SN sample in two: 
\begin{itemize}
\item $z_{\rm cut}=0.3$ approximately the time of dark matter/ dark energy equality;
\item $z_{\rm cut}=0.7$ approximately the time at which cosmic acceleration begins.
\end{itemize}
These two data splits are very different and allow us to illustrate all the possible outcomes of our type of tests.
The first split has almost equal weight in both parts with $630$ SN below $z=0.3$ and $418$ data points above.
The second split is heavily weighted toward the first part of the data set, with $924$ SN below $z=0.7$, compared with $124$ SN above $z=0.7$.

When analyzing the splits separately, the absolute magnitude measurement would be
applied to each half separately.  When joining the two data splits, we need to take into account
this double counting of data. This is equivalent to introducing two absolute magnitude 
measurements that are fully correlated between the splits.   As we shall see, this provides an extreme, albeit
trivial, example of fully correlated data points between datasets and their impact.

In all the following, cosmological predictions for the $\Lambda$CDM model are obtained with the CAMB~\cite{Lewis:1999bs} code.
The parameter posterior distributions are obtained with Markov chain Monte Carlo (MCMC) sampling with the CosmoMC~\cite{Lewis:2002ah} code and their analysis largely relies on the GetDist code~\cite{Lewis:2019xzd}. 

\section{Impact of correlations} \label{Sec:ImpactCorrelations}

The stronger the correlation between two data sets the more crucial it is to use statistical tools which take these correlations into account.

\begin{figure}[!ht]
\centering
\begin{subfigure}
  \centering
  \includegraphics[width=\columnwidth]{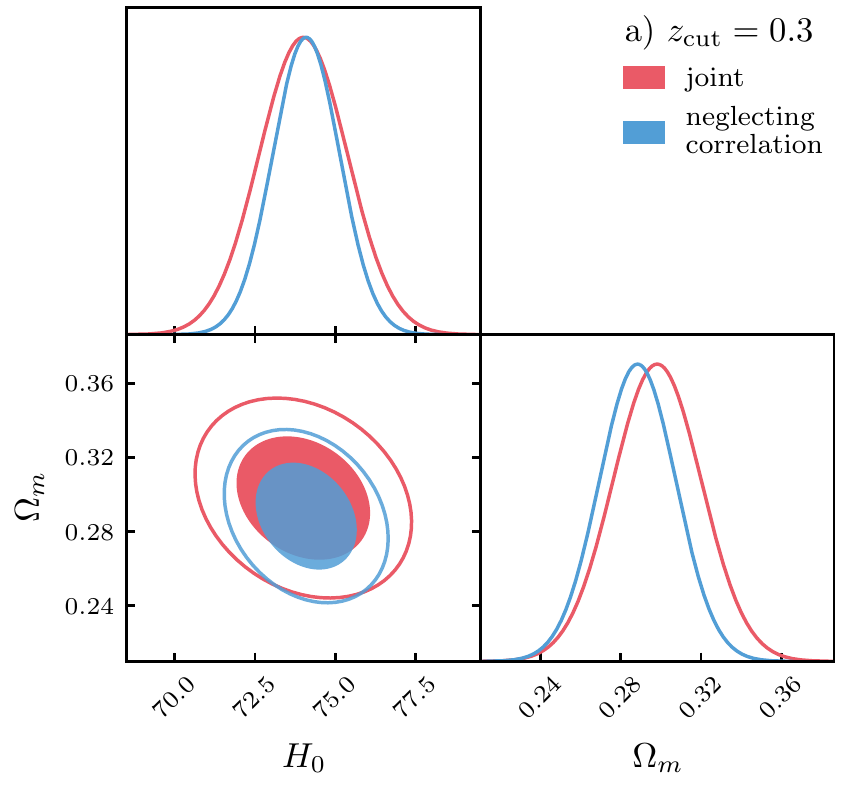}
\end{subfigure}
\begin{subfigure}
  \centering
  \includegraphics[width=\columnwidth]{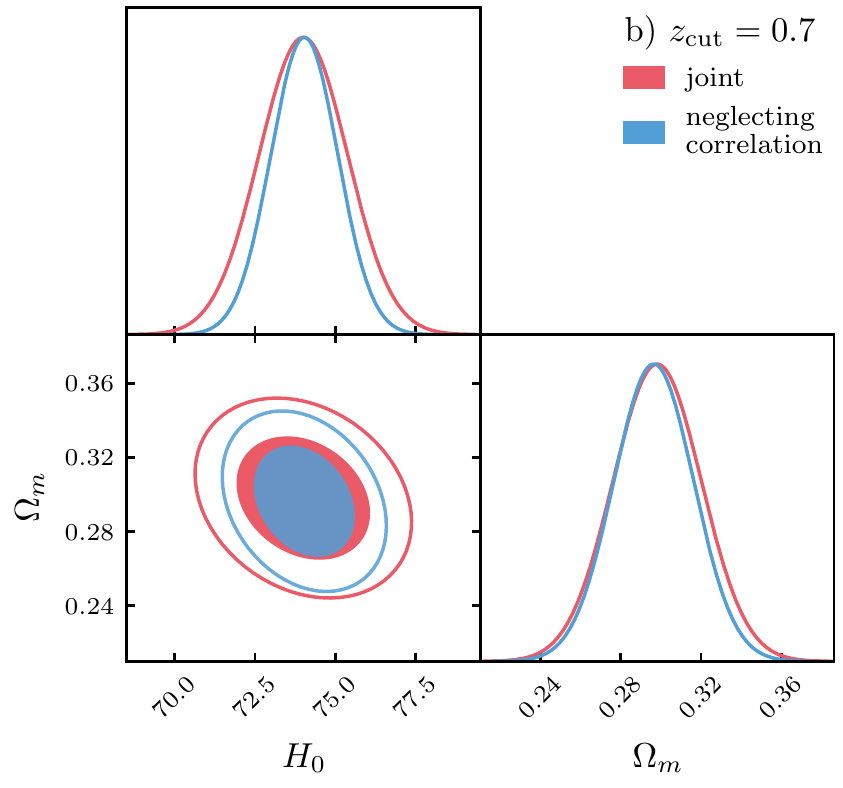}
\end{subfigure}
\caption{ \label{Fig:JointCorrelationPosteriorComparison}
The joint marginalized posterior of the full Pantheon SN data set compared to the results obtained neglecting the correlation between the $z>z_{\rm cut}$ and $z<z_{\rm cut}$ SN measurements for the two SN splits that we consider, in panels a) and b) respectively.
The filled contour corresponds to the 68\% C.L. region while the continuous contour shows the 95\% C.L. region. 
}
\end{figure}

We can see in Fig.~\ref{Fig:JointCorrelationPosteriorComparison} the impact that data correlations have on the joint SN posterior.
In both panels we show the full results taking into account all correlated modes, while we also show the results when we neglect the correlation between the data sets of the SN split, while keeping the correlation information within each subset of the split.
As we can clearly notice, the posterior is influenced in two different ways: the peak of the distribution shifts; also, the variance changes and looks more constraining when we neglect correlations, because we are not considering the part of the information in the two data sets that is redundant.

In multiple dimensions the correlation strength can be quantified by means of the Canonical Correlation Analysis (CCA)~\cite{CCA}.    
CCA allows us to understand the change in parameter variances as  we summarize here and fully discuss in App.~\ref{App:CCA}.

The two data splits that we consider have an almost completely correlated data mode due to the common calibration while the second correlated mode has a correlation coefficient of approximately $\rho_{12}=0.5$ for both splits.

The maximum error that we would make on the determination of the parameter variance, with respect to the full joint estimate, if we were to neglect correlations is discussed in App.~\ref{App:CCA} and is given by:
\begin{align}
{\rm max} \left( \frac{\sigma_{\rm no \, corr}^2}{\sigma_\joint^2} \right) = \frac{1}{1 \pm \rho_{12}} \;,
\end{align}
where $\sigma_{\rm no \, corr}^2$ is the wrong estimate of the variance of one parameter obtained neglecting correlations, $\sigma_\joint^2$ is the correct estimate of the variance, accounting for all data correlations, and $\rho_{12}$ is the maximum correlation coefficient.

In Fig.~\ref{Fig:JointCorrelationPosteriorComparison} we can clearly see that neglecting the duplication of the SN amplitude leads, when joining the two SN splits, to a false $\sqrt{2}$ improvement in the determination of the Hubble constant.  This is because the two determinations from the splits are nearly fully correlated because of the absolute magnitude
calibration that they share. 
The second correlated mode is then responsible for the residual underestimate of the error in the determination of $\Omega_m$ at about the $20\%$ level and reflects correlations in 
the measurement of SN magnitudes which resemble that parameter. 

We can understand the shift in the posterior in Fig.~\ref{Fig:JointCorrelationPosteriorComparison} as a difference in the SN maximum posterior due to data correlations.   Underlying this difference is the impact of correlations on  the likelihood at
a given parameter point.   
As discussed in~\cite{Raveri:2018wln} the value of the likelihood at MAP, $\hat{\theta}_p$, can be used as a goodness of fit measure to test the consistency of a data set with the model at hand.
The MAP measure for goodness of fit is then given by:
\begin{equation}\label{Eq:QMAPDef}
Q_{\rm MAP} \equiv -2 \ln \mathcal{L} (\hat{\theta}_p) + 2 \left\langle \ln \mathcal{L}(\hat{\theta}_p) \right\rangle_D + \left\langle Q_{\rm MAP} \right\rangle_D
\end{equation}
where the average is over data realizations. 
For Gaussian priors $Q_{\rm MAP}$ is distributed as a sum of Gamma distributed variables which can be (conservatively) approximated by a chi-square distribution of $d-N_{\rm eff}$ degrees of freedom, $Q_{\rm MAP} \sim \chi^2 (d -N_{\rm eff})$, where 
\begin{equation}\label{Eq:NeffDef}
N_{\rm eff} = N - {\rm tr}[\mathcal{C}_\Pi^{-1}\mathcal{C}_p]
\end{equation}
is the number of effective parameters that are being constrained by the data over the prior, with $N$ being the total number of model parameters.
Notice that there might be cases where the data covariance matrix is singular. In these cases one needs to compute the number of data points as the rank of the covariance, $d={\rm rank}(\Sigma)$.

The results of the application of this goodness of fit statistics to the SN data are summarized in Tab.~\ref{Tab:GOFresults}.
\begin{table}[t!]
\begin{ruledtabular}
\begin{tabular}{ l c c c c c c }
\textrm{Data} & $-2\ln \mathcal{L}_{\rm MAP}$ & \textrm{$N_{\rm eff}$} & \textrm{$d$} & $P(Q_{\rm MAP} > Q_{\rm MAP}^{\rm obs})$ \\
\hline \hline
$z<0.3$ & $633.63$ & $1.95+1$ & $630+1$ & $43.0\%$ \\
$z>0.3$ & $388.63$ & $1.96+1$ & $418+1$ & $82.9\%$ \\
\colrule	
$z<0.7$ & $926.92$  & $1.99+1$ & $924+1$ & $44.8\%$ \\
$z>0.7$ & $99.64$    & $1.49+1$ & $124+1$ & $93.6\%$ \\
\colrule
joint        & $1026.86$ & $1.99+1$ & $1048+1$ & $65.8\%$ \\			
\end{tabular}
\end{ruledtabular}
\caption{\label{Tab:GOFresults}
The results of the application of the $Q_{\rm MAP}$ goodness of fit statistics to the SN data sets that we consider.
The first column specifies the data set that is considered, the second one the value of likelihood at maximum posterior.
The third column shows the number of effective parameters, as in Eq.~(\ref{Eq:NeffDef}), highlighting that one parameter is constrained by the measurement of the absolute magnitude of SN.
The fourth column reports the number of data points used, highlighting the extra measurement of the SN absolute calibration.
The fifth column shows the probability to exceed the $Q_{\rm MAP}$ goodness of fit statistic.
}
\end{table}
As we can see the full SN catalog is a reasonably good fit. Notice that the joint data set contains two fully correlated measurements of $\hat\M$ so that the data covariance is singular and the number of data points is computed as the rank of the data covariance.
If we were to neglect correlations both the position of the maximum posterior and the likelihood value at MAP would change.
In the case of the $z_{\rm cut}=0.3$ split these changes are at about $2 \Delta\ln\mathcal{L} =0.2$, which corresponds to about half a sigma shift, which we see in the posterior of Fig.~\ref{Fig:JointCorrelationPosteriorComparison}a. We also find that the corresponding results for the $z_{\rm cut}=0.7$ split are smaller.
Neglecting correlations, in both cases, has also the effect of overestimating the number of degrees of freedom of the $Q_{\rm MAP}$ distribution. These would be $d_{SN}+2$ given that the full correlation between the $\hat\M$ measurements is neglected and the covariance matrix becomes non-singular.
In this case then, the goodness of fit test is mostly insensitive to the presence of correlations between the two SN splits, given the small change in the likelihood and the large number of data points.

We conclude this section by discussing the implications of the goodness of fit values shown in Tab.~\ref{Tab:GOFresults}.
In the $z_{\rm cut}=0.3$ case, as with the joint data set, both splits contain enough SN measurements to measure the two relevant cosmological parameters. The fit to the data is reasonably good and does not indicate the presence of tension or confirmation of high statistical significance. 
When considering the split of the data at $z_{\rm cut}=0.7$, the vast majority of the data points falls in the low-redshift subset.
This means that the low-redshift part measures both parameters better than the prior while the high-redshift one starts being influenced by the prior, as reflected in $N_{\rm eff}$.
The first part of the data provides a reasonably good fit, while the second one leans toward the fit being too good at a probability level that, however, does not exceed $95\%$.

\section{Data splits and parameter splits} \label{Sec:DataParamSplits}

%
\begin{figure*}[!ht]
\centering
\begin{subfigure}
  \centering
  \includegraphics[width=\columnwidth]{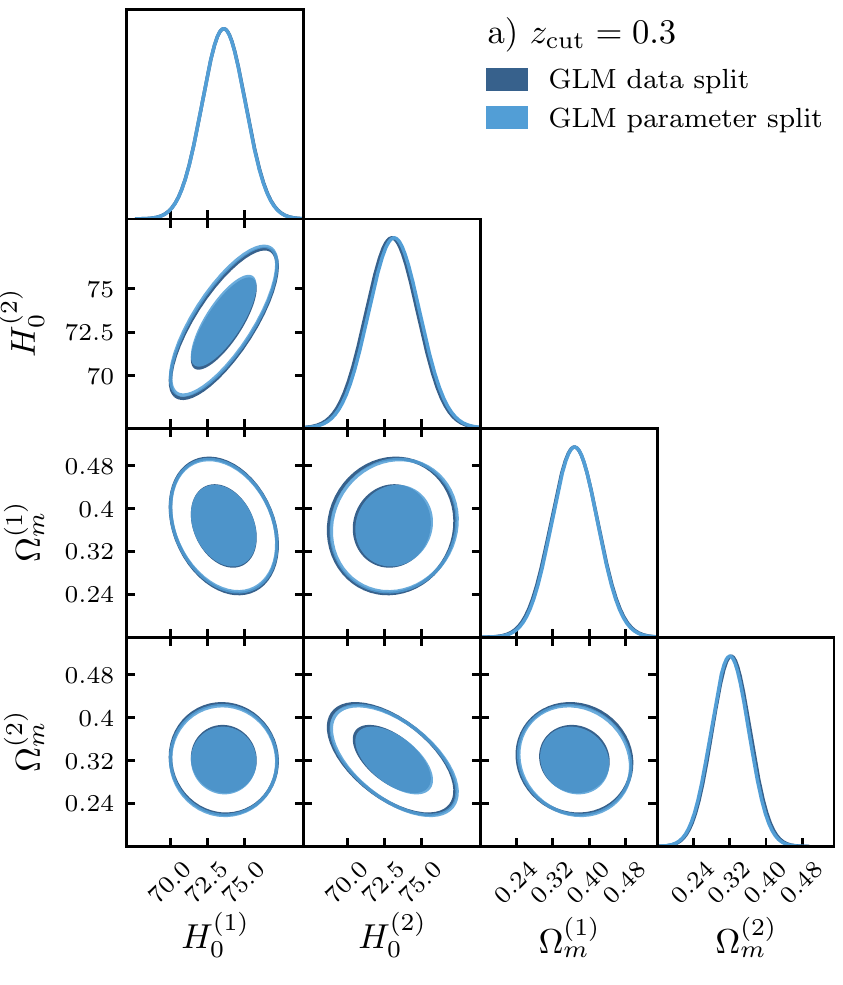}
\end{subfigure}
\begin{subfigure}
  \centering
  \includegraphics[width=\columnwidth]{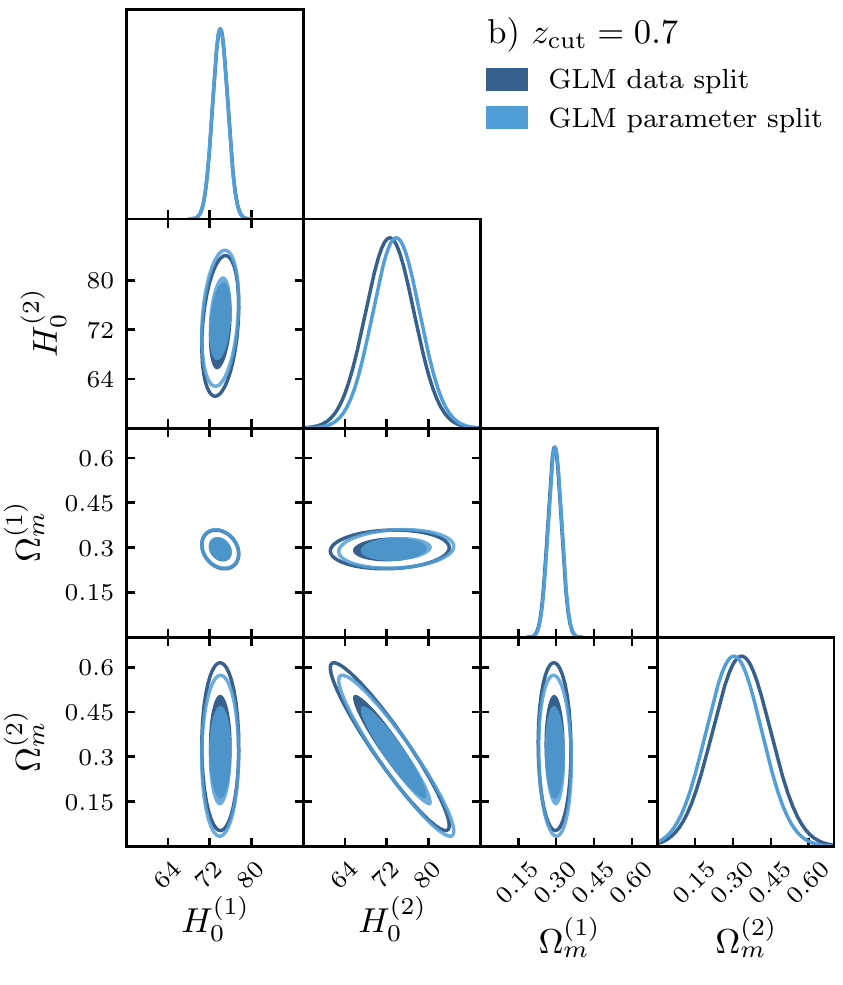}
\end{subfigure}
\caption{ \label{Fig:GLMParameterComparisonTriangle}
The GLM joint marginalized posterior parameter distribution for SN data splits and parameter splits. 
Different panels show different SN redshift cuts.
The filled contour corresponds to the 68\% C.L. region while the continuous contour shows the 95\% C.L. region.
}
\end{figure*}

We can split a correlated data set by taking partitions of the joint dataset, that we denote as $D_{\joint}$, in two pieces  that we indicate with $D_1$ and $D_2$.
Hereafter we denote quantities that refer to the joint data set with $\joint$ and quantities referring to the first and second data sets with the subscript $1$ and $2$ respectively.
In Appendix~\ref{App:UnsplitMultisplit} we discuss the natural generalization to an arbitrary number of splits.
We indicate the joint $D_\joint = D_1 \cup D_2$ data covariance as
\begin{equation} \label{Eq:JointCovariance}
\Sigma_\joint \equiv \left(
\begin{tabular}{cc}
$\Sigma_1$ & $\Sigma_{12}$ \\
$\Sigma_{21}$ & $\Sigma_2$
\end{tabular}
\right) \;,
\end{equation}
which is in general not block diagonal. 
Since the full covariance has to be symmetric then $\Sigma_1=\Sigma_1^T$, $\Sigma_2=\Sigma_2^T$ and $\Sigma_{21} = \Sigma_{12}^T$.
Notice that $\Sigma_1$, $\Sigma_2$ and $\Sigma_\joint$ have all to be symmetric and positive definite.

Within the GLM this data separation projects onto parameter space through the single and joint Jacobian matrices that are related by $M_\joint^T= (\partial m_1^T/\partial \theta, \partial m_2^T/\partial \theta) \equiv (M_1^T,M_2^T)$. 
The GLM estimate of the ML parameters in this case is given by:
\begin{align}
\theta^{\rm ML}_{a \single} =\,& \mathcal{C}_{a \single} M_a^T\Sigma_a^{-1} (x_a-m_{\Pi a} +M_a\theta_\Pi)  \,,\nonumber \\
\theta^{\rm ML}_\joint =\,& \mathcal{C}_\joint M_\joint^T \Sigma_\joint^{-1} (x_\joint -m_{\Pi\joint} +M_\joint\theta_\Pi) \,,
\end{align}
with $a \in \{1,2\}$.
Hereafter we denote quantities that are obtained within the single analysis of the split data sets with $\single$.
The maximum posterior parameters are obtained by adding on top of these estimates the prior, as in Eq.~(\ref{Eq:GLMMaximumPosterior}).
As we can see the inference of the parameters for the separate data splits depends only on the given data set, while their joint inference is influenced by the complementary set, through the correlation between the two.
For this reason it is not possible, in presence of data correlations, to reconstruct the joint ML parameters as a linear combination of parameter quantities that live in the single parameter space.

We can still, however, compute the covariance between different data split parameters both at the ML and MAP level as:
\begin{align} \label{Eq:DataSplitParameterCovariance}
{\rm cov}( \theta^{\rm ML}_{1\single}, \theta^{\rm ML}_{2\single} ) =\,&  \mathcal{C}_{1\single} M_1^T\Sigma_1^{-1}\Sigma_{12}\Sigma_2^{-1} M_2 \mathcal{C}_{2\single} \;, \nonumber \\
{\rm cov}( \theta^{p}_{1\single}, \theta^{p}_{2\single} ) =\,&  
\mathcal{C}_{p1\single} \mathcal{C}_{\Pi}^{-1}  \mathcal{C}_{p2\single} \nonumber \\
& +\mathcal{C}_{p1\single} M_1^T\Sigma_1^{-1}\Sigma_{12}\Sigma_2^{-1} M_2 \mathcal{C}_{p2\single}
\;.
\end{align}
As we can see these depend on both parameter space and data space quantities to take into account that the single parameter covariances do not include correlation contributions.

As an alternative strategy we can think of the  split as originating in parameter space, describing the two parts of the joint data set with different sets of parameters and always fitting the joint likelihood.
We denote with the subscript $\pcopy$ quantities that refer to this strategy of parameter duplication and, for example, we work with a $2N$ parameter vector that is defined by $\theta_\pcopy \equiv (\theta_{1\pcopy}, \theta_{2\pcopy})^T$. Similarly, we can define the duplicated prior parameter vector by $\theta_{\Pi\pcopy} = (\theta_{\Pi},\theta_{\Pi})^T$.  
One subtlety is that our null hypothesis is still that the data is drawn from the prior
distribution of a single parameter.  We shall see that this difference between the split analysis
and statistical properties of the data causes minor issues when counting these parameters if they are partially, but not fully constrained, by the prior.

In the GLM the block structure of the covariance in Eq.~(\ref{Eq:JointCovariance}) then projects on the two parameter copies with the Jacobian given by:
\begin{align}\label{Eq:CopyJacobian}
M_\pcopy = \left(
\begin{tabular}{c c}
$\partial m_1 / \partial \theta_1$ & $\mathbb{O}$ \\
$\mathbb{O}$ & $\partial m_2 / \partial \theta_2$
\end{tabular}
\right)  =
\left(
\begin{tabular}{c c}
$M_{1}$ & $\mathbb{O}$ \\
$\mathbb{O}$ & $M_{2}$
\end{tabular}
\right) \;. 
\end{align}
The maximum likelihood estimate of the copy parameters is given by Eq.~(\ref{Eq:GLMMaximumLikeParams}) and explicitly reads:
\begin{align} \label{Eq:CopyMLParameters}
\theta^{\rm ML}_{\pcopy} ={}&
\left(
\begin{tabular}{c}
$\theta^{\rm ML}_{1\pcopy}$ \\
$\theta^{\rm ML}_{2\pcopy}$
\end{tabular}
\right) \\
={}& \mathcal{C}_\pcopy M_\pcopy^T \Sigma_\joint^{-1} \left( x_\joint-m_{\Pi\pcopy}+M_\pcopy \theta_{\Pi\pcopy} \right) \;,  \nonumber
\end{align}
where we have used the definition of the parameter copies covariance $\mathcal{C}_\pcopy^{-1}=M_\pcopy^T \Sigma^{-1}_{\joint} M_\pcopy$ and we have defined the duplicate prior center model prediction $m_{\Pi\pcopy} = (m_{\Pi},m_{\Pi})^T$.
The parameter copy ML is then Gaussian distributed, over the space of data draws, with $\theta^{\rm ML}_{\pcopy} \sim \mathcal{N}(\theta_\pcopy; \theta_{\Pi\pcopy}, \mathcal{C}_\pcopy)$.
The maximum posterior parameters are obtained by adding copies of the Gaussian priors on top of the ML result.
We write explicitly the block structure of the parameter copies posterior covariance as:
\begin{equation}\label{Eq:PosteriorCopyParameterCovariance}
\mathcal{C}_{p\pcopy} \equiv 
\left(
\begin{tabular}{cc}
$\mathcal{C}_{p1\pcopy}$ & $\mathcal{C}_{p12\pcopy}$ \\
$\mathcal{C}_{p21\pcopy}$ & $\mathcal{C}_{p2\pcopy}$
\end{tabular}
\right) \;,
\end{equation}
that allows us to write the posterior estimate for the first parameter copy as:
\begin{align}\label{Eq:PosteriorParameterEstimatesParamCopy}
\theta_{1\pcopy}^p = \theta^{\rm ML}_{1\pcopy} &-\mathcal{C}_{p1\pcopy} \mathcal{C}_{\Pi}^{-1}(\theta^{\rm ML}_{1\pcopy}-\theta_\Pi) \nonumber \\
&-\mathcal{C}_{p12\pcopy} \mathcal{C}_{\Pi}^{-1}(\theta^{\rm ML}_{2\pcopy}-\theta_\Pi) \,,
\end{align}
and similarly for the second parameter copy.

\begin{figure}[!h]
\centering
\begin{subfigure}
  \centering
  \includegraphics[width=\columnwidth]{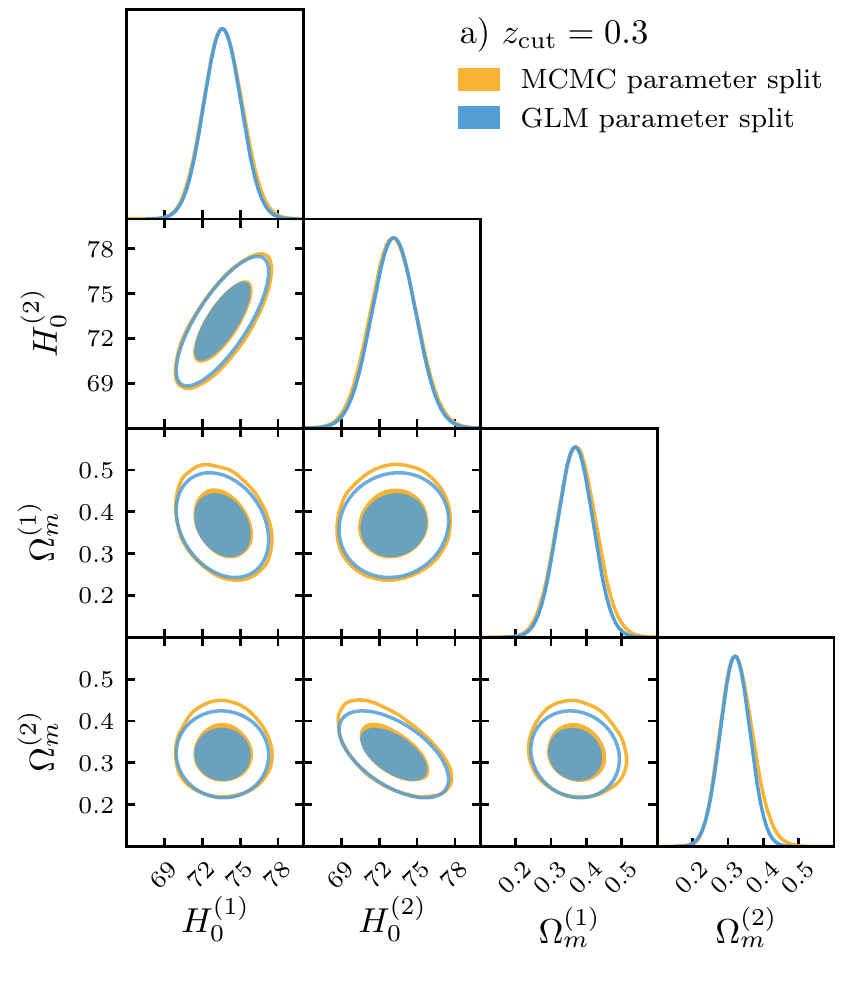}
\end{subfigure}
\begin{subfigure}
  \centering
  \includegraphics[width=\columnwidth]{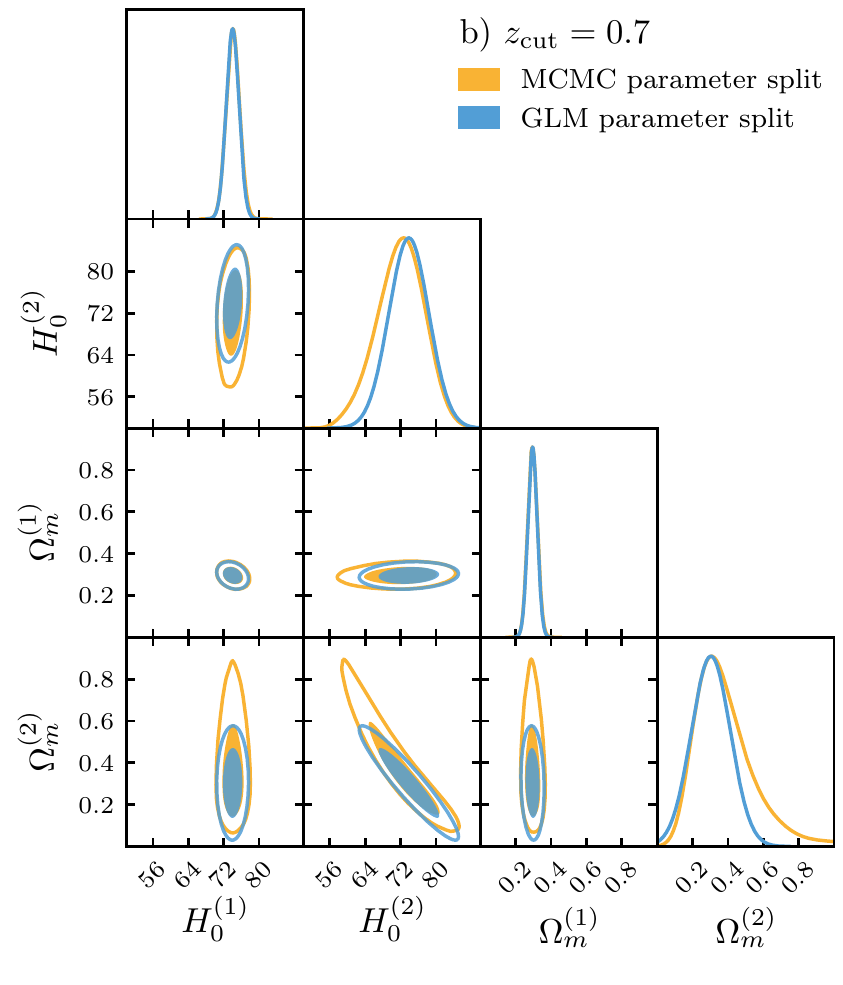}
\end{subfigure}
\caption{ \label{Fig:GLMMCMCComparison}
The comparison of the GLM and MCMC joint marginalized posterior parameter distribution for SN parameter splits. 
Different panels show different SN redshift cuts.
The filled contour corresponds to the 68\% C.L. region while the continuous contour shows the 95\% C.L. region.
}
\end{figure}
\begin{figure}[!ht]
\centering
\includegraphics[width=\columnwidth]{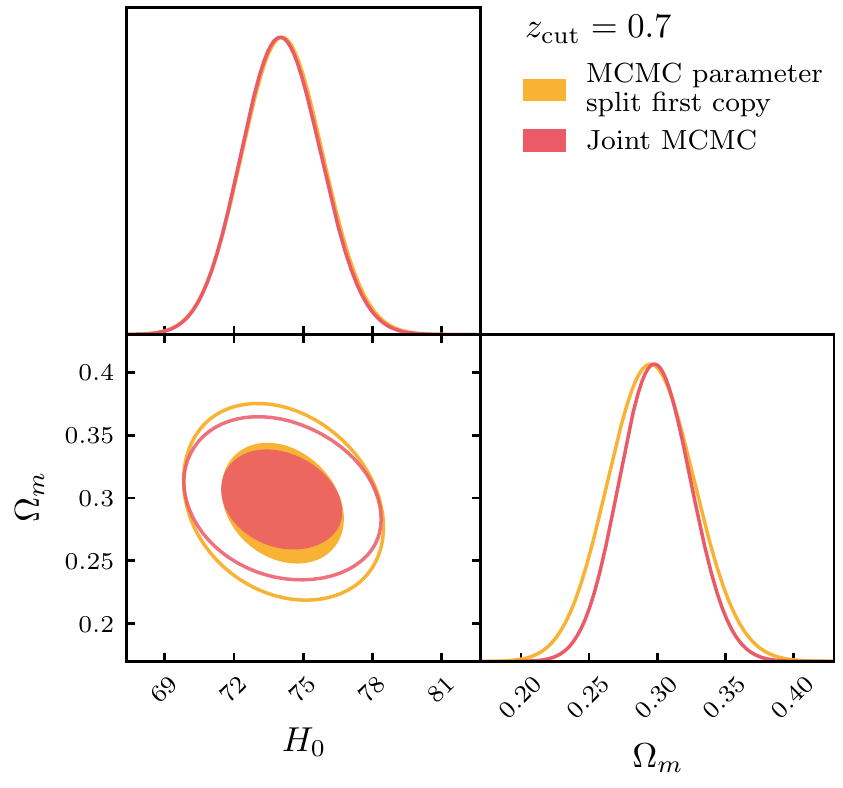}	
\caption{
The comparison of the MCMC joint marginalized posterior parameter distribution for the full SN data set and the low redshift end of the $z_{\rm cut}=0.7$ split.
The filled contour corresponds to the 68\% C.L. region while the continuous contour shows the 95\% C.L. region.
}
\label{Fig:MCMCCopyJointComparison}
\end{figure}

As we can see the GLM posterior distribution for one of the parameter copies is related to the parameters of the other copy in two ways: first indirectly in the ML estimate and then directly at the MAP level.
In particular, at the ML level, the parameters of one copy are related to the parameters of the other because of the shared data in
Eq.~(\ref{Eq:CopyMLParameters}). 
This is a natural consequence of the parameter duplication technique: since we always fit the joint data set, split parameters are influenced by data in both splits through their correlation.

In the posterior, there is a direct coupling between the ML and posterior estimators of the
copies.   In Eq.~\eqref{Eq:PosteriorParameterEstimatesParamCopy} this coupling is mediated by $\mathcal{C}_{p12\pcopy}$ in the last term. 
In the limit where the data is uncorrelated this term would vanish and, therefore, the  sets would not be able to communicate with each other; we would, therefore, retrieve the expressions in \cite{Raveri:2018wln}, which would also be the same as the corresponding expressions under the data split methodology.  With correlated copy parameters, the
maximization of the {\it joint} posterior depends on the ML values of each.
Contrast this with the case of the separate parameters of the data split.   
Even though the ML values are still correlated according to Eq.~(\ref{Eq:DataSplitParameterCovariance}), maximization of the posterior for each parameter is performed without reference or knowledge of its complement. 

The joint parameter results can be viewed as a subspace of the parameter copies where all the copies have the same value.
We define the projection matrix on this subspace as:
\begin{equation}\label{Eq:CopySameProj}
D_{\pcopy}^T = (\mathbb{I}_{N \times N}, \mathbb{I}_{N \times N})\; .
\end{equation}
When expressed as  $D_{\pcopy}$, it is known as the design matrix, which takes a single set of $N$ parameters and produces two separate parameters, i.e.\ the $2N$  copy parameters, to be estimated by the data.
Notice that the relation between the joint and copy Jacobian is given by $M_\joint^T = D_{\pcopy}^T M_\pcopy^T$. 
The joint parameter covariance is a linear combination of the copy parameter covariance given by
$\mathcal{C}_\joint^{-1} = D_{\pcopy}^T \mathcal{C}_\pcopy^{-1} D_{\pcopy}$.
The linear combination of the copy parameter estimates that forms the joint parameter estimate is:
\begin{equation}\label{Eq:JointFromCopy}
\theta_\joint^{\rm ML} = \mathcal{C}_\joint D_{\pcopy}^T \mathcal{C}_\pcopy^{-1} \theta^{\rm ML}_{\pcopy} \,.
\end{equation}
Likewise $D_{\pcopy}\mathcal{C}_\joint D_{\pcopy}^T \mathcal{C}_\pcopy^{-1}$ is the parameter projector that projects the copy parameters onto the space where they 
are the same $(\theta_{1C}^{\rm ML},\theta_{2C}^{\rm ML})^T \rightarrow (\theta_{\joint}^{\rm ML},\theta_{\joint}^{\rm ML})^T$.

The physical meaning of the parameter estimates between  the data split and parameter duplication approaches in principle differs in the presence of data correlations.  In App.~\ref{App:ExampleDataParamSplit} we present a simple example which illustrates these differences.
Here we would like to emphasize that 
 the data split strategy provides us with two distributions $P(\theta_{1\single}|D_1)$ and $P(\theta_{2\single}|D_2)$ that are interpreted as the posterior for the parameters of one data set with no knowledge of the other.
On the other hand, the second strategy provides the joint parameter distribution $P(\theta_{1\pcopy},\theta_{2\pcopy}|D_\joint)$ for both data sets.
 When marginalized over one of the parameter copies this gives $P(\theta_{1\pcopy}|D_\joint)$ which is the distribution of the parameters of the first data set, given that the full data set has been measured.

In our example we duplicate all the SN cosmological parameters but we do not duplicate the parameter describing the absolute magnitude. This would be fully correlated, since the corresponding measurements in the joint data set are fully correlated, closely matching the example case discussed in App.~\ref{App:ExampleDataParamSplit}.
As a consequence duplicating the SN calibration parameter and keeping track of its correlations gives the same results as not duplicating it and we omit its duplication for simplicity.
We have checked that all the results that we discuss are unchanged.

We find that the difference between the two ways of estimating parameters from the split SN data is minor.  
In Fig.~\ref{Fig:GLMParameterComparisonTriangle} we show the GLM prediction of the parameter posterior in the two split senses that we have discussed.

In the $z_{\rm cut}=0.3$ case we observe very little difference between the two parameter estimates.
This shows that, even though data correlations are relevant, for this data split, the two techniques do not strongly differ.
The $z_{\rm cut}=0.7$ case, on the other hand, shows somewhat more differences. 
In this case the high $z$ part of the SN catalog is significantly weaker than the low $z$ part and correlated data modes become more relevant making the two techniques more different. 
As a result we see a shift in the MAP estimate and a small decrease in their covariance.
In this case, in fact, the weaker split leverages the correlation with the strongest one to achieve slightly smaller error bars and parameter estimates that are closer to the ones of the strongest split.
We refer the reader to App.~\ref{App:ExampleDataParamSplit} for an in depth discussion of these effects, explained through a simple example.

Next we consider the extent to which the GLM model works overall to describe the redshift splits that we consider.
To this end, notice that the joint parameter distribution cannot be obtained with standard parameter estimation techniques in the data split case, for which we use only the GLM, while it can be straightforwardly obtained in the parameter split case.
We find that in the $z_{\rm cut}=0.3$ case the GLM works remarkably well, as shown in Fig.~\ref{Fig:GLMMCMCComparison} where we compare the GLM posterior to the MCMC posterior.
Both ends of the split have comparable constraining power and contain enough SN measurements to constrain both amplitude and shape parameters.
The high redshift end of the split has fewer SN measurements and hence shows hints of a slight non-Gaussian decay of the probability tail of $\Omega_m^{(2)}$.

The $z_{\rm cut}=0.7$ split on the other hand is different. The high redshift part of the data set contains few SN and the amplitude/shape degeneracy is far less constrained.
As we can see in Fig.~\ref{Fig:GLMMCMCComparison}, especially for the joint $\Omega_m^{(2)}$ and $H_0^{(2)}$ distribution the GLM contours are clearly different from the MCMC ones.
In particular we see high non-linearities in the model (i.e.~a markedly ``banana'' shaped degeneracy) that, when marginalized over, result in significant skewness of the 1D posteriors.
The $z_{\rm cut}=0.7$ case will then serve as a good example of how some tension estimators have built in strategies to mitigate  these types of non-Gaussianities.

Finally we remark that, in contrast, the posterior of the low redshift end and the joint SN posterior, shown in Fig.~\ref{Fig:MCMCCopyJointComparison}, is very close to Gaussian.
Both redshift splits, in fact, contain a large number of SN that is sufficient to shrink the measured errors so that the model does not explore its non-linear part.
For these two posteriors we expect the GLM to work well in the $z_{\rm cut}=0.7$ case too.

\section{Data Split CDEs} \label{Sec:CDEdatasplit}

In this section we discuss CDEs in case of data splits and show their application to the SN example.
Specifically, in Sec.~\ref{Sec:DataSplitParameterShift} we present parameter shift statistics and 
in Sec.~\ref{Sec:DataSplitGOFloss} we discuss Goodness of Fit loss.

\subsection{Parameter shifts} \label{Sec:DataSplitParameterShift}

Given two data sets we can compute the difference between the parameters obtained by considering the two data sets alone: $\Delta\theta_\single \equiv \theta_{1\single}^p-\theta_{2\single}^p$.
Within the GLM this is Gaussian distributed and it can be shown that its expectation value over data realizations is zero.
To form the optimal quadratic form to detect shifts in parameters,
\begin{equation}
Q_{\rm DM}^\single \equiv \Delta\theta_\single^T [\mathcal{C}(\Delta\theta_\single)]^{-1} \Delta\theta_\single,
\end{equation}
 we need to compute the parameter difference covariance $\mathcal{C}(\Delta\theta_\single)$.
For a discussion of optimal quadratic forms see App. D in~\cite{Raveri:2018wln}.
Within the GLM the shift covariance can be obtained starting from the covariance in data space and results in:
\begin{align} \label{Eq:SingleShiftCovariance}
\mathcal{C}(\Delta\theta_\single) =&\, \mathcal{C}_{p1\single} + \mathcal{C}_{p2\single} - \mathcal{C}_{p1\single}\mathcal{C}_\Pi^{-1} \mathcal{C}_{p2\single} - \mathcal{C}_{p2\single} \mathcal{C}_\Pi^{-1} \mathcal{C}_{p1\single} \nonumber \\
&\, - \mathcal{C}_{p1\single} M_1^T \Sigma_1^{-1} \Sigma_{12} \Sigma_2^{-1} M_2 \mathcal{C}_{p2\single} \nonumber \\
&\, - \mathcal{C}_{p2\single} M_2^T \Sigma_2^{-1} \Sigma_{21} \Sigma_1^{-1} M_1 \mathcal{C}_{p1\single} \,.
\end{align}
As we can see this expression agrees with~\cite{Raveri:2018wln} in the limit of uncorrelated data sets. It cannot be, however, expressed in terms of parameter space quantities only when data correlations are present.
In this case the parameter shift covariance depends on both the parameter and data covariance that are connected through the model Jacobian to account for the fact that data correlations are omitted from the single parameter estimates.

In addition to this, we can also write parameter shifts in update form, by comparing the parameters of one posterior (for simplicity $1$ here) to the joint parameter determination: $\Delta\theta^U_\single \equiv \theta_{1\single}^p - \theta_{\joint}^p$.
This is, again, Gaussian distributed with zero mean and covariance:
\begin{align} \label{Eq:SingleUpdateShiftCovariance}
\mathcal{C}(\Delta\theta^U_\single) =&\, \mathcal{C}_{p1\single} +\mathcal{C}_{p\joint} -\mathcal{C}_{p1\single} \mathcal{C}_\Pi^{-1} \mathcal{C}_{p\joint}
-\mathcal{C}_{p\joint} \mathcal{C}_\Pi^{-1}\mathcal{C}_{p1\single} \nonumber \\
&\, -\mathcal{C}_{p1\single} M_1^T \Sigma_1^{-1} (\Sigma_1, \Sigma_{21})^T \Sigma_\joint^{-1} M_{\joint} \mathcal{C}_{p\joint} \nonumber \\
&\, -\mathcal{C}_{p\joint} M_\joint^T \Sigma_\joint^{-1} (\Sigma_1, \Sigma_{12}) \Sigma_1^{-1}M_1\mathcal{C}_{p1\single} \,.
\end{align}
This agrees with~\cite{Raveri:2018wln} in the limit of uncorrelated data sets, where we recover $\mathcal{C}(\Delta\theta^U_\single) = \mathcal{C}_{p1\single} - \mathcal{C}_{p\joint}$, but becomes significantly more complicated in general due to the presence of data correlations.
We denote with 
\begin{equation}
Q_{\rm UDM}^\single \equiv (\Delta\theta^U_\single)^T [\mathcal{C}(\Delta\theta^U_\single)]^{-1} \Delta\theta^U_\single,
\end{equation}
 the optimal data split parameter shift statistics in update form.
Under the GLM, both $Q_{\rm DM}^\single$ and $Q_{\rm UDM}^\single$ are chi-squared distributed with number of degrees of freedom equivalent to the rank of their respective covariance matrix.

In case of uncorrelated data sets the statistical significance of parameter shifts in update form is the same as the statistical significance of the difference between $\theta_{1}^p-\theta_2^{\rm ML}$ since these two quantities are related by a linear transformation. 
However, in the presence of data correlations this is not the case since the single parameters do not contain the information on the data correlation that is contained in the joint parameter determination. 
In other words, it is not possible to write the update parameter shift as a linear combination of the shift in the two single parameters. Hence, we would expect to see some differences between the two estimates, related to the presence of correlated data and parameter modes.

From the previous discussion it appears clear that using the optimal, inverse covariance weighted, CDEs for data split parameter shifts is challenging in presence of data correlations.
Their covariances cannot be written in parameter space and depend on both the posterior and data covariance. These can be related to each other by projection operations involving derivatives of the observables that are cumbersome to compute accurately.
These considerations limit the applicability of these methods in practice.

The SN case is, however, simple enough that we can successfully apply these estimators within the GLM.
In the reminder of this section we present the challenges in applying them to the SN case and we comment on the results.

The SN data Jacobian, $M_\joint$, is estimated numerically by linear finite differences computed around the best fit of the joint SN data set. The finite difference parameter step is computed such that it would correspond to a SN chi-square difference of one, ensuring that the derivatives are estimated on the scale at which they are relevant and are not contaminated by numerical noise.
We assume that the model is fully linear so that the joint Jacobian determines the single data split Jacobian.

All the other quantities that are needed to compute $Q_{\rm DM}^\single$ and $Q_{\rm UDM}^\single$ are estimated from the GLM.
A numerically challenging aspect of  computing $Q_{\rm DM}^\single$ and $Q_{\rm UDM}^\single$ is identifying directions that can contribute to  parameter shifts and those that do not.
The latter parameter combinations can be either prior constrained or fully correlated, as can be seen from Eq.~(\ref{Eq:SingleShiftCovariance}) and Eq.~(\ref{Eq:SingleUpdateShiftCovariance}).
In practice, due to numerical noise, the parameter shift covariances are never exactly zero along these directions.  

In the uncorrelated case this problem is solved, at least for parameter shifts in update form, by computing the quadratic form using the Karhunen-Loeve (KL) decomposition of the covariances involved, as discussed in~\cite{Raveri:2018wln}.
In this case we select the directions that are used to compute the two parameter shift estimators based on the KL decomposition of the shift covariance matrices and the parameter covariance of the most constraining of the two data sets.
Once the KL decomposition is performed the spectrum of the KL eigenvalues can be examined to understand if there is a clear separation of modes with KL eigenvalues very close to zero and directions that are significantly different from zero.
This strategy also avoids problems with parameters having different units since the KL modes are invariant under changes of parameter basis. This also results in a wide separation between directions that can and cannot contribute to a shift making it easier to identify and remove the latter.
Once the directions that cannot contribute any shift are isolated and removed the parameter shifts and their covariance are both projected on the other directions and $Q_{\rm DM}^\single$ and $Q_{\rm UDM}^\single$ are computed.
The number of degrees of freedom of the two tests is given by the number of KL modes that are retained. 
In the SN example, this number is two, since the absolute magnitude constraint does not differ between the sets.

\begin{table}[ht!]
\begin{ruledtabular}
\begin{tabular}{ l c c c c c c }
\multicolumn{4}{l}{a) Data split GLM difference in means} \\
\hline \hline
\textrm{Redshift cut} & $Q_{\rm DM}^\single$ & dofs &  $P(Q_{\rm DM}^\single > Q_{\rm DM\,obs}^\single)$ \\
\colrule
$z_{\rm cut} = 0.3$ & $4.94$ & $2$ & $8.5\%$ \\
\colrule
$z_{\rm cut} = 0.7$ & $0.11$    & $2$ & $94.5\%$ \\
\hline \hline
\multicolumn{4}{l}{b) Data split GLM update difference in means} \\
\hline \hline
\textrm{Redshift cut} & $Q_{\rm UDM}^\single$ & dofs &  $P(Q_{\rm UDM}^\single > Q_{\rm UDM\,obs}^\single)$ \\
\colrule
$z_{\rm cut} = 0.3$ & $3.52$ & $2$ & $17.2\%$  \\
\colrule
$z_{\rm cut} = 0.7$ & $0.53$ & $2$ & $76.8\%$
\end{tabular}
\end{ruledtabular}
\caption{ \label{Tab:DataSplitParamShiftresults}
Results of the application of the data split parameter shift estimator in normal, $Q_{\rm DM}^\single$, and update form, $Q_{\rm UDM}^\single$, to the SN split that we consider.
The first column shows the SN split that is being considered while the second column reports the value of the $Q_{\rm DM}^\single$ 
and $Q_{\rm UDM}^\single$ parameter shifts statistics.
The third column shows the number of degrees of freedom of the two statistics and the fourth column the probability to exceed the measured value of $Q_{\rm DM}^\single$ and $Q_{\rm UDM}^\single$.
All quantities used to compute the results in this table are obtained with the GLM.
}
\end{table}

In Tab.~\ref{Tab:DataSplitParamShiftresults} we show the results of the application of the data split parameter shift statistics. 
Notice that with the data split strategy some quantities entering in the calculation of the results cannot be obtained from MCMC sampling so we  estimate the results with the GLM only.

When we consider parameter differences in update form we always pick the low redshift SN cut as the base parameters for the update since, among the two possibilities, it contains a larger number of SN measurements and is hence more Gaussian.

As we can see the statistical significance of the reported results differs for the two estimators, as expected because of non-negligible data correlations.
The $z_{\rm cut}=0.3$ split, in particular is found to be in agreement in both cases, with slightly different statistical significance.
The second SN split, at $z_{\rm cut}=0.7$, on the other hand, is found to have parameters that are too close, with respect to their covariance, at $94.5\%$ probability in normal form while in agreement in update form. 
Since both results are computed within the GLM, and the prior is only weakly informative, the difference between the two estimates is given by different weighting of correlated data modes and reflects the fact that, in presence of data correlations, 
$Q_{\rm DM}^\single$ and $Q_{\rm UDM}^\single$ are not expected to give the same results.

\subsection{Goodness of fit loss} \label{Sec:DataSplitGOFloss}

In addition to shifts in parameters we can use, as a CDE, the statistics of the ratio of the joint and single likelihoods at maximum posterior, $Q_{\rm DMAP}$~\cite{Raveri:2018wln}.
In the case where we consider data split we refer to this estimator as:
\begin{equation}
Q_{\rm DMAP}^\single \equiv 2 \ln \mathcal{L}_1(\theta_{p1}^\single) +2 \ln \mathcal{L}_2(\theta_{p2}^\single) -2 \ln \mathcal{L}_\joint(\theta_p^\joint)  \;,
\end{equation}
This quantifies goodness of fit loss as it corresponds to the degradation of the performances of the model when fitting two data sets jointly vs fitting the joint data.
When two data sets are considered separately the model can invest all its parameters in improving the fit to data.
On the other hand, when the two data sets are joined, the parameters have to compromise between the two and the joint fit will be worse.  However note that in the correlated case the two data sets are not independent so that the joint likelihood is not the product of the two independent likelihoods regardless of the parameters.
Consequently $Q_{\rm DMAP}^\single$ is not necessarily positive definite, complicating its interpretation as a goodness of fit loss.

Even computing the statistics of $Q_{\rm DMAP}^\single$ for correlated data sets, within the GLM,  proves extremely hard in case of data set splits.
In App.~\ref{App:QDMAPdataExact} we report its statistics and further elaborate on the technical difficulties in practically computing it.
Overall the algebraic expressions involved in its calculation are defined, as it happens for data split parameter shifts, in terms of quantities living in both parameter and data space.

Despite their complicated nature, in App.~\ref{App:QDMAPdataExact} we provide the full expressions that are necessary in order to compute the exact distribution of the GoF loss statistic with the data split technique. 
We have, furthermore, made use of these expressions to show that the distribution would not be, necessarily, well approximated with a simple chi-square distribution.

\section{Parameter Split CDEs} \label{Sec:CDEparamsplit}

In this section we follow the strategy of quantifying agreement and disagreement by duplicating model parameters. We first go through the analytic aspects of calculating the CDEs and then report the results of applying them to the SN example that we consider in this work.
In Sec.~\ref{Sec:ParamSplitParameterShift} we present parameter shift statistics, 
in Sec.~\ref{Sec:ParamSplitMCMCParameterShift} we discuss exact Monte Carlo parameter shift statistics, 
while in Sec.~\ref{Sec:ParamSplitGOFloss} we consider goodness of fit loss.

\subsection{Parameter shifts} \label{Sec:ParamSplitParameterShift}

We first consider the difference between the duplicate parameter posteriors, denoted by $\Delta\theta_\pcopy \equiv \theta_{1\pcopy}^p-\theta_{2\pcopy}^p$, in the case of parameter splits. 
To form the optimal estimator of the significance of the shifts, we construct the quadratic form:
\begin{equation}
Q_{\rm DM}^\pcopy \equiv (\Delta\theta_\pcopy)^T [\mathcal{C}(\Delta\theta_\pcopy)]^{-1} \Delta\theta_\pcopy
\label{eq:QDMform}
\end{equation}
using their covariance to weight shifts in different parameter space directions.
In this case the covariance reads:
\begin{equation} \label{Eq:CovarianceShiftCopy}
\mathcal{C}(\Delta\theta_\pcopy) = \mathcal{C}_{p1\pcopy} + \mathcal{C}_{p2\pcopy} - \mathcal{C}_{p12\pcopy} - \mathcal{C}_{p21\pcopy} \;.
\end{equation}
Notice that, in the uncorrelated limit $\mathcal{C}_{pij\pcopy} = \mathcal{C}_{piC} \mathcal{C}_\Pi^{-1} \mathcal{C}_{pjC}$ for $i,j\in [1,2]$.
Furthermore, Eq.~\eqref{Eq:CovarianceShiftCopy} implies that in the case of parameter duplication we can express the covariance of the parameter shifts using just the results from the MCMC chains. This is not true in the case of data splits, however, where the expression of the covariance includes terms related to the data covariance which account for the correlations.

We then calculate the covariance of parameter shifts in update form using one of the two parameter copies, namely $\theta_{1\pcopy}^p$, and the parameters inferred from the joint data set, $\theta_{\joint}^p$. Therefore, defining $\Delta\theta_\pcopy^U = \theta_{1\pcopy}^p-\theta_{\joint}^p$, the covariance of parameter shift in update form is written as
\begin{align} \label{Eq:CovarianceUpdateShiftCopy}
\mathcal{C}(\Delta\theta_\pcopy^U) = \mathcal{C}_{p1\pcopy} -\mathcal{C}_{p\joint} \,,
\end{align}
which is invariant in form to the one without correlations, which is discussed in~\cite{Raveri:2018wln}. 

We denote with 
\begin{equation}
Q_{\rm UDM}^\pcopy \equiv (\Delta\theta_\pcopy^U)^T [\mathcal{C}(\Delta\theta_\pcopy^U)]^{-1} \Delta\theta_\pcopy^U
\label{eq:QUDMform}
\end{equation}
 the optimal parameter-split parameter shift statistic in update form.
Notice that, since Eq.~(\ref{Eq:CovarianceUpdateShiftCopy}) is invariant in form with respect to the uncorrelated case considered in~\cite{Raveri:2018wln}, we can compute $Q_{\rm UDM}^\pcopy$ by means of the KL decomposition to filter out modes that are not improved by the data over the prior and hence subject to sampling noise. 
Under the GLM, both $Q_{\rm DM}^\pcopy$ and $Q_{\rm UDM}^\pcopy$ are chi-squared distributed with number of degrees of freedom equal to the rank of their covariances.

The statistical significance of the two $Q_{\rm DM}^\pcopy$ and $Q_{\rm UDM}^\pcopy$ estimators is the same for the maximum likelihood parameters while it might differ at the maximum posterior level in case of partially informative priors.
This difference stems from the fact that the update form of parameter shifts contains only one copy of the prior in the joint, whereas in the single parameter shift the prior is applied once to each data set.  Therefore, $\theta^p_J$ cannot be formed from a linear combination of
$\theta_{iC}^p$.   We can instead define 
a  joint parameter estimate that is so constructed
\begin{equation}
\tilde \theta_{\joint}^p = \tilde C_{p\joint} D_\pcopy^T \mathcal{C}_{p\pcopy}^{-1} ( \theta_{1\pcopy}^p ,\theta_{2\pcopy}^p)^T
\end{equation}
with covariance  $\tilde{\mathcal{C}}_{p\joint}^{-1} = \mathcal{C}_{p\joint}^{-1} + \mathcal{C}_\Pi^{-1} = D_\pcopy^T \mathcal{C}_{p\pcopy}^{-1} D_{\pcopy}$,
so that 
\begin{equation}\label{Eq:newParamSplitUpdateDifference}
\theta_{1\pcopy}^p - \tilde{\theta}_{\joint}^p = \tilde{\mathcal{C}}_{p\joint} D_\pcopy^T \mathcal{C}_{p\pcopy}^{-1} (\mathbb{O}, \theta_{1\pcopy}^p - \theta_{2\pcopy}^p)^T \; ,
\end{equation}
where the vector $\mathbb{O}$ has length $N_p$. 
This clearly shows that the statistical significance of $\theta_{1\pcopy}^p - \tilde{\theta}_{\joint}^p$ is the same as $\theta_{1\pcopy}^p - \theta_{2\pcopy}^p$ since the two are related by a linear, invertible transformation. 
We can then write the update parameter difference as:
\begin{equation}\label{Eq:ParamSplitUpdateShiftTilde}
\Delta\theta_C^U
= (\theta_{1\pcopy}^p - \tilde{\theta}_{\joint}^p) + (\tilde{\theta}_{\joint}^p - \theta_{\joint}^p) \;,
\end{equation}
which, in the uncorrelated limit reduces to $\theta_{1\pcopy}^p - \theta_2^{\rm ML}$.  This agrees with the discussion in \cite{Raveri:2018wln} of their Eq.~(47). 
More generally, the additional difference can be computed from $C_{p\joint}$ and $C_\Pi$ and can
cause $\Delta\theta_C^U$ to be larger than the difference implied by $\theta_{1\pcopy}^p - \theta_{2\pcopy}^p$ since the Gaussian priors in each copy tend to bring the posteriors closer together.  Note that for flat, range bound, priors as in our SN example the two copies do not lead to a stronger joint prior so that $\tilde \theta_J^p = \theta_J^p$.

Furthermore the difference between $Q_{\rm DM}^\pcopy$ and $Q_{\rm UDM}^\pcopy$ becomes relevant only if there is a non-negligible shift along partially prior constrained directions since the two estimators agree in the fully data and prior constrained limits.

It is clear at this point that making use of the parameter split methodology provides some advantages compared to the data splitting method.   Equations \eqref{Eq:CovarianceShiftCopy} and \eqref{Eq:CovarianceUpdateShiftCopy} for the covariances for the parameter 
split statistics should be compared with Eqs.~\eqref{Eq:SingleShiftCovariance} and \eqref{Eq:SingleUpdateShiftCovariance} for data split statistics.   Crucially the former can be simply calculated from
parameter covariances whereas the latter require manipulations of the data covariance.
We can also therefore check the GLM results using parameter covariances taken from the MCMC chain when evaluating Eqs.~(\ref{eq:QDMform}) and (\ref{eq:QUDMform}).

\begin{table}[t!]
\begin{ruledtabular}
\begin{tabular}{ l c c c c c c }
\multicolumn{4}{l}{a) Parameter split GLM difference in means} \\
\hline \hline
\textrm{Redshift cut} & $Q_{\rm DM}^\pcopy$ & dofs &  $P(Q_{\rm DM}^\pcopy > Q_{\rm DM\,obs}^\pcopy)$ \\
\colrule
$z_{\rm cut} = 0.3$ & $4.51$ & $2$ & $10.5\%$ \\
\colrule
$z_{\rm cut} = 0.7$ & $0.03$    & $2$ & $99.0\%$ \\
\hline \hline
\multicolumn{4}{l}{b) Parameter split GLM update difference in means} \\
\hline \hline
\textrm{Redshift cut} & $Q_{\rm UDM}^\pcopy$ & dofs &  $P(Q_{\rm UDM}^\pcopy > Q_{\rm UDM\,obs}^\pcopy)$ \\
\colrule
$z_{\rm cut} = 0.3$ & $4.52$ & $2$ & $10.4\%$ \\
\colrule
$z_{\rm cut} = 0.7$ & $0.03$    & $2$ & $99.0\%$ \\
\hline \hline
\multicolumn{4}{l}{c) Parameter split MCMC difference in means} \\
\hline \hline
Redshift cut & $Q_{\rm DM}^\pcopy$ & dofs & $P(Q_{\rm DM}^\pcopy > Q_{\rm DM\, obs}^\pcopy)$ \\
\hline \hline
$z_{\rm cut}=0.3$ &  $4.63$ & $2$ & $9.9\%$ \\			
\colrule
$z_{\rm cut}=0.7$ & $0.37$ & $2$ & $83.2\%$ \\
\hline \hline
\multicolumn{4}{l}{d) Parameter split MCMC update difference in means} \\
\hline \hline
Redshift cut & $Q_{\rm UDM}^\pcopy$ & dofs & $P(Q_{\rm UDM}^\pcopy > Q_{\rm UDM\, obs}^\pcopy)$ \\
\hline \hline
$z_{\rm cut}=0.3$ &  $4.75$ & $2$ & $9.3\%$ \\			
\colrule
$z_{\rm cut}=0.7$ & $0.06$ & $2$ & $96.9\%$ \\
\end{tabular}
\end{ruledtabular}
\caption{ \label{Tab:ParamSplitParamShiftresults}
Results of the application of the parameter split parameter shift estimator in normal, $Q_{\rm DM}^\pcopy$, and update form, $Q_{\rm UDM}^\pcopy$, to the SN split that we consider.
The first column shows the SN split that is being considered while the second column reports the value of the $Q_{\rm DM}^\pcopy$ and $Q_{\rm UDM}^\pcopy$ parameter shifts statistics.
The third column shows the number of degrees of freedom of the two statistics and the fourth column the probability to exceed the measured value of $Q_{\rm DM}^\pcopy$ and $Q_{\rm UDM}^\pcopy$.
All quantities used to compute the results in a) and b) are obtained from the GLM while in c) and d) they are obtained from the means and covariances estimated by an MCMC.
}
\end{table}
We now discuss the results obtained from applying the parameter shift estimators to the SN data considered in this work. 
We note here that when we consider parameter differences in update form we always use the low-redshift data part to compare with the joint data set since, compared with the other data subset, it contains a larger number of SNe and is therefore more Gaussian.

The summary of our results is presented in Tab.~\ref{Tab:ParamSplitParamShiftresults}. 
We show in the table the results obtained by doing a full GLM calculation of all covariances and parameter values and the results obtained using the parameter mean and  covariances from the MCMC sampling. 
As we can see in Tab.~\ref{Tab:ParamSplitParamShiftresults} from degree of freedom counting, the number of effective data constrained parameters in this case is the same as in the data-split applications in the previous section (summarized in Tab.~\ref{Tab:DataSplitParamShiftresults}), as expected. 

We can further see that, in the $z_{\rm cut}=0.3$ case, the parameter shift estimates in both standard form and update form agree very well within the GLM since the prior is only very weakly informative.
The difference in mean result also qualitatively agrees with the result of the data split calculation in the previous section.

The results obtained from the MCMC is slightly different from the GLM one because the parameter centers and covariances are computed from the samples and are influenced by slight non-Gaussianities in the distribution while the GLM parameters are obtained within the linear model. 

In the $z_{\rm cut}=0.7$ case the two GLM results, {as they should given the weak priors}, but point toward parameters that are too close to each other in units of their covariances.
Even though the high redshift part of this split is partially prior constrained in the tails of the posterior, the means and covariances are not substantially influenced by the prior.

As we can further see, the MCMC results, in the $z_{\rm cut}=0.7$ case, are significantly different reflecting the fact that non-Gaussianities are more relevant in this case.
In particular the $Q^\pcopy_{\rm DM}$ MCMC result is sensibly lower in statistical significance. 
Of the various estimators for the parameter means and covariances entering into $Q$'s, this the
only one that utilizes those of the high redshift part directly, rather than evaluated at a position  that is influenced by the stronger low redshift data.  In Fig.~\ref{Fig:GLMMCMCComparison}, 
we see that the slowly decaying tails of $H_0^{(2)}, \Omega_m^{(2)}$ increase the MCMC
covariance, separate the means, and misses the fact that the posterior peaks are
anomalously close given their local curvatures.   In update form, the impact of the long
tails is mitigated since it focuses on the peak region that is consistent with both data sets.
We shall see in the next section, this means that the $Q^\pcopy_{\rm UDM}$ MCMC results are more accurate than the purely GLM ones even in such a non-Gaussian case.

We also notice that the overall results in this case are qualitatively different from the data split ones that are less statistically significant.
This might happen because one part of the split is much weaker than the other and the parameter split estimate in this case are more heavily influenced by the strongest data set, as discussed in Sec.~\ref{Sec:DataParamSplits}.
When the data split estimates and parameter split estimates differ they might point toward a problem in the covariance matrix rather than the parameter mean.
This effect is compatible with our goodness of fit results, in Sec.~\ref{Sec:DataParamSplits}, that showed that the high redshift part of the $z_{\rm cut}=0.7$ split leans toward a fit which is too good, at about the same statistical significance.

While our general rule is to compute the update parameter difference by updating the stronger with the weaker data set, for the $z_{\rm cut}=0.3$ case their strengths are nearly
equal.   
We have checked that reversing the ordering to update the high redshift data with the low redshift data does not change the statistical significance appreciably in this case as 
expected.

The last aspect that we can quantify is the error in the assessment of statistical significance that we could have made if we were to neglect the correlation between the two data sets.

For the $z_{\rm cut}=0.3$ split both $Q_{\rm DM}$ and $Q_{\rm UDM}$ would largely misestimate statistical significance resulting in a probability to exceed of $23\%$ and $15\%$ respectively. Notice that the update form mitigates the error since all correlations are still accounted for in the joint estimate.
In the $z_{\rm cut}=0.7$ case, on the other hand, the two estimators neglecting correlations, would estimate a probability to exceed of $99\%$ and $56\%$ which is again largely wrong.

\subsection{Monte Carlo exact parameter shifts} \label{Sec:ParamSplitMCMCParameterShift}

Having an MCMC parameter estimation in the case of parameter duplication presents us with the additional possibility of computing parameter shifts as a Monte Carlo integral, as we discuss in this section.

We first consider the joint posterior probability density function of the two parameter copies $P(\theta_{1\pcopy}^p,\theta_{2\pcopy}^p)$.
We can then calculate the distribution of their difference, denoted by $\Delta \theta_\pcopy \equiv \theta_{1\pcopy}^p-\theta_{2\pcopy}^p$, as the $N$-dimensional convolution integral:
\begin{equation}\label{Eq:ParameterShiftConvolutionIntegral}
P(\Delta \theta_\pcopy) = \int_{V_p} P(\theta_{1\pcopy}^p,\theta_{1\pcopy}^p-\Delta \theta_\pcopy) \, d\theta_{1\pcopy}^p \;,
\end{equation}
over the whole parameter space volume $V_p$.
Note that this equation is general and describes the probability to observe a parameter shift $\Delta \theta_\pcopy$ without assuming the parameters to be independent. 
In the limit of uncorrelated data sets the joint probability distribution in the above expression reduces to $P(\theta_{1\pcopy}^p,\theta_{2\pcopy}^p) = P_1(\theta_{1\pcopy}^p)P_2(\theta_{2\pcopy}^p)$. 

To compute the statistical significance of a shift in parameters we then evaluate the integral:
\begin{equation}\label{Eq:MCMCSiftTension}
S = \int_{P(\Delta \theta_\pcopy) > P(0)}  P(\Delta \theta_\pcopy) \, d\Delta \theta_\pcopy \; ,
\end{equation}
where the volume of integration is defined as the region of parameter space where the probability to get a shift $\Delta \theta_\pcopy$ is above the isocontour of no shift, $\Delta \theta_\pcopy = 0$. 

To form the MCMC chain of parameter differences in the case of correlated data sets we can take, sample by sample, the difference between the first and second copy of the parameters, without changing the weights of the samples.
The result would be the MCMC estimate of the convolution integral in Eq.~(\ref{Eq:ParameterShiftConvolutionIntegral}).
Since the parameter duplication chain is run to convergence in the full $2N$ dimensional space the parameter difference chain is appropriately sampled.

Once we have the samples from the parameter difference probability we can compute the integral in Eq.~(\ref{Eq:MCMCSiftTension}) with a mixture of kernel density estimation (KDE) and MCMC techniques.

The probability of a difference in parameter, for every sample in the difference chain, is estimated through KDE with a Gaussian smoothing kernel that uses the scaled parameter difference covariance.
The smoothing kernel is explicitly given by:
\begin{align} \label{Eq:SmoothingKernel}
K(\Delta\theta_{1\pcopy},\Delta \theta_{2\pcopy}) = \mathcal{N}(\Delta\theta_{1\pcopy};\Delta \theta_{2\pcopy}, \Lambda\,\mathcal{C}(\Delta\theta_\pcopy)\, \Lambda ) \;,
\end{align}
where $\Lambda = {\rm diag}(\sqrt{\lambda})$ is a scaling matrix with $\lambda$ as the smoothing scaling parameter.

We fix this parameter using Silverman's rule of thumb~\cite{wand1994kernel} to:
\begin{align} \label{Eq:SilvermanROT}
\sqrt{\lambda} = \left( \frac{4}{n_{\rm s} (N+2)} \right)^{\frac{1}{N+4}} \,,
\end{align}
where $n_{\rm s}$ is the number of samples in the chain and $N$ is the number of parameters.

For a given MCMC sample $j$ the KDE probability of a shift is given by:
\begin{align} \label{Eq:KDEDifferenceProbability}
P(\Delta\theta_{j\pcopy}) = \frac{1}{\sum_{i=1}^{n_{\rm s}} w_i}\sum_{i=1}^{n_{\rm s}} w_i \, K(\Delta \theta_{j\pcopy}, \Delta \theta_{i\pcopy}) \;,
\end{align}
where $w_i$ denotes the weights of the samples and given that the smoothing kernel is normalized.
Eq.~(\ref{Eq:KDEDifferenceProbability}) is also computed for the zero shift so that the MCMC estimate of the integral in Eq.~(\ref{Eq:MCMCSiftTension}) is given by the number of samples that have a KDE probability of shift above the KDE probability of zero over the total number of samples.

This approach has several advantages.
First, the combination of MCMC and KDE makes the estimate weakly sensitive to the choice of the smoothing kernel.
The amount of over/under smoothing that the kernel might be doing is balanced by the fact that that would also happen for the zero shift estimate and would drop in the difference.
In other words we never just use directly the probability of a zero shift, as obtained from Eq.~(\ref{Eq:KDEDifferenceProbability}), that would largely depend on the smoothing kernel in general, but rather compute how many samples from the distribution are above that probability.
The second advantage is that this parameter shift estimate is now completely accounting for all possible non-Gaussianities in the parameter posterior.

The challenge in using this estimate is that, for statistically significant results, the estimate is likely to be noisy due to the fact that the MCMC chain would have very few samples in the tail of the distribution.

This sampling error can, however, be estimated in two ways.
The first is given by a shot noise estimate, by taking the square root of the number of MCMC samples in the smallest probability tail to account for both tensions and confirmation results.
The second is estimated as the variance of the result across multiple MCMC chains of the same distribution.
In this case we have $n_{\rm chains}$ chains and we compute the shift probability for each of them and then estimate the error as the ratio of the variance across chains, weighted by the number of chains, $\sigma_S^2 = {\rm var} (S)_{\rm chains} /n_{\rm chains}$, since the fiducial result uses all of them and given that different chains are independent.
The two error estimates are usually in good agreement for well converged chains.

\begin{figure*}[!ht]
\centering
\includegraphics[width=\textwidth]{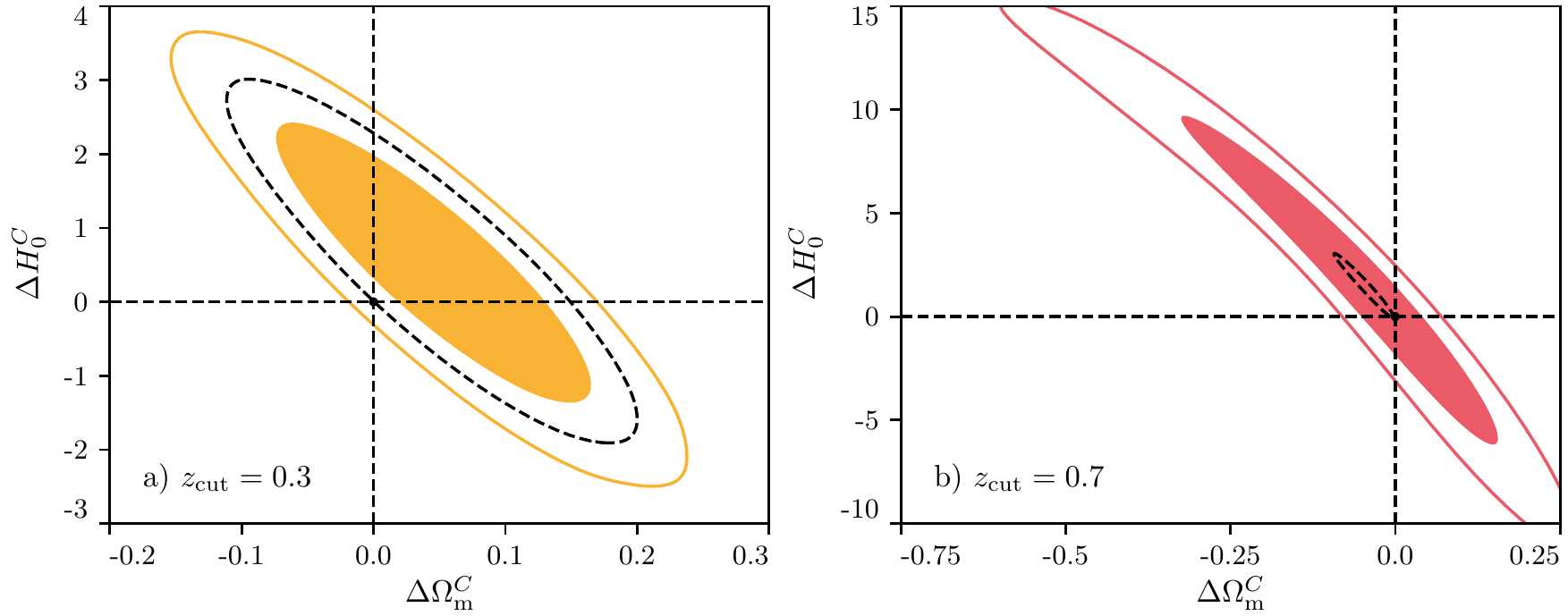}	
\caption{
The joint posterior of parameter split parameter differences for the two SN split that we consider.
The filled contour corresponds to the 68\% C.L. region while the continuous contour shows the 95\% C.L. region.
The dashed lines represent the position of zero shift while the dashed contour shows the probability level that intersects zero, as reported in Tab.~\ref{Tab:MCMCParamSplitParamShiftresults}.
}
\label{Fig:MCMCParameterDifference}
\end{figure*}
\begin{table}[ht!]
\begin{ruledtabular}
\begin{tabular}{ l c c }
\textrm{Redshift cut} & $S$  \\
\colrule
$z_{\rm cut} = 0.3$ & $9.9  \pm 0.4  \%$  \\
\colrule
$z_{\rm cut} = 0.7$ & $96.4 \pm 0.9 \%$ \\
\end{tabular}
\end{ruledtabular}
\caption{ \label{Tab:MCMCParamSplitParamShiftresults}
Results of the application of Monte Carlo exact parameter shift statistics, with the parameter split methodology, to the SN split that we consider. 
The first column shows the redshift of the data spit that we consider. 
The second column shows the result of Eq.~\eqref{Eq:MCMCSiftTension}, measuring the significance of tension or confirmation using exact MCMC techniques. 
The reported uncertainty is an estimate of the sampling error on a given quantity.
}
\end{table}

In Tab.~\ref{Tab:MCMCParamSplitParamShiftresults} we show the results of the MCMC calculation applied to the considered SN splits.
Since the SN parameter space is only two dimensional we can also show, in Fig.~\ref{Fig:MCMCParameterDifference}, the posterior distribution of the difference in parameters and use that to check the reliability of these estimates.

As we can see both results reported in Tab.~\ref{Tab:MCMCParamSplitParamShiftresults} match very well the posterior distribution even though the significance of the shift is not directly computed from that posterior estimate.   
We highlight that this graphical test, which is impossible in higher dimensions, still depends on the KDE smoothing that, in this case, is set to be the optimal one as described in~\cite{Lewis:2019xzd}.

As we can further see, the $z_{\rm cut}=0.3$ results matches the GLM result for $Q_{\rm DM}^\pcopy$ and $Q_{\rm UDM}^\pcopy$ in Tab.~\ref{Tab:ParamSplitParamShiftresults}. This is expected since we have shown that the parameter posterior, in this case, is very close to Gaussian, as also highlighted by the difference posterior in Fig.~\ref{Fig:MCMCParameterDifference}.

In the case of the $z_{\rm cut}=0.7$ split the MCMC result agrees, within sampling errors, with the parameter update result.
This case is, in fact, the most non-Gaussian that we consider, as can also be seen in Fig.~\ref{Fig:MCMCParameterDifference}, so $Q_{\rm DM}^\pcopy$ is expected to misestimate statistical significance.

We can also see from Tab.~\ref{Tab:MCMCParamSplitParamShiftresults} that the estimated sampling errors in the $z_{\rm cut}=0.7$ case are higher than the ones of the $z_{\rm cut}=0.3$ case.
Both chains were run to comparable convergence but the former result is higher in statistical significance. This means that the chain contains less sample in the tail, hence increasing the error estimate.
A smaller error could be achieved by running the second chain longer, at the expense of possibly significant computational resources.

These results also show that mitigation of non-Gaussianities by parameter update statistics computed from the MCMC samples, as in Tab.~\ref{Tab:ParamSplitParamShiftresults}d is working as expected and the two results are compatible within sampling errors.

In any case, when non-Gaussianities are suspected to be relevant, and there is reason to believe that their mitigation from the parameter shift in update form is not enough, the results can be checked with the MCMC techniques that we have just shown.
We, however, highlight that reaching an acceptable noise level in the MCMC estimate, for statistically significant results, requires very long chains to accurately sample the tails of the distribution.

\subsection{Goodness of fit loss} \label{Sec:ParamSplitGOFloss}

The last CDE that we discuss is goodness-of-fit loss with the parameter split approach.
In contrast to the data split case, when considering parameter splits $Q_{\rm DMAP}^C$ becomes easy to compute, as we discuss below.

At maximum likelihood level the statistics of goodness-of-fit loss is chi-squared distributed as a consequence of the fact that the parameter copies decompose the joint parameter estimate, as we show in App.~\ref{App:QDMAPcopyExact}. 

At the posterior level the goodness-of-fit loss statistics is defined as:
\begin{align}
Q_{\rm DMAP}^\pcopy \equiv  2 \ln \mathcal{L}_\joint(\theta^p_\pcopy) -2 \ln \mathcal{L}_\joint(\theta_\joint^p)  \,.
\end{align}
In App.~\ref{App:QDMAPcopyExact} we discuss in detail its exact distribution as a linear combination of chi-squared variables.
In practice the distribution of $Q_{\rm DMAP}^\pcopy$ can be approximated by that of a single chi-squared distribution, matching the mean of the exact distribution, with degrees of freedom:
\begin{align}\label{Eq:DMAPdofs}
\langle Q_{\rm DMAP}^\pcopy  \rangle =& N_{\rm eff}^\pcopy - N_{\rm eff}^\joint  \nonumber \\
& -{\rm tr} [ \mathcal{C}_{\Pi\pcopy}^{-1}\mathcal{C}_{p\pcopy}( \mathbb{I}_{2N}-\mathcal{C}_{\Pi\pcopy}^{-1}\mathcal{C}_{p\pcopy}) \mathbb{J}_{2N}] \;,
\end{align}
where we have defined $\mathbb{J}_{2N} \equiv D_{\pcopy}D_{\pcopy}^T - \mathbb{I}_{2N}$.

As we can see, the statistics of $Q_{\rm DMAP}^\pcopy$ can be easily computed from the posterior MCMC samples.
In the uncorrelated case it also reduces to the $Q_{\rm DMAP}$ statistics discussed in~\cite{Raveri:2018wln}.
Compared to the uncorrelated case, we notice that in the correlated case there is an extra term that is present in the mean of the exact $Q_{\rm DMAP}^\pcopy$ distribution, as shown in Eq.~\eqref{Eq:DMAPdofs}, in addition to the difference in the number of effective parameters.
Notice that this term vanishes for fully data or prior constrained directions.
Its appearance is associated with the mismatch of assuming the data is drawn from a single
parameter and prior but analyzed with split parameters and independent priors.

The results of the goodness of fit loss estimator for our SN analysis are shown in Tab.~\ref{Tab:DataSplitQMAPresults}. 

As we can see, for both SN split, results are in very good agreement with both the results for parameter shifts in update form in the previous sections and the MCMC exact calculations.

In App.~\ref{App:QDMAPcopyExact} we show that the exact distribution, in the SN case, is indeed very well approximated with a chi-square distribution and that results are largely unaltered.

We conclude the section discussing the misestimate of statistical significance that would be made if correlations between data sets were neglected. 
In both cases this would lead to large differences in the results with a probability to exceed the value of $Q_{\rm DMAP}^\pcopy$ at the 
$80\%$ and $25\%$ level for the $z_{\rm cut}=0.3$ and $z_{\rm cut}=0.7$ splits respectively.
In this example the correlation would be playing a crucial role in identifying a statistically significant discrepancy that would not be identified if correlations were not properly accounted for.

\begin{table}[t!]
\begin{ruledtabular}
\begin{tabular}{ l c c c }
Redshift cut & $Q_{\rm DMAP}^C$ & dofs & $P(Q_{\rm DMAP}^C > Q_{\rm DMAP\, obs}^C)$ \\
\hline \hline
$z_{\rm cut}=0.3$ &  $4.55$ & $1.94$ & $90.24\%$ \\			
\colrule
$z_{\rm cut}=0.7$ & $0.03$ & $1.58$ & $3.7\%$ \\
\end{tabular}
\end{ruledtabular}
\caption{\label{Tab:DataSplitQMAPresults}
Results of the goodness-of-fit loss statistic, using parameter duplication, to the SN split considered in this work. 
The first column shows the redshift of the data spit that we consider. 
In the second column we present the results for $Q_{\rm DMAP}^\pcopy$. 
The third column shows the number of degrees of freedom of the $Q_{\rm DMAP}^\pcopy$ statistic, while the fourth column shows the probability to exceed the measured value of $Q_{\rm DMAP}^\pcopy$.
}
\end{table}
%

\section{Conclusions} \label{Sec:Conclusions}

We have introduced, thoroughly discussed, and illustrated with supernovae data, a set of estimators of agreement and disagreement between cosmological data sets in presence of non-negligible data correlations.

In particular we have explored two different approaches. 
The first corresponds to considering separately different correlated data sets and building estimators based on the separate inference of both, while keeping track of data correlations in assessing agreement or disagreement. We called this a data split approach.
A complementary approach, that we refer to as parameter split, consists in splitting the parameter space, duplicating all relevant parameters, always analyzing the joint data set.

Both strategies are equivalent in the limit of vanishing data correlations but differ otherwise, as we have thoroughly explored.
Namely, we have studied and characterized the distribution of parameter shifts estimators and goodness of fit loss estimators with both strategies, discussing differences, their strengths and weaknesses.

We have found that in practical applications the parameter split strategy is easier to implement since it allows us to compute the statistical significance of both tensions and excess confirmation from the output of standard parameter estimation pipelines.

The parameter split strategy also provides a suite of estimators that have complementary properties.
This includes a parameter shift estimator in update form that is mitigated against possible non-Gaussianities of the parameter distributions, while maintaining the ease of application of a Gaussian estimator.
This can be complemented, as we have shown, by a fully MCMC estimator that quantifies parameter shifts regardless of the Gaussianity of the parameter distribution, at the expense of computational power due to the necessity of sampling well the tail of different distributions.
Goodness of fit loss with parameter duplicates further provides a check that the reported results are reliable in a completely different way.

In cases where the parameter posterior is Gaussian and contains parameter space directions that are either fully constrained by the data or the prior the three estimators are expected to give the same assessment of statistical significance, providing an essential cross check of the validity of these assumptions.
When this is true, the different estimators are also optimal having minimum variance among all the possible estimators that one can define.

We have applied, following our discussion, all estimators to the Pantheon SN data set split at two cosmologically relevant redshifts, $z_{\rm cut}=0.3$ and $z_{\rm cut}=0.7$, roughly corresponding to the times of DE-DM equality and the redshift at which cosmic acceleration begins.

We have shown that different data split estimators are not expected to recover the same results in presence of data correlations, even when the Gaussian approximation for the parameter posterior works well.
On the other hand we have found that the parameter split estimators all recover results that are in good agreement, as it is expected, when model parameters are either fully constrained by the data or the prior, as in the SN cases that we consider.

The parameter split estimators report that the two ends of the SN catalog, split at $z_{\rm cut}=0.3$, agree well and show no statistically significant indication of tensions nor excess confirmation.
On the other hand the two part of the high redshift split, at $z_{\rm cut}=0.7$, report excess confirmation at about $96\%$ probability. 
As we have discussed, this could be related either to the covariance of the low-high redshift SN being misestimated, or errors reported  too conservatively.
The latter explanation seems consistent with goodness of fit results indicating that the high redshift end of the $z_{\rm cut}=0.7$ SN split seems too good of a fit to the $\Lambda$CDM model, at about the $94\%$ confidence level.

The SN constraints on the shape of the distance-redshift relation are one of the reasons why late times resolutions of the Hubble constant tensions are not viable~\cite{Aylor:2018drw,Raveri:2019mxg,Knox:2019rjx} and it is hence important to understand the source of the excess goodness of fit in the high redshift part of the Hubble diagram that we report finding.

As the accuracy and complexity of different cosmological measurements grows and in preparation for the analysis of the current and future surveys it is important to solve remaining outstanding issues.
In particular the impact of non-Gaussianities on the behavior of different estimators needs to be understood and fully taken into account.
In addition our work opens the possibility of performing extensive tests of internal consistency of a single data set by splitting it into different parts that would naturally be correlated.
This raises the issue of look-elsewhere corrections for multiple tests being performed on the same data that needs to be fully quantified.
Nonetheless, this work provides important building blocks in this construction by providing estimators of agreement and disagreement between 
correlated cosmological data sets and quantifiable tests of non-Gaussianity in parameter posteriors.

\acknowledgments
We thank 
Dillon Brout, 
Rick Kessler and Dan Scolnic
for useful discussions.
MR was supported in part by NASA ATP Grant No. NNH17ZDA001N, and by funds provided by the Center for Particle Cosmology.
WH and MR were supported by U.S.~Dept.~of Energy contract DE-FG02-13ER41958 and the Simons Foundation.  
Computing resources were provided by the University of Chicago Research Computing Center through the Kavli Institute for Cosmological Physics at the University of Chicago. 

\appendix
%

\section{Canonical correlations} \label{App:CCA}

The strength of the correlation between two data sets corresponds to the magnitude of the cross covariance $\Sigma_{12}$ block in suitable units of $\Sigma_1$ and $\Sigma_2$,
which in general have dimensions $d_1 \times d_2$, $d_1 \times d_1$ and $d_2\times d_2$
respectively.
In one dimension ($d_1=d_2=1$) this would be quantified by the Pearson correlation coefficient:
\begin{align} \label{Eq:CorrelationCoeff1D}
\rho_{12} = \frac{\sigma_{12}^2}{\sigma_1\sigma_2} \,,
\end{align}
where we used the notation $\Sigma_1=\sigma_1^2$, $\Sigma_2=\sigma_2^2$, $\Sigma_{12}=\sigma_{12}^2$ for one dimensional quantities.
The quantity $\rho_{12}$ is bounded to be in $[-1,1]$ with zero meaning absence of correlation, while $1$ and $-1$ indicate complete correlation and anti-correlation respectively.

In multiple dimensions the equivalent procedure is often referred to as Canonical Correlation Analysis (CCA)~\cite{CCA} that we now discuss.

The idea is to look for the optimal linear combination of the two data vectors that maximize the correlation between them. 
If we take two vectors in the space of data, $x_1$ and $x_2$, we can build the quantity:
\begin{equation}\label{Eq:correlation}
\rho_{12} = \frac{x_1^T \Sigma_{12} x_2}{\sqrt{x_1^T \Sigma_1 x_1}\sqrt{x_2^T \Sigma_2 x_2}} \;,
\end{equation}
that we seek to maximize with respect to $x_1$ and $x_2$.

It can be shown that $x_1$, $x_2$ and $\rho_{12}$ are the solutions of the two eigenvalue problems:
\begin{align}\label{Eq:CCAequations}
\Sigma_1^{-1} \Sigma_{12} \Sigma_2^{-1} \Sigma_{21} x_1 = \rho_{12}^2 x_1 \,, \nonumber\\
\Sigma_2^{-1} \Sigma_{21} \Sigma_1^{-1} \Sigma_{12} x_2 = \rho_{12}^2 x_2 \;.
\end{align}
We refer to the solutions of Eq.~\eqref{Eq:CCAequations} as the CCA modes and $-1 < \rho_{12} < 1$ as their correlation coefficient.
Notice that the sign of $\rho_{12}$ is arbitrary and corresponds to a convention for
the relative sign of $x_1$ and $x_2$.
If given data modes $(x_1,x_2)$ are positively correlated, then $(-x_1,x_2)$ are negatively correlated.
When considered this way, as a pair of data vectors spanning the joint space, the CCA modes are
equivalent to the Karhunen-Loeve (KL) modes of $\Sigma_\joint$ and ${\rm diag}(\Sigma_1,\Sigma_2)$, which clarifies their implications for parameter estimation.  
These KL modes, $e^K$, are solutions to the generalized eigenvalue problem,
\begin{equation}
\left(
\begin{tabular}{cc}
$\Sigma_1$ & $\Sigma_{12}$ \\
$\Sigma_{12}$ & $\Sigma_2$
\end{tabular}
\right) e^K = \lambda_K
\left(
\begin{tabular}{cc}
$\Sigma_1$ & 0 \\
0 & $\Sigma_2$
\end{tabular}
\right) e^K \,.
\label{Eq:KLdefn}
\end{equation}
where $K$ indexes the modes.
The KL modes form a  complete and statistically independent basis for the joint data in that their amplitudes have no covariance for both $\Sigma_\joint$ and diag($\Sigma_1,\Sigma_2$).  The KL eigenvalue therefore is the ratio between the variances of these mode amplitudes with and without the $\Sigma_{12}$ correlations.  
Unlike an ordinary eigenvector decomposition, these modes are not orthogonal in the Euclidean sense, but rather orthogonal under the metrics provided by the covariance matrices.

The relationship to the CCA modes is that for each one of the $\min (d_1,d_2)$ unique $|\rho_{12}|$ eigenvalues with CCA solutions $x_1$ and $x_2$ there are two KL modes with $e^K = (x_1,x_2)$ and $(-x_1,x_2)$, and $\lambda_K = 1\pm | \rho_{12}|$.
The remaining  $|d_1-d_2|$ modes are uncorrelated, with $\lambda_K=1$, and have support only across the larger of the two data sets.  

The  impact of correlations on parameters is bounded by the largest correlation or the pair of modes for which $\lambda_K = 1\pm {\rm max}(|\rho_{12}|)$.   This maximal effect occurs if the parameter of interest is exactly one of these two KL mode amplitudes, where
the parameter variance would be misestimated by this factor, with the sign distinguishing an under and over estimate respectively.
Notice that since the maximum possible correlation is itself bounded by $|\rho_{12}| \leq 1$ then $0 \le \lambda_K \le 2$, meaning that the ratio of errors is likewise strictly bounded between $0$ and $\sqrt{2}$.  
Neglecting correlations can make parameter errors at most infinitely overestimated or underestimated by  $\sqrt{2}$.   
The former occurs when such correlations allow a zero noise measurement of a parameter.   
The latter occurs when correlations make the information in the two data sets completely redundant.
We illustrate these ideas with a simple example in App.~\ref{App:ExampleDataParamSplit}.

For a more general parameter of interest which is not exactly a KL mode itself, we can compute the impact of correlations by summing the parameter information in each mode independently.  
Given the sensitivity per KL mode to a parameter vector $\theta$ as
\begin{equation}
c^K \equiv \frac{\partial m}{\partial \theta} e^K = M^T e^K,
\end{equation}
the Fisher matrix is 
\begin{align} \label{Eq:CCAKLFisher}
 \mathcal{F}_{ij} \equiv \sum_K \frac{1}{\lambda_K}c_i^K c_j^K \;,
\end{align}
whereas falsely neglecting the correlations would give
\begin{align} \label{Eq:CCAKLFakeFisher}
\tilde{\mathcal{F}}_{ij}  \equiv \sum_K c_i^K c_j^K \;, 
\end{align}
where the index $K$ runs over the KL modes.  Note that if the KL modes with low
correlation dominate the information on a given parameter, the impact of correlations
on parameter variances decreases from the extreme of $1\pm {\rm max}(|\rho_{12}|)$.  
The Fisher estimate of the parameter covariance is then $\mathcal{C} = \mathcal{F}^{-1}$.

\begin{figure}[!ht]
\centering
\includegraphics[width=\columnwidth]{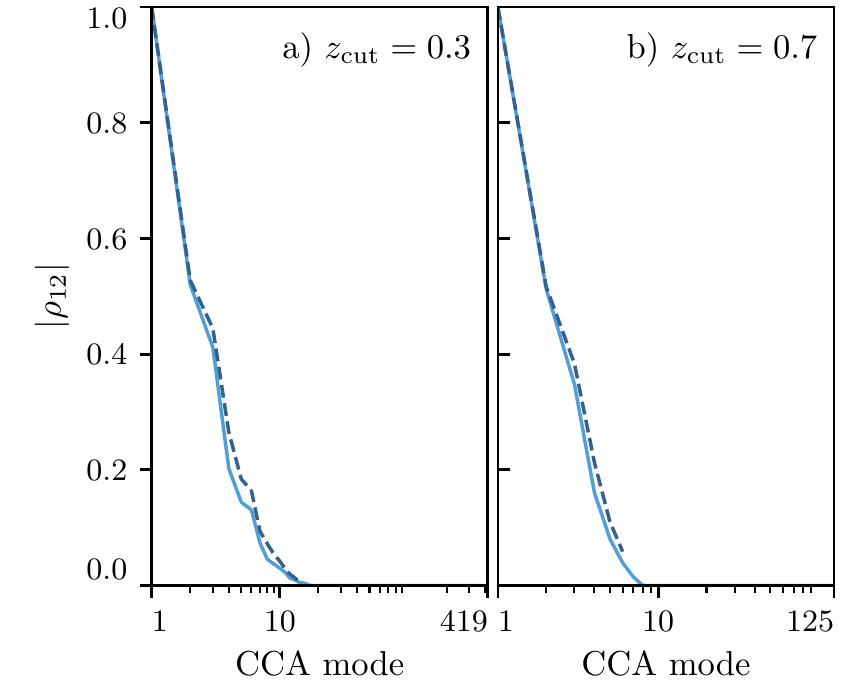}	
\caption{
The spectrum of the absolute value of the canonical correlation coefficients for the two SN data split considered.
The continuous and dashed lines show the results for the unbinned and binned SN data respectively.
}
\label{Fig:CCAspectrum}
\end{figure}

In the remainder of this section we comment on the correlated data modes of the two SN splits that we consider.

In Fig.~\ref{Fig:CCAspectrum} we show the spectra of the correlation coefficients for the two splits.
As we can see the spectra are similar and contain about ten correlated data modes, while the remaining ones are nearly uncorrelated.
In both cases, the first mode is completely correlated with $|\rho_{12}|=1$, and corresponds to the redundant measurement of the SN absolute calibration that is present in both splits.
The second mode has a correlation coefficient of $|\rho_{12}|=0.52$ for the $z_{\rm cut}=0.3$ split and $|\rho_{12}|=0.51$ for the $z_{\rm cut}=0.7$ one and corresponds to the first genuine SN data correlation.
In the same figure we also show the spectrum of canonical correlations for the redshift-binned SN measurements. 
As we can see there are fewer modes, corresponding to fewer data points, but the correlation coefficients are qualitatively unaltered showing that the correlations that we see are not due to  noise in the covariance matrix but rather comes from shared systematic correlations.

\begin{table}[t!]
\begin{ruledtabular}
\begin{tabular}{ c c c c c c c }
Principal component & $\Delta\Omega_m$ & $\Delta h$ & $\Delta \M$ \\
\colrule \hline
PC 1 & $0.99$ & $-0.13$ & $0.06$ \\
\colrule
PC 2 & $-0.14$    & $-0.94$ & $0.32$\\
\colrule
PC 3 & $-0.02$    & $0.33$ & $0.94$\\
\end{tabular}
\end{ruledtabular}
\caption{ \label{Tab:SNprincipalComponents}
The cosmological parameter combinations defining the principal components (PC) of the SN covariance.
}
\end{table}
\begin{figure}[!ht]
\centering
\includegraphics[width=\columnwidth]{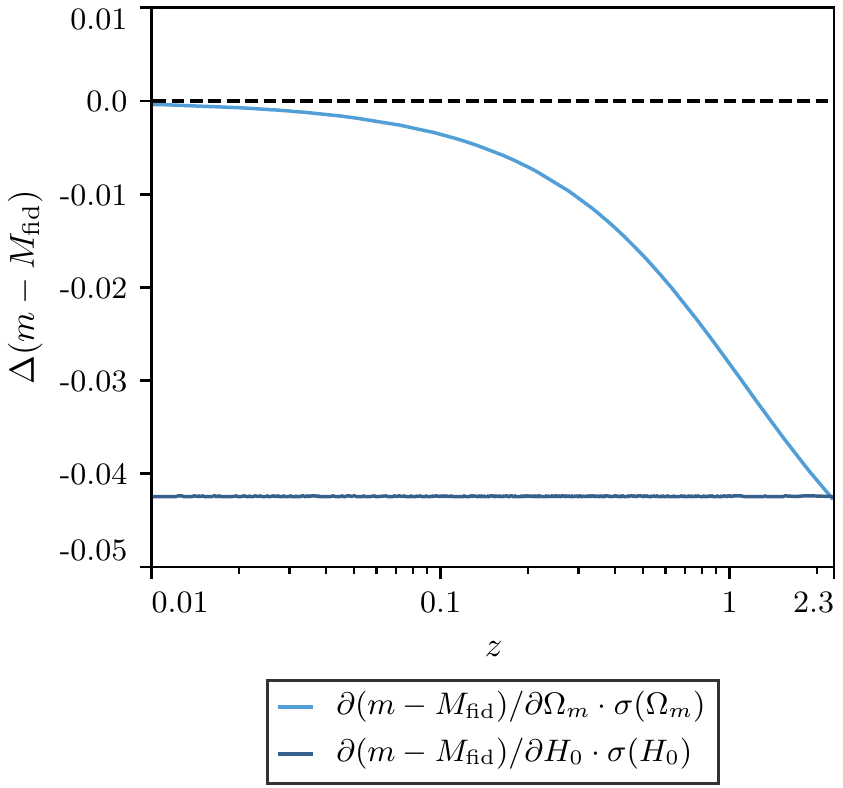}	
\caption{
The Jacobian of the SN magnitude with respect to different cosmological parameters.
Parameters are scaled to have unit variance so that the Jacobian is directly in magnitude units.
Different lines correspond to different parameter variations as shown in legend.
}
\label{Fig:SNParameterJacobian}
\end{figure}

In this regard, the CCA analysis shows that correlations must be kept since there are several
linear combinations of the data whose errors would be severely misestimated otherwise.
On the other hand this does not necessarily mean that cosmological parameters will be equally affected and we now want to quantify the impact that these data correlations have on the determination of cosmological parameters.

In the $\Lambda$CDM model SN magnitudes at different redshifts depend on cosmology through two parameters, $\Omega_m$ and $H_0$.
The effect of variations of these parameters on the SN magnitude is shown in Fig.~\ref{Fig:SNParameterJacobian}.
As we can see these describe variations in the amplitude and shape of $\m-\M_{\rm fid}$  as a function of redshift.
In addition to these parameters we have another parameter that describes the absolute magnitude of the SN with its corresponding data constraint which is required to make inferences
from the distance modulus $\m-\M$.

The inferred errors on the parameters $\Omega_m, H_0, \M$ are therefore 
correlated regardless of whether the SN magnitude measurements are themselves correlated.  In particular, the absolute magnitude calibration is degenerate with $H_0$ and both of these parameters are mildly degenerate with $\Omega_m$, as can also be seen in Fig.~\ref{Fig:JointCorrelationPosteriorComparison}.

To make the impact of correlations on parameters clear, and mostly unaltered by marginalization, we now perform a principal component (PC) analysis on the SN covariance.
Note that we compute the principal components of the covariance after transforming $H_0$ to $h\equiv H_0/(100 \; {\rm km} \; {\rm s}^{-1} \; {\rm Mpc}^{-1})$ to have a dimensionless set
of parameters with comparable scalings.  

The parameter coordinates of the SN principal components are reported in Tab.~\ref{Tab:SNprincipalComponents}.
As we can see the first PC is mostly influenced by changes in $\Omega_m$.
The second PC is not cosmologically interesting but is a highly constrained direction where changes to both $\M$ and cosmological parameters make all SN brighter or dimmer.
The third PC corresponds to the direction along the $\M$-$H_0$ degeneracy where measurements of $\M$ determine $H_0$.

\begin{figure}[!ht]
\centering
\includegraphics[width=\columnwidth]{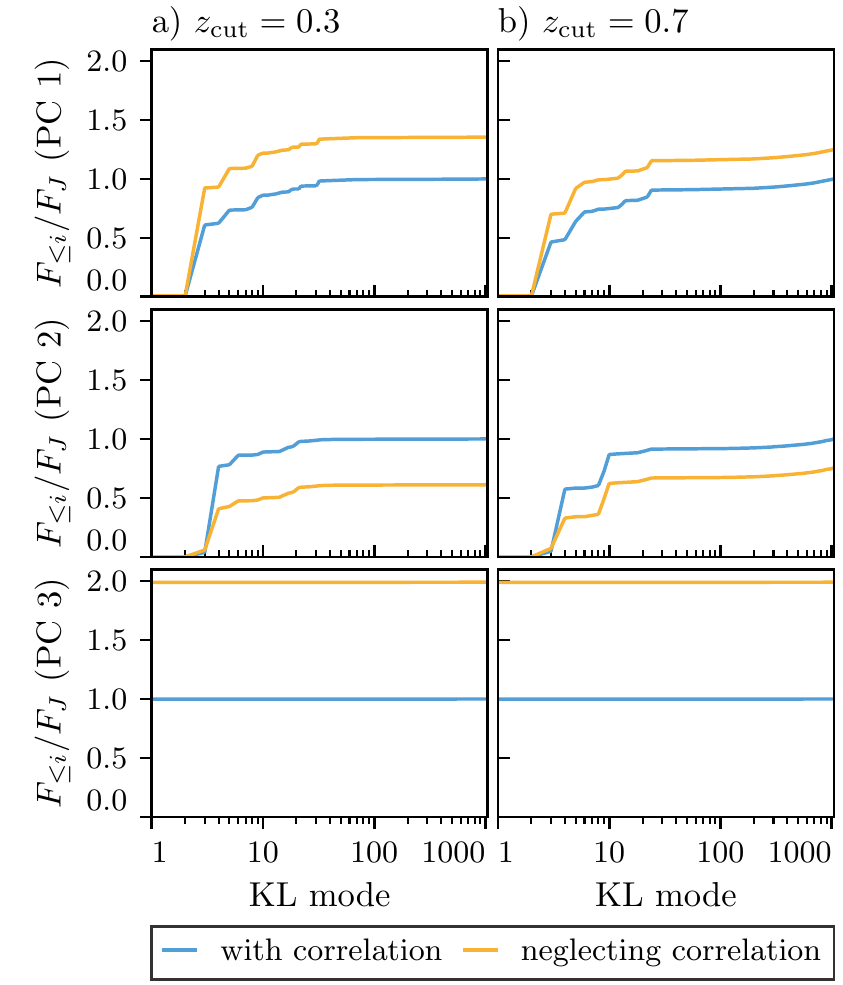}	
\caption{
The amplitude of the Fisher matrix elements corresponding to the three parameter principal components in units of their full data set values.
The parameter combinations defining the three principal components are shown in Table~\ref{Tab:SNprincipalComponents}.
All curves are obtained by summing the first $i$ KL modes ordered by absolute value of correlation coefficient and such that odd modes are correlated and even modes are anti-correlated.
The two colors correspond to the cases where we include or exclude the correlation between SN at $z>z_{\rm cut}$ and $z<z_{\rm cut}$, as shown in the legend.
}
\label{Fig:CCAparametercumulativeFisher}
\end{figure}

We can now employ Eq.~(\ref{Eq:CCAKLFisher}) to consider the cumulative impact of the KL modes in the PC parameter space, after ordering them by absolute value of correlation coefficient and such that odd modes are correlated and even modes are anti-correlated.  
Since we are using the PCs, the entries of the Fisher matrix provide a good representation of  the inverse PC variance.

In Fig.~\ref{Fig:CCAparametercumulativeFisher} we can see the Fisher matrix entries as we sum different KL modes, 
for the two cases where we keep and neglect the data correlations.
Both curves are shown in units of the full Fisher matrix including correlations.

As we can see, for the first PC and for both redshift splits, the dominant contribution to the difference between the two results 
comes from the third KL mode.
The third mode is correlated with a correlation coefficient of about $\rho_{12}=0.5$ and this corresponds, as in Eq.~(\ref{Eq:CCAKLFisher}), or an underestimate of the variance by a factor of
$1.5$.  This fractional underestimate is diluted somewhat by the sum of higher KL modes which have
smaller $\rho_{12}$, especially for the $z_{\rm cut}=0.7$ case where the lower redshift side
can measure shape changes from $\Omega_m$ on its own.   The net result is that
the variance of PC1 is underestimated by a factor of $1.4$ for $z_{\rm cut}=0.3$ and
$1.2$ for $z_{\rm cut}=0.7$.  
 This is also consistent with the underestimate of the variance of $\Omega_m$ shown in Fig.~\ref{Fig:JointCorrelationPosteriorComparison} when we neglect correlations between the two data set splits.

The dominant contribution to the difference in results for the second PC comes from the fourth mode which is anti-correlated with a correlation coefficient of about $\rho_{12}=-0.5$ corresponding to an overestimate of the PC variance of about $1.5$. 
This well constrained mode hardly influences cosmological results, which marginalize over $\M$.

The third PC
 gets most of its contribution from the first correlated mode. This is due to the
 shared absolute magnitude calibration which gives $\rho_{12}=1$ and hence a factor of $2$ underestimate.   Since the third PC involves the $H_0-\M$ degeneracy, it also explains 
 the underestimate of the variance of $H_0$ shown in  Fig.~\ref{Fig:JointCorrelationPosteriorComparison}.   
 
\begin{figure*}[htb]
\centering
\includegraphics[width=\textwidth]{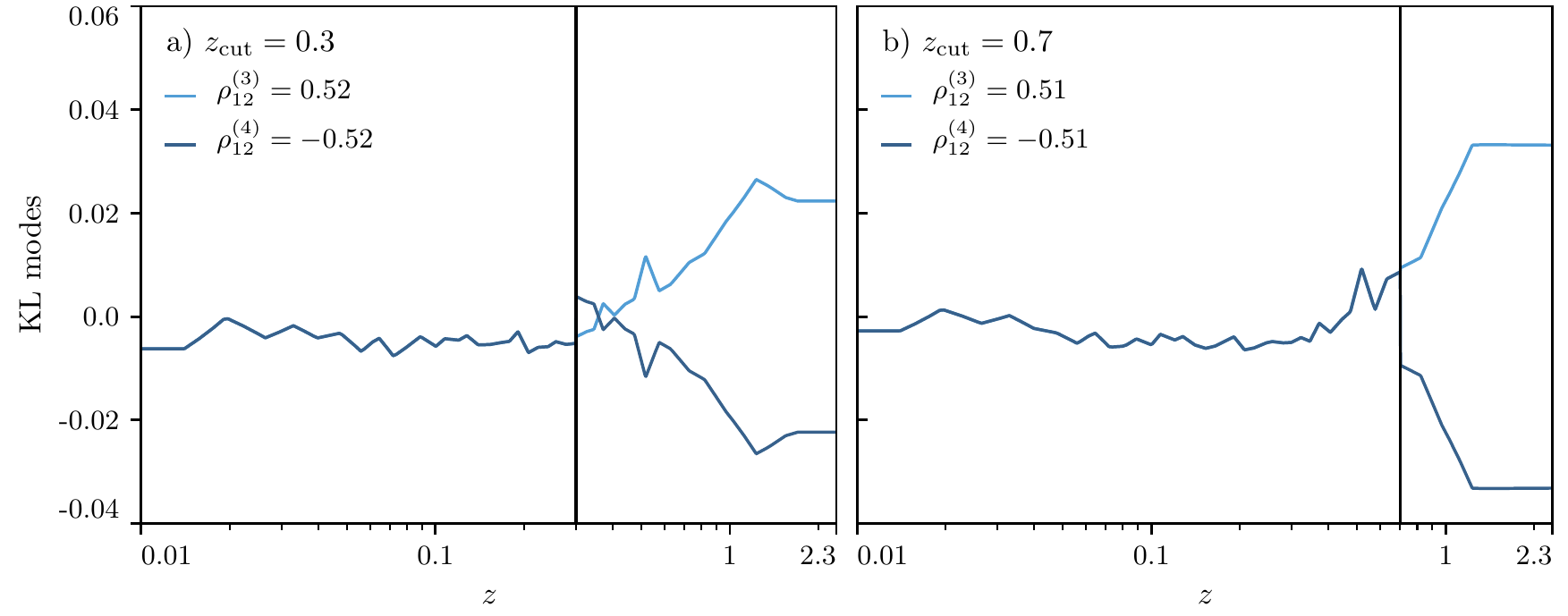}	
\caption{
The third and fourth most correlated modes for the two SN splits that we consider.
Different colors represent different modes, as shown in the legend.
The vertical line shows the redshift where we split the SN data.
}
\label{Fig:CCAmodes}
\end{figure*}

We can now look at the most relevant correlated data modes that are shown in Fig.~\ref{Fig:CCAmodes}.
These are obtained from the full SN data set, so that the discreteness of the data are evident, but exhibit far smoother trends in redshift than the scale of the individual redshifts themselves. The data split, in fact, highlights coherent effects across the redshift sample.

Comparing Fig.~\ref{Fig:CCAmodes} and Fig.~\ref{Fig:SNParameterJacobian} we can see that the third KL mode is qualitatively very similar to the effect of changing $\Omega_m$, possibly with a small amplitude component, and so it is not surprising that the data correlation corresponding to this mode reflects almost entirely on the parameter variance.
The fourth mode differs from the third one by a sign flip in the high redshift part and looks less like a smooth change in $\m-\M_{\rm fid}$, especially for $z_{\rm cut}=0.7$.

Overall we see that the CCA decomposition provides a powerful tool for quantifying and understanding
the impact of data correlations on parameter estimation. When the high ranked modes resemble the desired parameters themselves, this impact is maximal.

%

\section{Pedagogical example of data vs.~parameter splits} \label{App:ExampleDataParamSplit}

In this appendix we illustrate  the difference between splitting data and splitting parameters in the case of correlated data sets with a pedagogical example that can be fully treated analytically.

Take a $d$ dimensional data vector $x_i$ split as $(x_1\ldots x_{d_1})$ and $(x_{d_1+1} \ldots x_{d_1+d_2})$ with $d_1$ and $d_2$ elements respectively.
Let $\langle x_i \rangle =0$ and define a data covariance matrix to have all uncorrelated elements but for the last ($x_{d_1}$) and first  ($x_{d_1+1}$) data points of the two sets which have a correlation coefficient $R$:
\begin{equation} \label{Eq:ExampleGeneralCovariance}
\Sigma = 
\left(
\begin{array}{cccc}
\mathbb{I}  & \mathbb{O} & \mathbb{O} & \mathbb{O}  \\
\mathbb{O} & 1 & R & \mathbb{O}   \\
\mathbb{O} & R & 1 & \mathbb{O}  \\
\mathbb{O} & \mathbb{O} &\mathbb{O}  & \mathbb{I} \\
\end{array}
\right) \;,
\end{equation}
where, hereafter, $\mathbb{I}$ and $\mathbb{O}$ denote the identity and zero matrices of appropriate dimensions respectively.
In the language of CCA, there is only one pair of correlated modes  proportional to $(\mathbb{O},1,\pm 1,\mathbb{O})$ with
$\rho_{12}=\pm R$.

We use a simple model with a parameter that controls the mean of the data sets. The joint model Jacobian is then $M_\joint^T = (\mathbb{I},1,1,\mathbb{I})$ while the parameter split Jacobian is given by:
\begin{equation} \label{Eq:ExampleGeneralCopyJacobian}
M^T_{\pcopy} = 
\left(
\begin{array}{cccc}
\mathbb{I} & 1  & 0 & \mathbb{O} \\
\mathbb{O} & 0 & 1 & \mathbb{I}\\
\end{array}
\right) \;.
\end{equation}
Following Eq.~(\ref{Eq:GLMMaximumLikeParams}), the maximum likelihood parameter split estimates of the parameters are therefore:
\begin{align} \label{Eq:ExampleGeneralParameterCopy}
\theta_{1\pcopy} ={}&  \frac{1}{\alpha} \bigg[  d_2  \sum_{i=1}^{d_1} x_i + \left( \sum_{i=d_1+2}^{d_1+d_2} x_i - (d_2-1)x_{d_1+1} \right) R  \nonumber \\
& - (d_2-1) \sum_{i=1}^{d_1-1} x_i R^2 \bigg] \;, \nonumber\\
\theta_{2\pcopy} ={}&  \frac{1}{\alpha} \bigg[  d_1 \sum_{i=d_1+1}^{d_1+d_2} x_i   +\left( \sum_{i=1}^{d_1-1} x_i  - (d_1-1) x_{d_1}\right)R \nonumber \\
& - (d_1-1) \sum_{i=d_1+2}^{d_1+d_2} x_{i} R^2 \bigg] \;,
\end{align}
where we have defined $\alpha \equiv d_1 d_2 -(d_1-1)(d_2-1)  R^2$.
The parameter covariance matrix is then given by Eq.~(\ref{Eq:GLMMLCovariance}):
\begin{equation} \label{Eq:ExampleGeneralParameterCopyCovariance}
{\cal C}_\pcopy = \frac{1}{\alpha}
\left(
\begin{array}{cc}
{d_2 - (d_2-1) R^2}  &R  \\
{R} & d_1 -(d_1-1)R^2 
\end{array}
\right) \;.
\end{equation}
These results can now be compared to the data split estimators of the parameters:
\begin{eqnarray} \label{Eq:ExampleGeneralDataSplitParameters}
\theta_{1\single}  &=& \frac{1}{d_1} \sum_{i=1}^{d_1}  x_i  \;,\nonumber\\
\theta_{2\single} &=& \frac{1}{d_2} \sum_{i=d_1+1}^{d_1+d_2}  x_i  \;,
\end{eqnarray}
with covariance matrix given by:
\begin{equation} \label{Eq:ExampleGeneralDataSplitParameterCovariance}
{\cal C}_\single = \frac{1}{d_1 d_2}
\left(
\begin{array}{cc}
d_2  & R \\
R & d_1
\end{array}
\right) \;.
\end{equation}
As we can see, all results coincide in the uncorrelated limit, $R\rightarrow 0$ while generally differ when data correlations are present.
In particular, in case of the parameter copies, Eq.~(\ref{Eq:ExampleGeneralParameterCopy}), each parameter estimate depends on the full data set, even the uncorrelated pieces of the complementary set, where the weights are proportional to the correlation coefficient $R$.  This is because
those uncorrelated data still inform the mean of the correlated data point.  This example 
illustrates the fundamental difference between the two statistics: the data of set 2 influence
the parameters of set 1 and vice versa for parameter splits but not for data splits.

Similarly we can compute the parameter estimate for the joint data set:
\begin{align} \label{Eq:ExampleGeneralJointParameters}
 \theta_\joint ={} & \frac{1}{\alpha_J}\left[  
 (1+R) \sum_{i=1}^{d_1+d_2} x_i  -R (x_{d_1}+x_{d_1+1}) \right] \;, \nonumber\\
\alpha_J ={} & (1+R)(d_1+d_2-2)+2
\end{align}
with covariance:
\begin{align}
\mathcal{C}_J = \frac{1+R}{\alpha_J} \;,
\end{align}
and explicitly verify that we can decompose the joint parameter determination as a linear combination of the two parameter copies, as in Eq.~(\ref{Eq:JointFromCopy}) while the same does not apply for the data split parameter determinations.

It is now instructive to consider the two extreme cases of fully correlated and anti-correlated data sets, corresponding respectively to $R=1$ and $R=-1$.
In both cases the data covariance, Eq.~(\ref{Eq:ExampleGeneralCovariance}), becomes singular, with one data combination fixed with zero variance.
In the $R=1$ case the difference between $x_{d_1}$ and $x_{d_1+1}$ has zero variance and hence $x_{d_1}=x_{d_1+1}$.
In the $R=-1$ case the sum of $x_{d_1}$ and $x_{d_1+1}$ is fixed, so that $x_{d_1}=-x_{d_1+1}$.

The different parameter estimators, discussed above, then respond differently in these two cases, depending on how the correlated mode projects on the parameters of the model.
The data split parameters remain unaltered and do not respond in a particular way to the extreme correlation, since the presence of that correlation is ignored in the parameter fit in the first place.

From Eq.~(\ref{Eq:ExampleGeneralParameterCopy}), the parameter split parameters for $R=1$ become equal
\begin{align}
\theta_{1\pcopy} = \theta_{2\pcopy} = \theta_\joint = \frac{1}{\alpha} \left(  \sum_{i=1}^{d_1} x_i +\sum_{i=d_1+2}^{d_1+d_2} x_i \right) \;,
\end{align}
with $\alpha=d_1+d_2-1$, the number of independent data points.
It follows that the difference of the two copy parameters is fixed to zero while their sum has variance $2/\alpha$.  In this case the fully correlated data acts as a  bridge so that the best parameter estimator of each is always the joint estimator that uses all of the data optimally.   

In the opposite case, when $R=-1$, the two copy parameters are given by:
\begin{align}
\theta_{1\pcopy} = -\theta_{2\pcopy} = \frac{1}{\alpha} \left(  \sum_{i=1}^{d_1} x_i -\sum_{i=d_1+2}^{d_1+d_2} x_i \right)
\end{align}
while the joint parameter estimate is $\theta_\joint=0$.   This reflects the fact that
a shift in  $\theta_\joint$ reflects a shift in the mean of all points whereas a noise
fluctuation can only shift the difference between the correlated points, not their sum.
The sum of the two copy parameters is then fixed to zero while their difference has variance $2/\alpha$.   The former represents a parameter that can be measured free of noise
in the joint case when including correlations and is a simple example of saturating the $\lambda_K=0$
KL bound discussed in the previous section below Eq.~(\ref{Eq:KLdefn}).

As we can see, in these two extreme cases, if the two data sets share some linear combination of their data that can be measured free of noise, then information is fully shared between the parameter splits in a manner that depends on the projection of this linear
combination onto parameter space.  The case of partial correlation is analogous but 
in that case the two parameters are likewise no longer fully correlated.  

To further clarify the difference between the statistics for finite $R$, let us
take the simplest example where $d_1=1$ and $d=d_2+1$.
The estimator of parameter split parameter difference is then:
\begin{eqnarray} \label{Eq:ExampleParameterCopyDifference}
\Delta \theta_\pcopy \equiv \theta_{1\pcopy}-\theta_{2\pcopy} &=& x_1 -\frac{1+(d-2) R}{d-1}  x_2   + \frac{R-1}{d-1} \sum_{i=3}^d x_i  \;, \nonumber\\
{\rm var}(\Delta \theta_\pcopy )  &=& \frac{(1-R)(d+(d-2)R)}{d-1} \;,
\end{eqnarray}
while the data split estimator is:
\begin{eqnarray} \label{Eq:ExampleDataSplitParameterDifference}
\Delta \theta_\single \equiv \theta_{1\single} -\theta_{2\single}  &=& x_1- \frac{1}{d-1}\sum_{i=2}^d x_i \;, \nonumber\\
{\rm var}(\Delta \theta_\single)  &=& \frac{d-2 R}{d-1} \;.
\end{eqnarray}
Even though $\Delta\theta_C \rightarrow 0$ as $R\rightarrow 1$, its variance does
as well since the fluctuations in the uncorrelated data drop out of the difference.   
Thus for finite $R$, $\Delta\theta_C$ may be significantly anomalous
even though its magnitude is much less than $\Delta\theta_S$.  Nonetheless, for a given data realization, 
the two would  report different statistical significance in general.

To understand the difference in significance let us illustrate this with a simple example.   Suppose the anomalous aspect of the
data were an extreme fluctuation in the value of $x_1$ itself.    For a typical realization of $x_2$,
this would appear as an anomalous, but different,
 value for $\Delta\theta_C$ and $\Delta \theta_S$.   However the realization of $x_2$
 contains fluctuations from both the correlated noise and the uncorrelated noise.   
 A rare fluctuation in the  uncorrelated piece will separate the significance of the two statistics.  
For example if the uncorrelated piece separated $x_2$ from $x_1$ more than expected given $R$, it would affect the parameter split estimator more than the data split.  As  $R\rightarrow 1$, the former depends mainly on  $x_1 -x_2$, which is controlled by the uncorrelated piece
of the noise rather than the anomalous value of $x_1$ itself.  In this sense, it is more important
for the parameter split statistics that correlations are modeled accurately than it is 
for the data split statistics.

In this simplified case it is also easy to write the correlation between the two parameter shift estimates:
\begin{align}
{\rm corr}\left( \Delta \theta_\pcopy, \Delta \theta_\single \right) = \sqrt{\frac{(1-R)(d+R(d-2))}{d-2R}} \,,
\end{align}
which shows that in the $R=0$ case the two estimators are completely correlated and become gradually uncorrelated as $R$ increases.
As discussed above, the correlation of a single data point  reduces the
parameter split difference whereas the data split  difference still fluctuates because of all of the uncorrelated
data, thereby decorrelating the two estimators.

\section{Exact distribution of goodness of fit loss statistic with data split}\label{App:QDMAPdataExact}

In this section we discuss in detail the exact distribution of the $Q_{\rm DML}^\single$ and $Q_{\rm DMAP}^\single$ estimators, with the data split methodology.

We first consider the ratio of maximum likelihoods of the joint data set and the two subsets. 
By direct calculation it can be shown that, up to constant offsets that is irrelevant to the calculation of statistical significance this ratio can be written as the quadratic form in the data:
\begin{eqnarray} \label{Eq:AppDataSplitDML}
Q_{\rm DML}^\single &=& X_{\joint}^T \Big[  (\mathbb{I}_{\joint}-\mathbb{P}_{\joint})^T \Sigma^{-1}_{\joint}  (\mathbb{I}_{\joint}-\mathbb{P}_{\joint}) \nonumber \\
&&-\tiny \Big( 
\begin{array}{cc}
(\mathbb{I}_{1}-\mathbb{P}_{1})^T \Sigma^{-1}_{1}  (\mathbb{I}_{1}-\mathbb{P}_{1})  & \mathbb{O} \\
\mathbb{O}       &  (\mathbb{I}_{2}-\mathbb{P}_{2})^T \Sigma^{-1}_{2}  (\mathbb{I}_{2}-\mathbb{P}_{2})
\end{array}
\Big) \Big] X_{\joint} \nonumber \\
&\equiv& X_\joint^T A_{\rm DML}^\single X_\joint \; ,
\end{eqnarray}
where $X_{\joint}$ is the full data vector, distributed according to the evidence of the joint data set, and $A_{\rm DML}^\single$ is the matrix that defines $Q_{\rm DML}^\single$. The indices 1 and 2 denote the two subsets of the data after we split the joint set. 

Through explicit computation, the form of the joint projector is given by:
\begin{align*}
\mathbb{P}_\joint \equiv \left( 
\begin{tabular}{cc}
$p_{11}$ & $p_{12}$ \\
$p_{21}$ & $p_{22}$
\end{tabular}
\right)
\end{align*}
with
\begin{align*}
p_{11} &= M_1 \mathcal{C}_\joint (M_1^T - M_2^T \Sigma_2^{-1} \Sigma_{21}) K_1^{-1} \; ,\\
p_{12} &= M_1 \mathcal{C}_\joint (M_2^T - M_1^T \Sigma_1^{-1} \Sigma_{12}) K_2^{-1} \; , \\
p_{21} & = M_2 \mathcal{C}_\joint (M_1^T - M_2^T \Sigma_2^{-1} \Sigma_{21}) K_1^{-1} \; ,\\
p_{22} &= M_2 \mathcal{C}_\joint (M_2^T - M_1^T \Sigma_1^{-1} \Sigma_{12}) K_2^{-1} \; ,
\end{align*}
where we have defined:
\begin{align*}
K_1 &\equiv \Sigma_1 - \Sigma_{12} \Sigma_2^{-1} \Sigma_{21} \; , \\
K_2 &\equiv \Sigma_2 - \Sigma_{21} \Sigma_1^{-1} \Sigma_{12} \; .
\end{align*}
In order to calculate the distribution of Eq.~(\ref{Eq:AppDataSplitDML}) we follow the procedure discussed in App. A of \cite{Raveri:2018wln} and compute the eigenvalues, $\lambda$, of $A_{\rm DML}^\single\mathcal{S}_\joint$, where $\mathcal{S}_\joint$ is the covariance for the joint distribution of the data which, for Gaussian priors, is $\mathcal{S}_\joint = \Sigma_\joint + M_\joint \mathcal{C}_\Pi M_\joint^T$.
This allows to decompose $Q_{\rm DML}^\single$ in the following way:
\begin{align}\label{eqL12:Q_DLM decomposition def}
Q_{\rm DML}^\single = \sum\limits_{i} \lambda_i \mathcal{U}_i^2 \; ,
\end{align}
where each $\mathcal{U}_i \sim \mathcal{N}_\joint (x_\joint; \mathbb{O}, \mathbb{I})$ so that $Q_{\rm DML}^\single$ is a weighted sum of chi squared variables.

By direct calculation we have:
\begin{align}\label{Eq:AppDataSplitDMLASigma}
\mathcal{A}_{\rm DML}^S \equiv A_{\rm DML}^\single \mathcal{S}_\joint \equiv
\left( 
\begin{tabular}{cc}
$A$ & $B$ \\
$C$ & $D$
\end{tabular}
\right) \; ,
\end{align}
where:
\begin{align*}
A &&=& \left[ \Sigma_1^{-1} M_1 \mathcal{C}_1 - K_1^{-1} (M_1 - \Sigma_{12} \Sigma_2^{-1} M_2) \mathcal{C}_\joint \right] M_1^T \; , \\
B &&=& -K_1^{-1} (M_1 - \Sigma_{12} \Sigma_2^{-1} M_2) \mathcal{C}_\joint M_2^T \\
&&& - (\mathbb{I}_1 - \Sigma_1^{-1} M_1 \mathcal{C}_1 M_1^T) \Sigma_1^{-1} \Sigma_{12} \; , \\
C &&=& -K_2^{-1} (M_2 - \Sigma_{21} \Sigma_1^{-1} M_1) \mathcal{C}_\joint M_1^T \\
&&& - (\mathbb{I}_2 - \Sigma_2^{-1} M_2 \mathcal{C}_2 M_2^T) \Sigma_2^{-1} \Sigma_{21} \; , \\
D &&=& \left[ \Sigma_2^{-1} M_2 \mathcal{C}_2 - K_2^{-1} (M_2 - \Sigma_{21} \Sigma_1^{-1} M_1) \mathcal{C}_\joint \right] M_2^T \; .
\end{align*}
An analytic solution to the above eigenvalue problem is not easily obtained, but can be obtained numerically to evaluate the exact distribution of $Q_{\rm DML}^\single$.
We highlight that, similarly to what happens for data split parameter shifts, the calculation of the statistics involves quantities that are defined both at the parameter space and data space level.

Note that the expressions we derived above reduce to the corresponding ones in \cite{Raveri:2018wln} in the limit of uncorrelated data sets.

Furthermore we can notice that the quadratic form defined by $Q_{\rm DML}^\single$ is not necessarily positive definite. 
This is a consequence of the fact that the projector on the joint parameter space is not a sub-space of the span of the single data set projector.
This severely limits the possibility of approximating $Q_{\rm DML}^\single$ with a chi squared distribution, which is positive definite, especially for events in the confirmation tail that would be very close to $Q_{\rm DML}^\single=0$.

In addition, the fact that $Q_{\rm DML}^\single$ is not chi squared distributed means that correlated data fluctuations are not optimally weighted. 

We then consider the ratio of likelihoods at maximum posterior (DMAP) in the data split case.
To do so we add the extra terms that transform ML into MAP so that the matrix that controls $Q_{\rm DMAP}^\single$ is given by:
\begin{align}\label{Eq:AppDataSplitDMLASigmaFinal}
\mathcal{A}_{\rm DMAP}^\single =\;& \mathcal{A}_{\rm DML}^\single +\Sigma_\joint^{-1}M_\joint\mathcal{C}_\joint \mathcal{C}_\Pi^{-1}\mathcal{C}_{p\joint} M_\joint^T  \\
& -\left( 
\begin{tabular}{cc}
$\tilde{M}_1^T\mathcal{C}_\Pi^{-1}\mathcal{C}_{p1}M_1^T$ &  $\tilde{B}$ \\
 $\tilde{B}(1\leftrightarrow 2)$ & $\tilde{M}_2^T\mathcal{C}_\Pi^{-1}\mathcal{C}_{p2}M_2^T$
\end{tabular}
\right) \; , \nonumber
\end{align}
where, for compactness, we have defined $\tilde{B} \equiv \tilde{M}_1^T\mathcal{C}_\Pi^{-1}\mathcal{C}_{p1}\mathcal{C}_1^{-1}\mathcal{C}_{p1}(\mathcal{C}_\Pi^{-1}\tilde{M}_1\Sigma_{12}+M_2^T )$.

These results, for both $Q_{\rm DML}^\single$ and $Q_{\rm DMAP}^\single$, can be used to compute the respective exact distributions.
The trace of these distributions coincides with the results obtained in the uncorrelated case, but we notice that it is problematic to approximate them with simpler distributions because both of them are not positive definite.

The fact that both $Q_{\rm DML}^\single$ and $Q_{\rm DMAP}^\single$ are not positive definite means that there are aspects of the data where the joint likelihood is better than the product of the separate likelihoods.
This can never happen for uncorrelated data sets and is a consequence of the presence of correlated data modes.
In particular, the data modes that are fit by the model separately are the ones that would contribute to the positive definiteness of the above statistics, since they can zero out different data fluctuations, whereas the correlated data modes that are left out can contribute negatively. Fitting the data jointly, however, always takes correlated modes into account, so that the contribution to the chi-square from them is considered.

\begin{figure}[!ht]
\centering
\includegraphics[width=\columnwidth]{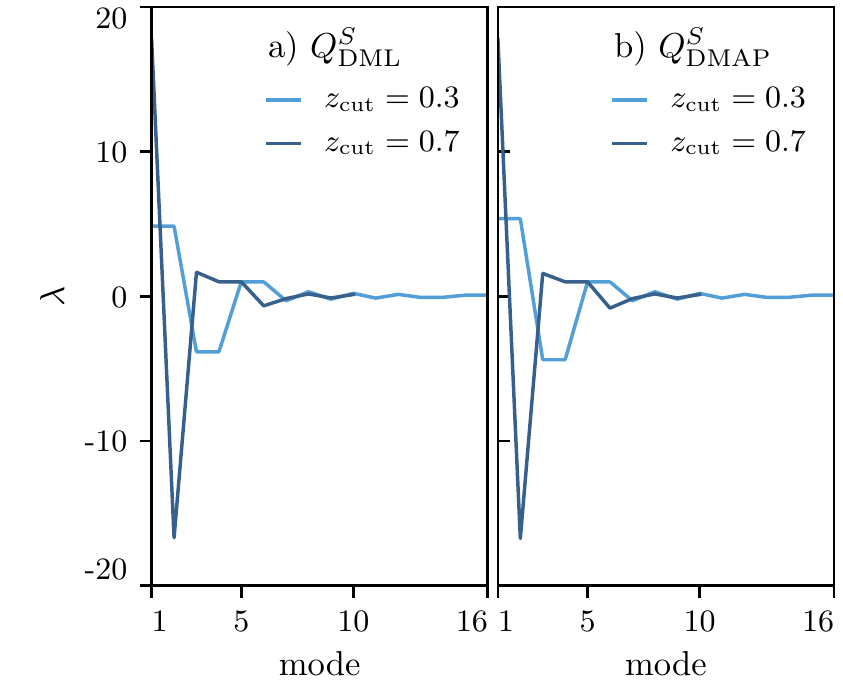}	
\caption{
Spectrum of the exact distribution of $Q_{\rm DML}^\single$ in panel a and $Q_{\rm DMAP}^\single$ in panel b.
Different lines correspond to different SN split, as shown in legend.
Modes that have eigenvalues below $0.05$ are not shown.
}
\label{Fig:ExactDataSplitSpectrum}
\end{figure}

We close this appendix by discussing the application of these exact statistics to our SN example.
In Fig.~\ref{Fig:ExactDataSplitSpectrum} we show the eigenvalues of the matrix in Eq.~(\ref{Eq:AppDataSplitDMLASigma})
for $Q_{\rm DML}^\single$ and the eigenvalues of the matrix in Eq.~(\ref{Eq:AppDataSplitDMLASigmaFinal}) for $Q_{\rm SMAP}^\single$.
This clearly shows that both estimators are not positive definite which means that an approximation of the exact distribution with a 
$\chi^2$ distribution would not be appropriate.
We can also see that the number of non-zero components largely exceeds the number of parameters as a consequence of the fact that for this data split estimator data and parameter modes are mixed.
Notice also that both distributions are similar since the prior is not informative.

\section{Exact distribution of goodness of fit loss statistic with parameter split}\label{App:QDMAPcopyExact}

In this section we discuss in detail the exact distribution of the $Q_{\rm DML}^\pcopy$ and $Q_{\rm DMAP}^\pcopy$ estimators, as presented in Sec.~\ref{Sec:ParamSplitGOFloss}, and how they can be approximated.

We first consider the statistics of ML ratios (DML). We focus on the distribution of the DML statistic between the joint chain and the one with the duplicated parameter space. To do so, we begin by considering the ML parameter split determination, $\theta_{\pcopy}^{\rm ML} = (\theta^{\rm ML}_{1 \pcopy},\theta^{\rm ML}_{2 \pcopy})^T$, and the joint ML parameters, $\theta_{\joint}^{\rm ML}$. 
We then use them to define the difference in joint log-likelihood at the ML point as:
\begin{align}\label{Eq:CorrelatedDML}
Q_{\rm DML}^C \equiv -2 \ln \mathcal{L}_\joint(\theta_{\joint}^{\rm ML}) + 2 \ln \mathcal{L}_\joint(\theta_{\pcopy}^{\rm ML}) \; .
\end{align}
Note that in the limit of uncorrelated data this reduces to $Q_{\rm DML}^C = -2 \ln \mathcal{L}_\joint(\theta_{\joint}^{\rm ML}) + 2 \ln \mathcal{L}_1(\theta^{\rm ML}_{1}) + 2 \ln \mathcal{L}_2(\theta^{\rm ML}_{2})$ which is similar to the expressions used in~\cite{Raveri:2018wln}.

In the GLM, it can be shown that, up to constants which are not important for our purpose, we get the following quadratic form in data space:
\begin{eqnarray}\label{eqDMAP:likelihood at MAP in GLM}
Q_{\rm DML}^C &=& X^T \left[ \left( \mathbb{I} - \mathbb{P}_\joint \right)^T \Sigma_\joint^{-1} \left( \mathbb{I} - \mathbb{P}_\joint \right) \right. \nonumber \\
&&\left. - \left( \mathbb{I} - \mathbb{P}_\pcopy \right)^T \Sigma_\joint^{-1} \left( \mathbb{I} - \mathbb{P}_\pcopy \right) \right] X \nonumber \\
& \equiv & X^T A_{\rm DML}^\pcopy X \; ,
\end{eqnarray}
where we have used the joint projector $\mathbb{P}_\joint = M_\joint \mathcal{C}_\joint M_\joint^T \Sigma_J^{-1}$ and the projector under parameter duplication written as $\mathbb{P}_\pcopy = M_\pcopy \mathcal{C}_\pcopy M_\pcopy^T \Sigma_J^{-1}$.
Now, we can rewrite the joint projector in the following way:
\begin{align}\label{Eq:JointProjectorExpression}
\mathbb{P}_\joint = M_\pcopy D_{\pcopy} (D_{\pcopy}^T \mathcal{C}_\pcopy^{-1} D_{\pcopy})^{-1} D_{\pcopy}^T M_\pcopy^T \Sigma_\joint^{-1} \; ,
\end{align}
while the projector in the case of parameter duplication can be expressed as:
\begin{align}\label{Eq:CopyProjectorExpression}
\mathbb{P}_\pcopy = M_\pcopy (M_\pcopy^T \Sigma_\joint^{-1} M_\pcopy)^{-1} M_\pcopy^T \Sigma_J^{-1} \; .
\end{align}
Then, using the above expressions it is straightforward to show that the joint set of parameters is a subset of the duplicate set, since $\mathbb{P}_\joint \mathbb{P}_\pcopy = \mathbb{P}_\pcopy \mathbb{P}_\joint = \mathbb{P}_\joint$. Therefore, we can use theorem (5.2.5) in~\cite{Mathai:Quadraticforms} to show that, at the ML level,
\begin{eqnarray}\label{eqDMAP:DMAP chi-square}
Q_{\rm DML}^\pcopy &\sim& \chi^2({\rm rank}(\mathbb{I}-\mathbb{P}_\joint) - {\rm rank}(\mathbb{I}-\mathbb{P}_\pcopy)) \nonumber \\
&=& \chi^2(N_\pcopy-N_\joint) \; ,
\end{eqnarray}
where $N_\pcopy$ and $N_\joint$ are the number of parameter duplicates and the number of joint parameters respectively.
Note that, in the limit of uncorrelated data sets $N_\pcopy = N_1 + N_2$, where $N_1$ and $N_2$ are the number of relevant parameters for the first and second data sets respectively.

In contrast with the case of data split the exact statistics of the parameter split DML estimator is a chi square, which also means that $Q_{\rm DML}^\pcopy$ is optimal.

The exact statistics of $Q_{\rm DML}^\pcopy$ can also be obtained by explicitly computing the eigenvalues of $A_{\rm DML}^\pcopy \mathcal{S}_\joint = \mathbb{P}_\pcopy^T - \mathbb{P}_\joint^T$, where $\mathcal{S}_\joint = \Sigma_\joint + M_\joint \mathcal{C}_\Pi M_\joint^T$.
Notice that $M_\joint \mathcal{C}_\Pi M_\joint^T$ represents a prior that is fully correlated between the split parameters, whereas our parameter split analysis assumes separate priors that are uncorrelated.   This is necessary since otherwise the split parameters would be expected to vary according to $\mathcal{C}_\Pi$ leading to a counterfactually large 
expected improvement from fitting them separately.  Conversely, the split parameter technique cannot employ fully correlated priors because no matter how weak such a prior is, it would
force the split parameter posterior means to the same values (see App.~\ref{App:ExampleDataParamSplit}).

We can now turn to the distribution of $Q_{\rm DMAP}$, with parameter copies.
This can be written as:
\begin{eqnarray}\label{Eq:AppDQDMAPGaussian}
Q_{\rm DMAP}^\pcopy &\equiv & -2 \ln \mathcal{L}_\joint(\theta_\joint^p) + 2 \ln \mathcal{L}_\joint(\theta^p_\pcopy) \nonumber \\
&=& Q_{\rm DML}^\pcopy + X_\joint^T \left[ \tilde{M}_\joint^T \mathcal{C}_\Pi^{-1} \mathcal{C}_{p\joint} \mathcal{C}_\joint^{-1} \mathcal{C}_{p\joint} \mathcal{C}_\Pi^{-1} \tilde{M}_\joint \right. \nonumber \\
&& \left. - \tilde{M}_\pcopy^T \mathcal{C}_{\Pi\pcopy}^{-1} \mathcal{C}_{p\pcopy} \mathcal{C}_\pcopy^{-1} \mathcal{C}_{p\pcopy}\mathcal{C}_{\Pi\pcopy}^{-1} \tilde{M}_\pcopy \right] X_\joint \nonumber \\
&\equiv & X_\joint^T A_{\rm DMAP}^\pcopy X_\joint \;, 
\end{eqnarray}
where we have used that for Gaussian priors the likelihood at the point of maximum posterior is given by $-2\ln \mathcal{L}(\theta_{p\joint}) = X^T[ (\mathbb{I}-\mathbb{P}_\joint)^T \Sigma_\joint^{-1} (\mathbb{I}-\mathbb{P}_\joint) + \tilde{M}_\joint^T \mathcal{C}_\Pi^{-1} \mathcal{C}_{p\joint} \mathcal{C}_\joint^{-1} \mathcal{C}_{p\joint} \mathcal{C}_\Pi^{-1} \tilde{M}_\joint ]X$ for the joint, and similarly for the parameter copy case. 
In the above, the copy prior covariance is defined as $\mathcal{C}_{\Pi \pcopy} = {\rm diag}(\mathcal{C}_\Pi,\mathcal{C}_\Pi)$.

To calculate the exact distribution of $Q_{\rm DMAP}^\pcopy$, we follow the same procedure as in the case of data splits in App. \ref{App:QDMAPdataExact}. Therefore, we start with the computation of the matrix $A_{\rm DMAP}^\pcopy  \mathcal{S}_\joint$ whose spectrum completely specifies the distribution of $Q_{\rm DMAP}^\pcopy$ as a sum of independent Gamma distributed variables. 
It can be shown that this matrix reduces to: 
\begin{eqnarray}\label{Eq:AppDMAPexactASigma}
&& A_{\rm DMAP}^\pcopy  \mathcal{S}_\joint = \mathbb{P}_\pcopy^T - \mathbb{P}_\joint^T + \tilde{M}_\joint^T \mathcal{C}_\Pi^{-1} \mathcal{C}_{p\joint} M_\joint^T  \\
&& \hspace{0.5cm} - \tilde{M}_\pcopy^T \mathcal{C}_{\Pi\pcopy}^{-1} \mathcal{C}_{p\pcopy} \mathcal{C}_{\pcopy}^{-1} \mathcal{C}_{p\pcopy}\left( \mathcal{C}_{\Pi\pcopy}^{-1}\mathcal{C}_\pcopy + D_{\pcopy} D_{\pcopy}^T \right) M_\pcopy^T \;. \nonumber 
\end{eqnarray}
It can then be also show that the non-zero eigenvalues of Eq.~(\ref{Eq:AppDMAPexactASigma}) are also the eigenvalues of the matrix:
\begin{eqnarray}\label{Eq:AppDMAPexactASigmaParamSpace}
\mathcal{A}_{\rm DMAP}^\pcopy &=& \mathbb{I}_{2N} +D_{\pcopy}\mathcal{C}_\Pi^{-1}\mathcal{C}_{p\joint}D_{\pcopy}^T -\mathcal{C}_{p\pcopy}^{-1} D_{\pcopy} \mathcal{C}_{p\joint}D_{\pcopy}^T \nonumber \\
&& -\mathcal{C}_{\Pi\pcopy}^{-1}\mathcal{C}_{p\pcopy}D_{\pcopy}D_{\pcopy}^T + \mathcal{C}_{\Pi\pcopy}^{-1}\mathcal{C}_{p\pcopy}\mathcal{C}_{\Pi\pcopy}^{-1}\mathcal{C}_{p\pcopy} \mathbb{J}_{2N} \;, \nonumber \\
\end{eqnarray}
where we have defined, for convenience, the exchange matrix $\mathbb{J}_{2N} \equiv D_{\pcopy}D_{\pcopy}^T - \mathbb{I}_{2N}$
that exchanges the off diagonal blocks with the diagonal ones.
Note that the above expression is written in terms of quantities that can be obtained from MCMC samples of the posterior of both the parameter copy and joint chains.

Either one can use Eq. (\ref{Eq:AppDMAPexactASigmaParamSpace}) to compute the exact distribution or one can approximate it by a chi squared distribution matching the mean of the exact distribution as a first order Patnaiks' approximation~\cite{10.2307/2332149}.
The mean of the exact distribution and the number of degrees of freedom of the chi squared approximation is given by: 
\begin{eqnarray}\label{Eq:AppDMAPexactASigmaTrace}
{\rm tr}[\mathcal{A}_{\rm DMAP}^\pcopy] &=& N + {\rm tr}[\mathcal{C}_\Pi^{-1} \mathcal{C}_{p\joint}] - {\rm tr}[\mathcal{C}_{\Pi\pcopy}^{-1} \mathcal{C}_{p\pcopy}]  \\ 
&& - {\rm tr} [ \mathcal{C}_{\Pi\pcopy}^{-1}\mathcal{C}_{p\pcopy}( \mathbb{I}_{2N}-\mathcal{C}_{\Pi\pcopy}^{-1}\mathcal{C}_{p\pcopy})  \mathbb{J}_{2N}] \nonumber \\
&=& N_{\rm eff}^\pcopy - N_{\rm eff}^\joint \nonumber \\
&&\qquad\qquad\quad \mathclap{+ {\rm tr}[\mathcal{C}_\Pi^{-1}( \mathcal{C}_{p1\pcopy} +\mathcal{C}_{p2\pcopy} -\mathcal{C}_\Pi ) \mathcal{C}_\Pi^{-1} (\mathcal{C}_{p12\pcopy} + \mathcal{C}_{p21\pcopy})] \; .} \nonumber 
\end{eqnarray}
We can furthermore calculate the variance of the distribution as it is proportional to the trace of the matrix $(\mathcal{A}_{\rm DMAP}^\pcopy)^2$. This, however, does not significantly simplify and in practical applications it is significantly easier to compute the variance numerically.

All the results in this appendix agree, in the uncorrelated limit, with the results in~\cite{Raveri:2018wln}.

\begin{figure}[!ht]
\centering
\includegraphics[width=\columnwidth]{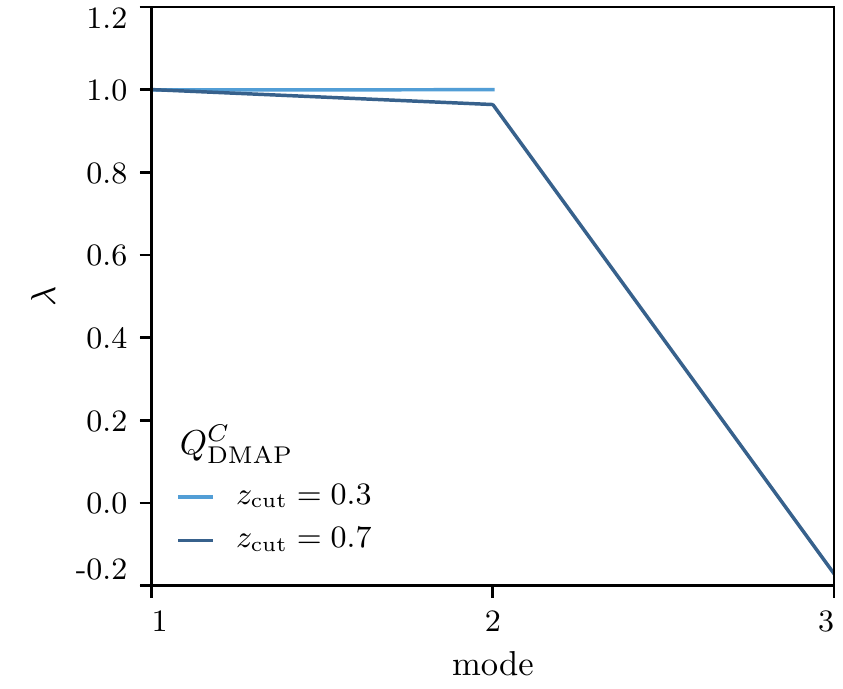}	
\caption{
Spectrum of the exact distribution of $Q_{\rm DMAP}^\pcopy$.
Different lines correspond to different SN split, as shown in legend.
Modes that have eigenvalues below $0.05$ are not shown.
}
\label{Fig:DMAPexactParamSplitSpectrum}
\end{figure}

We now compute the two exact distributions for our SN example considering only $Q_{\rm DMAP}^\pcopy$ as $Q_{\rm DML}^\pcopy$ is chi square distributed.
The eigenvalues of both the $Q_{\rm DMAP}^\pcopy$ matrix, as in Eq.~(\ref{Eq:AppDMAPexactASigmaParamSpace}), are shown in Fig.~\ref{Fig:DMAPexactParamSplitSpectrum}.

As we can see, since the prior is not informative for the $z_{\rm cut}=0.3$ case, the eigenvalues result in a chi squared distribution with two degrees of freedom.
On the other hand, the presence of a mildly informative prior for the $z_{\rm cut}=0.7$ case makes one zero eigenvalue for DML different from zero and slightly negative.
We, however notice that, since the distribution for $Q_{\rm DMAP}^\pcopy$ is exact for directions that are either fully data and fully prior constrained, contrarily to what happens in the data split case, negative eigenvalues that arise from our mild inconsistency in accounting for the priors on split parameters are usually a small correction.

It is possible to use in practice these eigenvalues to check whether there is a difference in the statistical significance of the $Q_{\rm DMAP}^\pcopy$ exact and approximate distribution.
We find that, in this case, the misestimate of statistical significance is sub-percent.

\section{Arbitrarily split parameters}\label{App:UnsplitMultisplit}

In this appendix we generalize the discussion of the parameter split estimators to the case where we consider parameters that are multiply split or not split at all.

Quantities associated with the split and unsplit part of the parameter space will be denoted by the subscripts ``$C$" and ``$U$" respectively.
We denote the unsplit posterior parameters with $\theta_{\unsplitcopy}^p$ and the $n$ posterior parameter copies with $\theta_{\splitcopy}^p = (\theta_{1\pcopy}^p, \theta_{2\pcopy}^p, \dots, \theta_{n\pcopy}^p)^T$. Therefore, the full posterior parameter vector can be written as $\theta_\pcopy^p = (\theta_{\splitcopy}^p, \theta_{\unsplitcopy}^p)^T$. 
In what follows, the total number of copy parameters will be $nN_\pcopy$, where $N_\pcopy$ is the number of split parameters, and the number of unsplit parameters will be $N_\Nunsplit$; therefore, $N_p =  n N_\pcopy + N_\Nunsplit$ is the total number of parameters in the final parameter vector. 

Note also that the joint analysis deals with the original $N_\joint = N_\pcopy + N_\Nunsplit$ parameters in total. The joint parameter vector will be denoted as $\theta_\joint^p = (\theta_{\splitjoint}^p, \theta_\unsplitjoint^p)^T$, where the two parts $\theta_\splitjoint$ and $\theta_\unsplitjoint$ correspond to the parameter subspaces that are split and unsplit in the parameter split methodology, respectively.

The design matrix $D_\pcopy$ relates the joint quantities with the copy ones. Constructing this appropriately is then enough to generalize our analysis as described in the previous sections. 
Let $D_\splitcopy$ be the $nN_\Nsplit \times N_\Nsplit$ dimensional design matrix related to the part of parameter space that is being copied $n$ times; thus, $D_\splitcopy^T = (\mathbb{I}_\Nsplit, \dots, \mathbb{I}_\Nsplit)$ with $n$ instances of the identity matrix $\mathbb{I}_\Nsplit$ of dimensions $N_\Nsplit \times N_\Nsplit$. Let also $\mathbb{I}_\Nunsplit$ be the $N_\Nunsplit \times N_\Nunsplit$ identity matrix related to the $N_\Nunsplit$ unsplit parameters. 
Then, the design matrix takes the form
\begin{equation}\label{AppMultisplitUnslit:DesignMatrix}
D_\pcopy = \left( 
\begin{tabular}{cc}
$D_\splitcopy$ & $\mathbb{O}$ \\
$\mathbb{O}$ & $\mathbb{I}_\Nunsplit$
\end{tabular}
\right) \; ,
\end{equation}
where $\mathbb{O}$ is the vector with the appropriate number of zeros in each case. Thus, the full design matrix has dimensions of $(nN_\Nsplit + N_\Nunsplit) \times (N_\Nsplit + N_\Nunsplit)$. 

We first consider parameter shifts of the form $\theta_{i\pcopy}^p - \theta_{j\pcopy}^p$ between the $i$-th and $j$-th copies, where $i,j = 1, 2, \dots n$ run over all the $n$ parameter copies. 
Then, we can express the general form of the covariance between the two parameter differences $\theta_{i\pcopy}^p - \theta_{j\pcopy}^p$ and $\theta_{k\pcopy}^p - \theta_{l\pcopy}^p$, with $k,l=1,2,\dots,n$, as
\begin{equation}
\langle (\theta_{i\pcopy}^p - \theta_{j\pcopy}^p) (\theta_{k\pcopy}^p - \theta_{l\pcopy} ^p)^T \rangle 
= \mathcal{C}_{pik\pcopy} + \mathcal{C}_{pjl\pcopy} - \mathcal{C}_{pil\pcopy} - \mathcal{C}_{pjk\pcopy} \; .
\end{equation}
These matrices then construct the covariance that is associated with the split copy part of the parameter space. 
Since there is no shift in the unsplit parameters, the parameter differences and covariances associated with the unsplit part of the parameter space is zero.

We now turn to the discussion of update parameter differences.
In this case, we consider differences between the posterior parameters from a joint analysis, namely $\theta_\joint^p$, and the copy parameter vector $(\theta_{i\pcopy}^p, \theta_{\unsplitcopy}^p)^T$ which includes the unsplit copy parameters as well as the $i$-th copy parameter set. 
We thus form the parameter differences in update form as
\begin{align}\label{AppMultisplitUnslit:UpdateDifferenceVector}
\Delta \theta^U_\pcopy &\equiv (\theta_{i\pcopy}^p, \theta_{\unsplitcopy}^p)^T - \theta_{\joint}^p \nonumber \\ 
&= (\theta_{i\pcopy}^p - \theta_{\splitjoint}^p,\theta_{\unsplitcopy}^p - \theta_{\unsplitjoint}^p)^T \; .
\end{align}
Note that $\theta_{\unsplitjoint}^p$ are generally different from $\theta_{\unsplitcopy}^p$, and that the unsplit parameters can be correlated with the split parameters.
We can then explicitly calculate the parts of the covariance between such update parameter differences.

We begin by considering the covariance of the split parameter differences, which results in:
\begin{equation}
\langle (\theta_{i\pcopy}^p-\theta_{\splitjoint}^p) (\theta_{j\pcopy}^p-\theta_{\splitjoint}^p)^T \rangle = \mathcal{C}_{pij\pcopy} -  \mathcal{C}_{p\splitjoint} \; ,
\end{equation}
where $\mathcal{C}_{p\splitjoint} = \langle (\theta_{\splitjoint}^p) (\theta_{\splitjoint}^p)^T \rangle
-
\langle (\theta_{\splitjoint}^p) \rangle \langle (\theta_{\splitjoint}^p)^T \rangle
$ is the covariance of the parameters in the split part of the joint set. We have used the fact that $\langle (\theta_{i\pcopy}^p) (\theta_{\splitjoint}^p)^T \rangle 
-\langle (\theta_{i\pcopy}^p) \rangle\langle (\theta_{\splitjoint}^p)^T \rangle
= \mathcal{C}_{p\splitjoint}$.

Similarly to the above, we can calculate the covariance of the unsplit parameter differences as:
\begin{equation}
\langle (\theta_{\unsplitcopy}^p-\theta_{\unsplitjoint}^p) (\theta_{\unsplitcopy}^p-\theta_{\unsplitjoint}^p)^T \rangle = \mathcal{C}_{p\unsplitcopy} -  \mathcal{C}_{p\unsplitjoint} \; ,
\end{equation}
where the covariance matrices $\mathcal{C}_{p\unsplitcopy}$
 and $\mathcal{C}_{p\unsplitjoint}$
correspond to the unsplit part of the copy and joint parameter sets respectively in the same manner as for the split parameters above.
Finally, we can calculate the covariance between split and unsplit parameter differences, which yields:
\begin{equation}
\langle (\theta_{i\pcopy}^p-\theta_{\splitjoint}^p) (\theta_{\unsplitcopy}^p-\theta_{\unsplitjoint}^p)^T \rangle = \mathcal{C}_{pi\unsplitcopy} -  \mathcal{C}_{p\splitunsplitjoint} \; .
\end{equation}
In the above we have defined the covariance $\mathcal{C}_{pi\unsplitcopy}$ between the copy $i$ and unsplit copy parameters and $\mathcal{C}_{p\splitunsplitjoint}$ between the split and unsplit joint parameters, again as above.

We can now comment on the relation between the $Q_{\rm DM}^\pcopy$ and $Q_{\rm UDM}^\pcopy$ estimators and their statistical significance. 
As in the case of two parameter copies without unsplit parameters, which is discussed in Sec.~\ref{Sec:ParamSplitParameterShift}, their significance is the same for the maximum likelihood parameters, since   
\begin{align*}
\left( 
\begin{tabular}{c}
$\theta_{i\pcopy}^{\rm ML}$ \\
$\theta_\unsplitcopy^{\rm ML}$
\end{tabular}
\right) -  
\theta_{\joint}^{\rm ML} = \mathcal{C}_\joint D_\pcopy^T \mathcal{C}_\pcopy^{-1}
\left( 
\begin{tabular}{c}
$D_\splitcopy \theta_{i\pcopy}^{\rm ML} - \theta_{\splitcopy}^{\rm ML}$ \\
$\mathbf{O}$
\end{tabular}
\right)
\end{align*}
where $\theta_{\splitcopy}^{\rm ML} = (\theta_{1\pcopy}^{\rm ML}, \theta_{2\pcopy}^{\rm ML}, \dots, \theta_{n\pcopy}^{\rm ML})^T$ and the zero vector $\mathbf{O}$ has length $N_\Nunsplit$.

At the maximum posterior level the two statistics can differ, however, since the update parameter shifts contain only one copy of the prior in the joint but the prior is applied once to each set in the split analysis.
As we did in Sec.~\ref{Sec:ParamSplitParameterShift} here we can also define the joint parameter estimate $\tilde{\theta}_{\joint}^p = (\tilde{\theta}_{\splitjoint}^p,\theta_{\unsplitjoint}^p)^T$ that counts the prior $n$ times and has covariance $\tilde{\mathcal{C}}_{p\joint}^{-1} = \mathcal{C}_{\joint}^{-1} +  D_\pcopy^T \mathcal{C}_{\Pi\pcopy}^{-1}  D_\pcopy = D_\pcopy^T \mathcal{C}_{p\pcopy}^{-1} D_{\pcopy}$.
Then, the update parameter shifts would be defined as 
\begin{equation}\label{App:newParamSplitUpdateDifference}
\left( 
\begin{tabular}{c}
$\theta_{i\pcopy}^p$ \\
$\theta_\unsplitcopy^p$
\end{tabular}
\right) -  
\tilde{\theta}_{\joint}^p = \tilde{\mathcal{C}}_{p\joint} D_\pcopy^T \mathcal{C}_{p\pcopy}^{-1}
\left( 
\begin{tabular}{c}
$D_\splitcopy \theta_{i\pcopy}^p - \theta_{\splitcopy}^p$ \\
$\mathbf{O}$
\end{tabular}
\right)
\end{equation}
where $\mathbf{O}$ is the zero vector of length $N_\Nunsplit$. 
Therefore, the statistical significance of the update differences $(\theta_{i\pcopy}^p, \theta_\unsplitcopy^p)^T - \tilde{\theta}_{\joint}^p$ is the same as that of the parameter shifts $\theta_\pcopy^p - D_\splitcopy \theta_{i\pcopy}^p$ since they are related by a linear and invertible transformation.
We can then always use $\tilde{\theta}_\joint^p$ to rewrite the update parameter difference as:
\begin{equation}\label{App:ParamSplitUpdateShiftTilde}
\left( 
\begin{tabular}{c}
$\theta_{i\pcopy}^p$ \\
$\theta_\unsplitcopy^p$
\end{tabular}
\right)  - \theta_{\joint}^p = 
\left[
\left( 
\begin{tabular}{c}
$\theta_{i\pcopy}^p$ \\
$\theta_\unsplitcopy^p$
\end{tabular}
\right) - 
\tilde{\theta}_{\joint}^p 
\right] + (\tilde{\theta}_{\joint}^p - \theta_{\joint}^p) \; .
\end{equation}

To complete the generalization in the case of $n$ parameter copies with unsplit parameters, we now discuss how the statistics of goodness-of-fit loss both at the ML level, through $Q_{\rm DML}^{\pcopy} \equiv X_\joint^T A_{\rm DML}^{\pcopy} X_\joint$, and at the level of MAP, through $Q_{\rm DMAP}^{\pcopy} \equiv X_\joint^T A_{\rm DMAP}^{\pcopy} X_\joint$, can be computed.
To do so we can directly follow the discussion in App.~\ref{App:QDMAPcopyExact} to construct the matrices $A_{\rm DML}^{\pcopy} \mathcal{S}_\joint$ and $A_{\rm DMAP}^{\pcopy} \mathcal{S}_\joint$, respectively for ML and MAP, where we define the covariance matrix $\mathcal{S}_\joint \equiv \Sigma_\joint + M_{\joint} \mathcal{C}_\Pi M_{\joint}^T$ for Gaussian priors. 

Doing so it can be shown that the expressions for both ML and MAP goodness-of-fit loss statistics remain invariant compared to the results in App.~\ref{App:QDMAPcopyExact}, provided that one uses the design matrix in Eq.~\eqref{AppMultisplitUnslit:DesignMatrix}. 
Therefore, it is still true that the joint projector is a subset of the copy one, and thus $Q_{\rm DML}^\pcopy$ is chi-squared distributed with ${\rm rank}(\mathbb{I}-\mathbb{P}_{\joint}) - {\rm rank}(\mathbb{I}-\mathbb{P}_{\pcopy})$ degrees of freedom. 

Furthermore, $Q_{\rm DMAP}$ can be approximated by a chi-square distribution by matching moments of the approximate and exact distributions; the mean will then be given by the equivalent of Eq.~\eqref{Eq:AppDMAPexactASigmaTrace} if we define $N_{\rm eff}^{\pcopy} = nN_\Nsplit + N_\Nunsplit - {\rm tr}[\mathcal{C}_{\Pi\pcopy}^{-1} \mathcal{C}_{p\pcopy}]$. 

At last we highlight that, with the MCMC chain of multiple parameter copies we can easily construct the distribution of parameter differences and proceed with the statistical significance calculation as in Sec.~\ref{Sec:ParamSplitMCMCParameterShift} to compute the overall statistical significance of multiple parameter shifts.

\bibliographystyle{apsrev4-1}
\bibliography{biblio}

\begin{thebibliography}{48}%
\makeatletter
\providecommand \@ifxundefined [1]{%
 \@ifx{#1\undefined}
}%
\providecommand \@ifnum [1]{%
 \ifnum #1\expandafter \@firstoftwo
 \else \expandafter \@secondoftwo
 \fi
}%
\providecommand \@ifx [1]{%
 \ifx #1\expandafter \@firstoftwo
 \else \expandafter \@secondoftwo
 \fi
}%
\providecommand \natexlab [1]{#1}%
\providecommand \enquote  [1]{``#1''}%
\providecommand \bibnamefont  [1]{#1}%
\providecommand \bibfnamefont [1]{#1}%
\providecommand \citenamefont [1]{#1}%
\providecommand \href@noop [0]{\@secondoftwo}%
\providecommand \href [0]{\begingroup \@sanitize@url \@href}%
\providecommand \@href[1]{\@@startlink{#1}\@@href}%
\providecommand \@@href[1]{\endgroup#1\@@endlink}%
\providecommand \@sanitize@url [0]{\catcode `\\12\catcode `\$12\catcode
  `\&12\catcode `\#12\catcode `\^12\catcode `\_12\catcode `\%12\relax}%
\providecommand \@@startlink[1]{}%
\providecommand \@@endlink[0]{}%
\providecommand \url  [0]{\begingroup\@sanitize@url \@url }%
\providecommand \@url [1]{\endgroup\@href {#1}{\urlprefix }}%
\providecommand \urlprefix  [0]{URL }%
\providecommand \Eprint [0]{\href }%
\providecommand \doibase [0]{http://dx.doi.org/}%
\providecommand \selectlanguage [0]{\@gobble}%
\providecommand \bibinfo  [0]{\@secondoftwo}%
\providecommand \bibfield  [0]{\@secondoftwo}%
\providecommand \translation [1]{[#1]}%
\providecommand \BibitemOpen [0]{}%
\providecommand \bibitemStop [0]{}%
\providecommand \bibitemNoStop [0]{.\EOS\space}%
\providecommand \EOS [0]{\spacefactor3000\relax}%
\providecommand \BibitemShut  [1]{\csname bibitem#1\endcsname}%
\let\auto@bib@innerbib\@empty
\bibitem [{\citenamefont {Verde}\ \emph {et~al.}(2019)\citenamefont {Verde},
  \citenamefont {Treu},\ and\ \citenamefont {Riess}}]{Verde:2019ivm}%
  \BibitemOpen
  \bibfield  {author} {\bibinfo {author} {\bibfnamefont {L.}~\bibnamefont
  {Verde}}, \bibinfo {author} {\bibfnamefont {T.}~\bibnamefont {Treu}}, \ and\
  \bibinfo {author} {\bibfnamefont {A.~G.}\ \bibnamefont {Riess}},\ }in\ \href
  {\doibase 10.1038/s41550-019-0902-0} {\emph {\bibinfo {booktitle} {{Nature
  Astronomy 2019}}}}\ (\bibinfo {year} {2019})\ \Eprint
  {http://arxiv.org/abs/1907.10625} {arXiv:1907.10625 [astro-ph.CO]}
  \BibitemShut {NoStop}%
\bibitem [{\citenamefont {Marshall}\ \emph {et~al.}(2006)\citenamefont
  {Marshall}, \citenamefont {Rajguru},\ and\ \citenamefont
  {Slosar}}]{Marshall:2004zd}%
  \BibitemOpen
  \bibfield  {author} {\bibinfo {author} {\bibfnamefont {P.}~\bibnamefont
  {Marshall}}, \bibinfo {author} {\bibfnamefont {N.}~\bibnamefont {Rajguru}}, \
  and\ \bibinfo {author} {\bibfnamefont {A.}~\bibnamefont {Slosar}},\ }\href
  {\doibase 10.1103/PhysRevD.73.067302} {\bibfield  {journal} {\bibinfo
  {journal} {Phys. Rev.}\ }\textbf {\bibinfo {volume} {D73}},\ \bibinfo {pages}
  {067302} (\bibinfo {year} {2006})},\ \Eprint
  {http://arxiv.org/abs/astro-ph/0412535} {arXiv:astro-ph/0412535 [astro-ph]}
  \BibitemShut {NoStop}%
\bibitem [{\citenamefont {Feroz}\ \emph {et~al.}(2008)\citenamefont {Feroz},
  \citenamefont {Allanach}, \citenamefont {Hobson}, \citenamefont {AbdusSalam},
  \citenamefont {Trotta},\ and\ \citenamefont {Weber}}]{Feroz:2008wr}%
  \BibitemOpen
  \bibfield  {author} {\bibinfo {author} {\bibfnamefont {F.}~\bibnamefont
  {Feroz}}, \bibinfo {author} {\bibfnamefont {B.~C.}\ \bibnamefont {Allanach}},
  \bibinfo {author} {\bibfnamefont {M.}~\bibnamefont {Hobson}}, \bibinfo
  {author} {\bibfnamefont {S.~S.}\ \bibnamefont {AbdusSalam}}, \bibinfo
  {author} {\bibfnamefont {R.}~\bibnamefont {Trotta}}, \ and\ \bibinfo {author}
  {\bibfnamefont {A.~M.}\ \bibnamefont {Weber}},\ }\href {\doibase
  10.1088/1126-6708/2008/10/064} {\bibfield  {journal} {\bibinfo  {journal}
  {JHEP}\ }\textbf {\bibinfo {volume} {10}},\ \bibinfo {pages} {064} (\bibinfo
  {year} {2008})},\ \Eprint {http://arxiv.org/abs/0807.4512} {arXiv:0807.4512
  [hep-ph]} \BibitemShut {NoStop}%
\bibitem [{\citenamefont {March}\ \emph {et~al.}(2011)\citenamefont {March},
  \citenamefont {Trotta}, \citenamefont {Amendola},\ and\ \citenamefont
  {Huterer}}]{March:2011rv}%
  \BibitemOpen
  \bibfield  {author} {\bibinfo {author} {\bibfnamefont {M.~C.}\ \bibnamefont
  {March}}, \bibinfo {author} {\bibfnamefont {R.}~\bibnamefont {Trotta}},
  \bibinfo {author} {\bibfnamefont {L.}~\bibnamefont {Amendola}}, \ and\
  \bibinfo {author} {\bibfnamefont {D.}~\bibnamefont {Huterer}},\ }\href
  {\doibase 10.1111/j.1365-2966.2011.18679.x} {\bibfield  {journal} {\bibinfo
  {journal} {Mon. Not. Roy. Astron. Soc.}\ }\textbf {\bibinfo {volume} {415}},\
  \bibinfo {pages} {143} (\bibinfo {year} {2011})},\ \Eprint
  {http://arxiv.org/abs/1101.1521} {arXiv:1101.1521 [astro-ph.CO]} \BibitemShut
  {NoStop}%
\bibitem [{\citenamefont {Amendola}\ \emph {et~al.}(2013)\citenamefont
  {Amendola}, \citenamefont {Marra},\ and\ \citenamefont
  {Quartin}}]{Amendola:2012wc}%
  \BibitemOpen
  \bibfield  {author} {\bibinfo {author} {\bibfnamefont {L.}~\bibnamefont
  {Amendola}}, \bibinfo {author} {\bibfnamefont {V.}~\bibnamefont {Marra}}, \
  and\ \bibinfo {author} {\bibfnamefont {M.}~\bibnamefont {Quartin}},\ }\href
  {\doibase 10.1093/mnras/stt008} {\bibfield  {journal} {\bibinfo  {journal}
  {Mon. Not. Roy. Astron. Soc.}\ }\textbf {\bibinfo {volume} {430}},\ \bibinfo
  {pages} {1867} (\bibinfo {year} {2013})},\ \Eprint
  {http://arxiv.org/abs/1209.1897} {arXiv:1209.1897 [astro-ph.CO]} \BibitemShut
  {NoStop}%
\bibitem [{\citenamefont {Verde}\ \emph {et~al.}(2013)\citenamefont {Verde},
  \citenamefont {Protopapas},\ and\ \citenamefont {Jimenez}}]{Verde:2013wza}%
  \BibitemOpen
  \bibfield  {author} {\bibinfo {author} {\bibfnamefont {L.}~\bibnamefont
  {Verde}}, \bibinfo {author} {\bibfnamefont {P.}~\bibnamefont {Protopapas}}, \
  and\ \bibinfo {author} {\bibfnamefont {R.}~\bibnamefont {Jimenez}},\ }\href
  {\doibase 10.1016/j.dark.2013.09.002} {\bibfield  {journal} {\bibinfo
  {journal} {Phys. Dark Univ.}\ }\textbf {\bibinfo {volume} {2}},\ \bibinfo
  {pages} {166} (\bibinfo {year} {2013})},\ \Eprint
  {http://arxiv.org/abs/1306.6766} {arXiv:1306.6766 [astro-ph.CO]} \BibitemShut
  {NoStop}%
\bibitem [{\citenamefont {Bennett}\ \emph {et~al.}(2014)\citenamefont
  {Bennett}, \citenamefont {Larson}, \citenamefont {Weiland},\ and\
  \citenamefont {Hinshaw}}]{Bennett:2014tka}%
  \BibitemOpen
  \bibfield  {author} {\bibinfo {author} {\bibfnamefont {C.~L.}\ \bibnamefont
  {Bennett}}, \bibinfo {author} {\bibfnamefont {D.}~\bibnamefont {Larson}},
  \bibinfo {author} {\bibfnamefont {J.~L.}\ \bibnamefont {Weiland}}, \ and\
  \bibinfo {author} {\bibfnamefont {G.}~\bibnamefont {Hinshaw}},\ }\href
  {\doibase 10.1088/0004-637X/794/2/135} {\bibfield  {journal} {\bibinfo
  {journal} {Astrophys. J.}\ }\textbf {\bibinfo {volume} {794}},\ \bibinfo
  {pages} {135} (\bibinfo {year} {2014})},\ \Eprint
  {http://arxiv.org/abs/1406.1718} {arXiv:1406.1718 [astro-ph.CO]} \BibitemShut
  {NoStop}%
\bibitem [{\citenamefont {Martin}\ \emph {et~al.}(2014)\citenamefont {Martin},
  \citenamefont {Ringeval}, \citenamefont {Trotta},\ and\ \citenamefont
  {Vennin}}]{Martin:2014lra}%
  \BibitemOpen
  \bibfield  {author} {\bibinfo {author} {\bibfnamefont {J.}~\bibnamefont
  {Martin}}, \bibinfo {author} {\bibfnamefont {C.}~\bibnamefont {Ringeval}},
  \bibinfo {author} {\bibfnamefont {R.}~\bibnamefont {Trotta}}, \ and\ \bibinfo
  {author} {\bibfnamefont {V.}~\bibnamefont {Vennin}},\ }\href {\doibase
  10.1103/PhysRevD.90.063501} {\bibfield  {journal} {\bibinfo  {journal} {Phys.
  Rev.}\ }\textbf {\bibinfo {volume} {D90}},\ \bibinfo {pages} {063501}
  (\bibinfo {year} {2014})},\ \Eprint {http://arxiv.org/abs/1405.7272}
  {arXiv:1405.7272 [astro-ph.CO]} \BibitemShut {NoStop}%
\bibitem [{\citenamefont {Karpenka}\ \emph {et~al.}(2015)\citenamefont
  {Karpenka}, \citenamefont {Feroz},\ and\ \citenamefont
  {Hobson}}]{Karpenka:2014moa}%
  \BibitemOpen
  \bibfield  {author} {\bibinfo {author} {\bibfnamefont {N.~V.}\ \bibnamefont
  {Karpenka}}, \bibinfo {author} {\bibfnamefont {F.}~\bibnamefont {Feroz}}, \
  and\ \bibinfo {author} {\bibfnamefont {M.~P.}\ \bibnamefont {Hobson}},\
  }\href {\doibase 10.1093/mnras/stv415} {\bibfield  {journal} {\bibinfo
  {journal} {Mon. Not. Roy. Astron. Soc.}\ }\textbf {\bibinfo {volume} {449}},\
  \bibinfo {pages} {2405} (\bibinfo {year} {2015})},\ \Eprint
  {http://arxiv.org/abs/1407.5496} {arXiv:1407.5496 [astro-ph.IM]} \BibitemShut
  {NoStop}%
\bibitem [{\citenamefont {Larson}\ \emph {et~al.}(2015)\citenamefont {Larson},
  \citenamefont {Weiland}, \citenamefont {Hinshaw},\ and\ \citenamefont
  {Bennett}}]{Larson:2014roa}%
  \BibitemOpen
  \bibfield  {author} {\bibinfo {author} {\bibfnamefont {D.}~\bibnamefont
  {Larson}}, \bibinfo {author} {\bibfnamefont {J.~L.}\ \bibnamefont {Weiland}},
  \bibinfo {author} {\bibfnamefont {G.}~\bibnamefont {Hinshaw}}, \ and\
  \bibinfo {author} {\bibfnamefont {C.~L.}\ \bibnamefont {Bennett}},\ }\href
  {\doibase 10.1088/0004-637X/801/1/9} {\bibfield  {journal} {\bibinfo
  {journal} {Astrophys. J.}\ }\textbf {\bibinfo {volume} {801}},\ \bibinfo
  {pages} {9} (\bibinfo {year} {2015})},\ \Eprint
  {http://arxiv.org/abs/1409.7718} {arXiv:1409.7718 [astro-ph.CO]} \BibitemShut
  {NoStop}%
\bibitem [{\citenamefont {Addison}\ \emph {et~al.}(2016)\citenamefont
  {Addison}, \citenamefont {Huang}, \citenamefont {Watts}, \citenamefont
  {Bennett}, \citenamefont {Halpern}, \citenamefont {Hinshaw},\ and\
  \citenamefont {Weiland}}]{Addison:2015wyg}%
  \BibitemOpen
  \bibfield  {author} {\bibinfo {author} {\bibfnamefont {G.~E.}\ \bibnamefont
  {Addison}}, \bibinfo {author} {\bibfnamefont {Y.}~\bibnamefont {Huang}},
  \bibinfo {author} {\bibfnamefont {D.~J.}\ \bibnamefont {Watts}}, \bibinfo
  {author} {\bibfnamefont {C.~L.}\ \bibnamefont {Bennett}}, \bibinfo {author}
  {\bibfnamefont {M.}~\bibnamefont {Halpern}}, \bibinfo {author} {\bibfnamefont
  {G.}~\bibnamefont {Hinshaw}}, \ and\ \bibinfo {author} {\bibfnamefont
  {J.~L.}\ \bibnamefont {Weiland}},\ }\href {\doibase
  10.3847/0004-637X/818/2/132} {\bibfield  {journal} {\bibinfo  {journal}
  {Astrophys. J.}\ }\textbf {\bibinfo {volume} {818}},\ \bibinfo {pages} {132}
  (\bibinfo {year} {2016})},\ \Eprint {http://arxiv.org/abs/1511.00055}
  {arXiv:1511.00055 [astro-ph.CO]} \BibitemShut {NoStop}%
\bibitem [{\citenamefont {Raveri}(2016)}]{Raveri:2015maa}%
  \BibitemOpen
  \bibfield  {author} {\bibinfo {author} {\bibfnamefont {M.}~\bibnamefont
  {Raveri}},\ }\href {\doibase 10.1103/PhysRevD.93.043522} {\bibfield
  {journal} {\bibinfo  {journal} {Phys. Rev.}\ }\textbf {\bibinfo {volume}
  {D93}},\ \bibinfo {pages} {043522} (\bibinfo {year} {2016})},\ \Eprint
  {http://arxiv.org/abs/1510.00688} {arXiv:1510.00688 [astro-ph.CO]}
  \BibitemShut {NoStop}%
\bibitem [{\citenamefont {Seehars}\ \emph {et~al.}(2014)\citenamefont
  {Seehars}, \citenamefont {Amara}, \citenamefont {Refregier}, \citenamefont
  {Paranjape},\ and\ \citenamefont {Akeret}}]{Seehars:2014ora}%
  \BibitemOpen
  \bibfield  {author} {\bibinfo {author} {\bibfnamefont {S.}~\bibnamefont
  {Seehars}}, \bibinfo {author} {\bibfnamefont {A.}~\bibnamefont {Amara}},
  \bibinfo {author} {\bibfnamefont {A.}~\bibnamefont {Refregier}}, \bibinfo
  {author} {\bibfnamefont {A.}~\bibnamefont {Paranjape}}, \ and\ \bibinfo
  {author} {\bibfnamefont {J.}~\bibnamefont {Akeret}},\ }\href {\doibase
  10.1103/PhysRevD.90.023533} {\bibfield  {journal} {\bibinfo  {journal} {Phys.
  Rev.}\ }\textbf {\bibinfo {volume} {D90}},\ \bibinfo {pages} {023533}
  (\bibinfo {year} {2014})},\ \Eprint {http://arxiv.org/abs/1402.3593}
  {arXiv:1402.3593 [astro-ph.CO]} \BibitemShut {NoStop}%
\bibitem [{\citenamefont {Seehars}\ \emph {et~al.}(2016)\citenamefont
  {Seehars}, \citenamefont {Grandis}, \citenamefont {Amara},\ and\
  \citenamefont {Refregier}}]{Seehars:2015qza}%
  \BibitemOpen
  \bibfield  {author} {\bibinfo {author} {\bibfnamefont {S.}~\bibnamefont
  {Seehars}}, \bibinfo {author} {\bibfnamefont {S.}~\bibnamefont {Grandis}},
  \bibinfo {author} {\bibfnamefont {A.}~\bibnamefont {Amara}}, \ and\ \bibinfo
  {author} {\bibfnamefont {A.}~\bibnamefont {Refregier}},\ }\href {\doibase
  10.1103/PhysRevD.93.103507} {\bibfield  {journal} {\bibinfo  {journal} {Phys.
  Rev.}\ }\textbf {\bibinfo {volume} {D93}},\ \bibinfo {pages} {103507}
  (\bibinfo {year} {2016})},\ \Eprint {http://arxiv.org/abs/1510.08483}
  {arXiv:1510.08483 [astro-ph.CO]} \BibitemShut {NoStop}%
\bibitem [{\citenamefont {Grandis}\ \emph {et~al.}(2016)\citenamefont
  {Grandis}, \citenamefont {Rapetti}, \citenamefont {Saro}, \citenamefont
  {Mohr},\ and\ \citenamefont {Dietrich}}]{Grandis:2016fwl}%
  \BibitemOpen
  \bibfield  {author} {\bibinfo {author} {\bibfnamefont {S.}~\bibnamefont
  {Grandis}}, \bibinfo {author} {\bibfnamefont {D.}~\bibnamefont {Rapetti}},
  \bibinfo {author} {\bibfnamefont {A.}~\bibnamefont {Saro}}, \bibinfo {author}
  {\bibfnamefont {J.~J.}\ \bibnamefont {Mohr}}, \ and\ \bibinfo {author}
  {\bibfnamefont {J.~P.}\ \bibnamefont {Dietrich}},\ }\href {\doibase
  10.1093/mnras/stw2028} {\bibfield  {journal} {\bibinfo  {journal} {Mon. Not.
  Roy. Astron. Soc.}\ }\textbf {\bibinfo {volume} {463}},\ \bibinfo {pages}
  {1416} (\bibinfo {year} {2016})},\ \Eprint {http://arxiv.org/abs/1604.06463}
  {arXiv:1604.06463 [astro-ph.CO]} \BibitemShut {NoStop}%
\bibitem [{\citenamefont {Addison}\ \emph {et~al.}(2018)\citenamefont
  {Addison}, \citenamefont {Watts}, \citenamefont {Bennett}, \citenamefont
  {Halpern}, \citenamefont {Hinshaw},\ and\ \citenamefont
  {Weiland}}]{Addison:2017fdm}%
  \BibitemOpen
  \bibfield  {author} {\bibinfo {author} {\bibfnamefont {G.~E.}\ \bibnamefont
  {Addison}}, \bibinfo {author} {\bibfnamefont {D.~J.}\ \bibnamefont {Watts}},
  \bibinfo {author} {\bibfnamefont {C.~L.}\ \bibnamefont {Bennett}}, \bibinfo
  {author} {\bibfnamefont {M.}~\bibnamefont {Halpern}}, \bibinfo {author}
  {\bibfnamefont {G.}~\bibnamefont {Hinshaw}}, \ and\ \bibinfo {author}
  {\bibfnamefont {J.~L.}\ \bibnamefont {Weiland}},\ }\href {\doibase
  10.3847/1538-4357/aaa1ed} {\bibfield  {journal} {\bibinfo  {journal}
  {Astrophys. J.}\ }\textbf {\bibinfo {volume} {853}},\ \bibinfo {pages} {119}
  (\bibinfo {year} {2018})},\ \Eprint {http://arxiv.org/abs/1707.06547}
  {arXiv:1707.06547 [astro-ph.CO]} \BibitemShut {NoStop}%
\bibitem [{\citenamefont {Nicola}\ \emph {et~al.}(2019)\citenamefont {Nicola},
  \citenamefont {Amara},\ and\ \citenamefont {Refregier}}]{Nicola:2018rcd}%
  \BibitemOpen
  \bibfield  {author} {\bibinfo {author} {\bibfnamefont {A.}~\bibnamefont
  {Nicola}}, \bibinfo {author} {\bibfnamefont {A.}~\bibnamefont {Amara}}, \
  and\ \bibinfo {author} {\bibfnamefont {A.}~\bibnamefont {Refregier}},\ }\href
  {\doibase 10.1088/1475-7516/2019/01/011} {\bibfield  {journal} {\bibinfo
  {journal} {JCAP}\ }\textbf {\bibinfo {volume} {1901}},\ \bibinfo {pages}
  {011} (\bibinfo {year} {2019})},\ \Eprint {http://arxiv.org/abs/1809.07333}
  {arXiv:1809.07333 [astro-ph.CO]} \BibitemShut {NoStop}%
\bibitem [{\citenamefont {Weiland}\ \emph {et~al.}(2018)\citenamefont
  {Weiland}, \citenamefont {Osumi}, \citenamefont {Addison}, \citenamefont
  {Bennett}, \citenamefont {Watts}, \citenamefont {Halpern},\ and\
  \citenamefont {Hinshaw}}]{Weiland:2018kon}%
  \BibitemOpen
  \bibfield  {author} {\bibinfo {author} {\bibfnamefont {J.~L.}\ \bibnamefont
  {Weiland}}, \bibinfo {author} {\bibfnamefont {K.}~\bibnamefont {Osumi}},
  \bibinfo {author} {\bibfnamefont {G.~E.}\ \bibnamefont {Addison}}, \bibinfo
  {author} {\bibfnamefont {C.~L.}\ \bibnamefont {Bennett}}, \bibinfo {author}
  {\bibfnamefont {D.~J.}\ \bibnamefont {Watts}}, \bibinfo {author}
  {\bibfnamefont {M.}~\bibnamefont {Halpern}}, \ and\ \bibinfo {author}
  {\bibfnamefont {G.}~\bibnamefont {Hinshaw}},\ }\href {\doibase
  10.3847/1538-4357/aad18b} {\bibfield  {journal} {\bibinfo  {journal}
  {Astrophys. J.}\ }\textbf {\bibinfo {volume} {863}},\ \bibinfo {pages} {161}
  (\bibinfo {year} {2018})},\ \Eprint {http://arxiv.org/abs/1801.01226}
  {arXiv:1801.01226 [astro-ph.CO]} \BibitemShut {NoStop}%
\bibitem [{\citenamefont {Huang}\ \emph {et~al.}(2018)\citenamefont {Huang},
  \citenamefont {Addison}, \citenamefont {Weiland},\ and\ \citenamefont
  {Bennett}}]{Huang:2018xle}%
  \BibitemOpen
  \bibfield  {author} {\bibinfo {author} {\bibfnamefont {Y.}~\bibnamefont
  {Huang}}, \bibinfo {author} {\bibfnamefont {G.~E.}\ \bibnamefont {Addison}},
  \bibinfo {author} {\bibfnamefont {J.~L.}\ \bibnamefont {Weiland}}, \ and\
  \bibinfo {author} {\bibfnamefont {C.~L.}\ \bibnamefont {Bennett}},\ }\href
  {\doibase 10.3847/1538-4357/aaeb1f} {\bibfield  {journal} {\bibinfo
  {journal} {Astrophys. J.}\ }\textbf {\bibinfo {volume} {869}},\ \bibinfo
  {pages} {38} (\bibinfo {year} {2018})},\ \Eprint
  {http://arxiv.org/abs/1804.05428} {arXiv:1804.05428 [astro-ph.CO]}
  \BibitemShut {NoStop}%
\bibitem [{\citenamefont {Raveri}\ and\ \citenamefont
  {Hu}(2019)}]{Raveri:2018wln}%
  \BibitemOpen
  \bibfield  {author} {\bibinfo {author} {\bibfnamefont {M.}~\bibnamefont
  {Raveri}}\ and\ \bibinfo {author} {\bibfnamefont {W.}~\bibnamefont {Hu}},\
  }\href {\doibase 10.1103/PhysRevD.99.043506} {\bibfield  {journal} {\bibinfo
  {journal} {Phys. Rev.}\ }\textbf {\bibinfo {volume} {D99}},\ \bibinfo {pages}
  {043506} (\bibinfo {year} {2019})},\ \Eprint
  {http://arxiv.org/abs/1806.04649} {arXiv:1806.04649 [astro-ph.CO]}
  \BibitemShut {NoStop}%
\bibitem [{\citenamefont {Motloch}\ and\ \citenamefont
  {Hu}(2018)}]{Motloch:2018pjy}%
  \BibitemOpen
  \bibfield  {author} {\bibinfo {author} {\bibfnamefont {P.}~\bibnamefont
  {Motloch}}\ and\ \bibinfo {author} {\bibfnamefont {W.}~\bibnamefont {Hu}},\
  }\href {\doibase 10.1103/PhysRevD.97.103536} {\bibfield  {journal} {\bibinfo
  {journal} {Phys. Rev.}\ }\textbf {\bibinfo {volume} {D97}},\ \bibinfo {pages}
  {103536} (\bibinfo {year} {2018})},\ \Eprint
  {http://arxiv.org/abs/1803.11526} {arXiv:1803.11526 [astro-ph.CO]}
  \BibitemShut {NoStop}%
\bibitem [{\citenamefont {Motloch}\ and\ \citenamefont
  {Hu}(2019)}]{Motloch:2018pyt}%
  \BibitemOpen
  \bibfield  {author} {\bibinfo {author} {\bibfnamefont {P.}~\bibnamefont
  {Motloch}}\ and\ \bibinfo {author} {\bibfnamefont {W.}~\bibnamefont {Hu}},\
  }\href {\doibase 10.1103/PhysRevD.99.023506} {\bibfield  {journal} {\bibinfo
  {journal} {Phys. Rev.}\ }\textbf {\bibinfo {volume} {D99}},\ \bibinfo {pages}
  {023506} (\bibinfo {year} {2019})},\ \Eprint
  {http://arxiv.org/abs/1810.09347} {arXiv:1810.09347 [astro-ph.CO]}
  \BibitemShut {NoStop}%
\bibitem [{\citenamefont {Adhikari}\ and\ \citenamefont
  {Huterer}(2019)}]{Adhikari_2019}%
  \BibitemOpen
  \bibfield  {author} {\bibinfo {author} {\bibfnamefont {S.}~\bibnamefont
  {Adhikari}}\ and\ \bibinfo {author} {\bibfnamefont {D.}~\bibnamefont
  {Huterer}},\ }\href {\doibase 10.1088/1475-7516/2019/01/036} {\bibfield
  {journal} {\bibinfo  {journal} {Journal of Cosmology and Astroparticle
  Physics}\ }\textbf {\bibinfo {volume} {2019}},\ \bibinfo {pages} {036}
  (\bibinfo {year} {2019})}\BibitemShut {NoStop}%
\bibitem [{\citenamefont {Kerscher}\ and\ \citenamefont
  {Weller}(2019)}]{Kerscher:2019pzk}%
  \BibitemOpen
  \bibfield  {author} {\bibinfo {author} {\bibfnamefont {M.}~\bibnamefont
  {Kerscher}}\ and\ \bibinfo {author} {\bibfnamefont {J.}~\bibnamefont
  {Weller}},\ }\href {\doibase 10.21468/SciPostPhysLectNotes.9} {\  (\bibinfo
  {year} {2019}),\ 10.21468/SciPostPhysLectNotes.9},\ \Eprint
  {http://arxiv.org/abs/1901.07726} {arXiv:1901.07726 [astro-ph.CO]}
  \BibitemShut {NoStop}%
\bibitem [{\citenamefont {Handley}\ and\ \citenamefont
  {Lemos}(2019)}]{Handley:2019wlz}%
  \BibitemOpen
  \bibfield  {author} {\bibinfo {author} {\bibfnamefont {W.}~\bibnamefont
  {Handley}}\ and\ \bibinfo {author} {\bibfnamefont {P.}~\bibnamefont
  {Lemos}},\ }\href {\doibase 10.1103/PhysRevD.100.043504} {\bibfield
  {journal} {\bibinfo  {journal} {Phys. Rev.}\ }\textbf {\bibinfo {volume}
  {D100}},\ \bibinfo {pages} {043504} (\bibinfo {year} {2019})},\ \Eprint
  {http://arxiv.org/abs/1902.04029} {arXiv:1902.04029 [astro-ph.CO]}
  \BibitemShut {NoStop}%
\bibitem [{\citenamefont {Huang}\ \emph {et~al.}(2019)\citenamefont {Huang},
  \citenamefont {Addison},\ and\ \citenamefont {Bennett}}]{Huang:2019nor}%
  \BibitemOpen
  \bibfield  {author} {\bibinfo {author} {\bibfnamefont {Y.}~\bibnamefont
  {Huang}}, \bibinfo {author} {\bibfnamefont {G.}~\bibnamefont {Addison}}, \
  and\ \bibinfo {author} {\bibfnamefont {C.}~\bibnamefont {Bennett}},\ }\href
  {\doibase 10.3847/1538-4357/ab3654} {\  (\bibinfo {year} {2019}),\
  10.3847/1538-4357/ab3654},\ \Eprint {http://arxiv.org/abs/1904.10521}
  {arXiv:1904.10521 [astro-ph.CO]} \BibitemShut {NoStop}%
\bibitem [{\citenamefont {Lin}\ and\ \citenamefont
  {Ishak}(2019)}]{Lin:2019zdn}%
  \BibitemOpen
  \bibfield  {author} {\bibinfo {author} {\bibfnamefont {W.}~\bibnamefont
  {Lin}}\ and\ \bibinfo {author} {\bibfnamefont {M.}~\bibnamefont {Ishak}},\
  }\href@noop {} {\  (\bibinfo {year} {2019})},\ \Eprint
  {http://arxiv.org/abs/1909.10991} {arXiv:1909.10991 [astro-ph.CO]}
  \BibitemShut {NoStop}%
\bibitem [{\citenamefont {Zhang}\ \emph {et~al.}(2005)\citenamefont {Zhang},
  \citenamefont {Hui},\ and\ \citenamefont {Stebbins}}]{Zhang:2003ii}%
  \BibitemOpen
  \bibfield  {author} {\bibinfo {author} {\bibfnamefont {J.}~\bibnamefont
  {Zhang}}, \bibinfo {author} {\bibfnamefont {L.}~\bibnamefont {Hui}}, \ and\
  \bibinfo {author} {\bibfnamefont {A.}~\bibnamefont {Stebbins}},\ }\href
  {\doibase 10.1086/497676} {\bibfield  {journal} {\bibinfo  {journal}
  {Astrophys. J.}\ }\textbf {\bibinfo {volume} {635}},\ \bibinfo {pages} {806}
  (\bibinfo {year} {2005})},\ \Eprint {http://arxiv.org/abs/astro-ph/0312348}
  {arXiv:astro-ph/0312348 [astro-ph]} \BibitemShut {NoStop}%
\bibitem [{\citenamefont {Chu}\ and\ \citenamefont {Knox}(2005)}]{Chu:2004qx}%
  \BibitemOpen
  \bibfield  {author} {\bibinfo {author} {\bibfnamefont {M.}~\bibnamefont
  {Chu}}\ and\ \bibinfo {author} {\bibfnamefont {L.}~\bibnamefont {Knox}},\
  }\href {\doibase 10.1086/427064} {\bibfield  {journal} {\bibinfo  {journal}
  {Astrophys. J.}\ }\textbf {\bibinfo {volume} {620}},\ \bibinfo {pages} {1}
  (\bibinfo {year} {2005})},\ \Eprint {http://arxiv.org/abs/astro-ph/0407198}
  {arXiv:astro-ph/0407198 [astro-ph]} \BibitemShut {NoStop}%
\bibitem [{\citenamefont {Wang}\ \emph {et~al.}(2007)\citenamefont {Wang},
  \citenamefont {Hui}, \citenamefont {May},\ and\ \citenamefont
  {Haiman}}]{Wang:2007fsa}%
  \BibitemOpen
  \bibfield  {author} {\bibinfo {author} {\bibfnamefont {S.}~\bibnamefont
  {Wang}}, \bibinfo {author} {\bibfnamefont {L.}~\bibnamefont {Hui}}, \bibinfo
  {author} {\bibfnamefont {M.}~\bibnamefont {May}}, \ and\ \bibinfo {author}
  {\bibfnamefont {Z.}~\bibnamefont {Haiman}},\ }\href {\doibase
  10.1103/PhysRevD.76.063503} {\bibfield  {journal} {\bibinfo  {journal} {Phys.
  Rev.}\ }\textbf {\bibinfo {volume} {D76}},\ \bibinfo {pages} {063503}
  (\bibinfo {year} {2007})},\ \Eprint {http://arxiv.org/abs/0705.0165}
  {arXiv:0705.0165 [astro-ph]} \BibitemShut {NoStop}%
\bibitem [{\citenamefont {Abate}\ and\ \citenamefont
  {Lahav}(2008)}]{Abate:2008au}%
  \BibitemOpen
  \bibfield  {author} {\bibinfo {author} {\bibfnamefont {A.}~\bibnamefont
  {Abate}}\ and\ \bibinfo {author} {\bibfnamefont {O.}~\bibnamefont {Lahav}},\
  }\href {\doibase 10.1111/j.1745-3933.2008.00519.x} {\bibfield  {journal}
  {\bibinfo  {journal} {Mon. Not. Roy. Astron. Soc.}\ }\textbf {\bibinfo
  {volume} {389}},\ \bibinfo {pages} {47} (\bibinfo {year} {2008})},\ \Eprint
  {http://arxiv.org/abs/0805.3160} {arXiv:0805.3160 [astro-ph]} \BibitemShut
  {NoStop}%
\bibitem [{\citenamefont {Ruiz}\ and\ \citenamefont
  {Huterer}(2015)}]{Ruiz:2014hma}%
  \BibitemOpen
  \bibfield  {author} {\bibinfo {author} {\bibfnamefont {E.~J.}\ \bibnamefont
  {Ruiz}}\ and\ \bibinfo {author} {\bibfnamefont {D.}~\bibnamefont {Huterer}},\
  }\href {\doibase 10.1103/PhysRevD.91.063009} {\bibfield  {journal} {\bibinfo
  {journal} {Phys. Rev.}\ }\textbf {\bibinfo {volume} {D91}},\ \bibinfo {pages}
  {063009} (\bibinfo {year} {2015})},\ \Eprint {http://arxiv.org/abs/1410.5832}
  {arXiv:1410.5832 [astro-ph.CO]} \BibitemShut {NoStop}%
\bibitem [{\citenamefont {Bernal}\ \emph {et~al.}(2016)\citenamefont {Bernal},
  \citenamefont {Verde},\ and\ \citenamefont {Cuesta}}]{Bernal:2015zom}%
  \BibitemOpen
  \bibfield  {author} {\bibinfo {author} {\bibfnamefont {J.~L.}\ \bibnamefont
  {Bernal}}, \bibinfo {author} {\bibfnamefont {L.}~\bibnamefont {Verde}}, \
  and\ \bibinfo {author} {\bibfnamefont {A.~J.}\ \bibnamefont {Cuesta}},\
  }\href {\doibase 10.1088/1475-7516/2016/02/059} {\bibfield  {journal}
  {\bibinfo  {journal} {JCAP}\ }\textbf {\bibinfo {volume} {1602}},\ \bibinfo
  {pages} {059} (\bibinfo {year} {2016})},\ \Eprint
  {http://arxiv.org/abs/1511.03049} {arXiv:1511.03049 [astro-ph.CO]}
  \BibitemShut {NoStop}%
\bibitem [{\citenamefont {Köhlinger}\ \emph {et~al.}(2019)\citenamefont
  {Köhlinger}, \citenamefont {Joachimi}, \citenamefont {Asgari}, \citenamefont
  {Viola}, \citenamefont {Joudaki},\ and\ \citenamefont
  {Tröster}}]{Kohlinger:2018sxx}%
  \BibitemOpen
  \bibfield  {author} {\bibinfo {author} {\bibfnamefont {F.}~\bibnamefont
  {Köhlinger}}, \bibinfo {author} {\bibfnamefont {B.}~\bibnamefont
  {Joachimi}}, \bibinfo {author} {\bibfnamefont {M.}~\bibnamefont {Asgari}},
  \bibinfo {author} {\bibfnamefont {M.}~\bibnamefont {Viola}}, \bibinfo
  {author} {\bibfnamefont {S.}~\bibnamefont {Joudaki}}, \ and\ \bibinfo
  {author} {\bibfnamefont {T.}~\bibnamefont {Tröster}},\ }\href {\doibase
  10.1093/mnras/stz132} {\bibfield  {journal} {\bibinfo  {journal} {Mon. Not.
  Roy. Astron. Soc.}\ }\textbf {\bibinfo {volume} {484}},\ \bibinfo {pages}
  {3126} (\bibinfo {year} {2019})},\ \Eprint {http://arxiv.org/abs/1809.01406}
  {arXiv:1809.01406 [astro-ph.CO]} \BibitemShut {NoStop}%
\bibitem [{\citenamefont {Lemos}\ \emph {et~al.}(2019)\citenamefont {Lemos},
  \citenamefont {Köhlinger}, \citenamefont {Handley}, \citenamefont
  {Joachimi}, \citenamefont {Whiteway},\ and\ \citenamefont
  {Lahav}}]{Lemos:2019txn}%
  \BibitemOpen
  \bibfield  {author} {\bibinfo {author} {\bibfnamefont {P.}~\bibnamefont
  {Lemos}}, \bibinfo {author} {\bibfnamefont {F.}~\bibnamefont {Köhlinger}},
  \bibinfo {author} {\bibfnamefont {W.}~\bibnamefont {Handley}}, \bibinfo
  {author} {\bibfnamefont {B.}~\bibnamefont {Joachimi}}, \bibinfo {author}
  {\bibfnamefont {L.}~\bibnamefont {Whiteway}}, \ and\ \bibinfo {author}
  {\bibfnamefont {O.}~\bibnamefont {Lahav}},\ }\href@noop {} {\  (\bibinfo
  {year} {2019})},\ \Eprint {http://arxiv.org/abs/1910.07820} {arXiv:1910.07820
  [astro-ph.CO]} \BibitemShut {NoStop}%
\bibitem [{\citenamefont {Scolnic}\ \emph {et~al.}(2018)\citenamefont {Scolnic}
  \emph {et~al.}}]{Scolnic:2017caz}%
  \BibitemOpen
  \bibfield  {author} {\bibinfo {author} {\bibfnamefont {D.~M.}\ \bibnamefont
  {Scolnic}} \emph {et~al.},\ }\href {\doibase 10.3847/1538-4357/aab9bb}
  {\bibfield  {journal} {\bibinfo  {journal} {Astrophys. J.}\ }\textbf
  {\bibinfo {volume} {859}},\ \bibinfo {pages} {101} (\bibinfo {year}
  {2018})},\ \Eprint {http://arxiv.org/abs/1710.00845} {arXiv:1710.00845
  [astro-ph.CO]} \BibitemShut {NoStop}%
\bibitem [{\citenamefont {Riess}\ \emph {et~al.}(2019)\citenamefont {Riess},
  \citenamefont {Casertano}, \citenamefont {Yuan}, \citenamefont {Macri},\ and\
  \citenamefont {Scolnic}}]{Riess:2019cxk}%
  \BibitemOpen
  \bibfield  {author} {\bibinfo {author} {\bibfnamefont {A.~G.}\ \bibnamefont
  {Riess}}, \bibinfo {author} {\bibfnamefont {S.}~\bibnamefont {Casertano}},
  \bibinfo {author} {\bibfnamefont {W.}~\bibnamefont {Yuan}}, \bibinfo {author}
  {\bibfnamefont {L.~M.}\ \bibnamefont {Macri}}, \ and\ \bibinfo {author}
  {\bibfnamefont {D.}~\bibnamefont {Scolnic}},\ }\href {\doibase
  10.3847/1538-4357/ab1422} {\bibfield  {journal} {\bibinfo  {journal}
  {Astrophys. J.}\ }\textbf {\bibinfo {volume} {876}},\ \bibinfo {pages} {85}
  (\bibinfo {year} {2019})},\ \Eprint {http://arxiv.org/abs/1903.07603}
  {arXiv:1903.07603 [astro-ph.CO]} \BibitemShut {NoStop}%
\bibitem [{\citenamefont {Guy}\ \emph {et~al.}(2005)\citenamefont {Guy},
  \citenamefont {Astier}, \citenamefont {Nobili}, \citenamefont {Regnault},\
  and\ \citenamefont {Pain}}]{Guy:2005me}%
  \BibitemOpen
  \bibfield  {author} {\bibinfo {author} {\bibfnamefont {J.}~\bibnamefont
  {Guy}}, \bibinfo {author} {\bibfnamefont {P.}~\bibnamefont {Astier}},
  \bibinfo {author} {\bibfnamefont {S.}~\bibnamefont {Nobili}}, \bibinfo
  {author} {\bibfnamefont {N.}~\bibnamefont {Regnault}}, \ and\ \bibinfo
  {author} {\bibfnamefont {R.}~\bibnamefont {Pain}} (\bibinfo {collaboration}
  {SNLS}),\ }\href {\doibase 10.1051/0004-6361:20053025} {\bibfield  {journal}
  {\bibinfo  {journal} {Astron. Astrophys.}\ }\textbf {\bibinfo {volume}
  {443}},\ \bibinfo {pages} {781} (\bibinfo {year} {2005})},\ \Eprint
  {http://arxiv.org/abs/astro-ph/0506583} {arXiv:astro-ph/0506583 [astro-ph]}
  \BibitemShut {NoStop}%
\bibitem [{\citenamefont {Lewis}\ \emph {et~al.}(2000)\citenamefont {Lewis},
  \citenamefont {Challinor},\ and\ \citenamefont {Lasenby}}]{Lewis:1999bs}%
  \BibitemOpen
  \bibfield  {author} {\bibinfo {author} {\bibfnamefont {A.}~\bibnamefont
  {Lewis}}, \bibinfo {author} {\bibfnamefont {A.}~\bibnamefont {Challinor}}, \
  and\ \bibinfo {author} {\bibfnamefont {A.}~\bibnamefont {Lasenby}},\ }\href
  {\doibase 10.1086/309179} {\bibfield  {journal} {\bibinfo  {journal}
  {Astrophys. J.}\ }\textbf {\bibinfo {volume} {538}},\ \bibinfo {pages} {473}
  (\bibinfo {year} {2000})},\ \Eprint {http://arxiv.org/abs/astro-ph/9911177}
  {arXiv:astro-ph/9911177 [astro-ph]} \BibitemShut {NoStop}%
\bibitem [{\citenamefont {Lewis}\ and\ \citenamefont
  {Bridle}(2002)}]{Lewis:2002ah}%
  \BibitemOpen
  \bibfield  {author} {\bibinfo {author} {\bibfnamefont {A.}~\bibnamefont
  {Lewis}}\ and\ \bibinfo {author} {\bibfnamefont {S.}~\bibnamefont {Bridle}},\
  }\href {\doibase 10.1103/PhysRevD.66.103511} {\bibfield  {journal} {\bibinfo
  {journal} {Phys. Rev.}\ }\textbf {\bibinfo {volume} {D66}},\ \bibinfo {pages}
  {103511} (\bibinfo {year} {2002})},\ \Eprint
  {http://arxiv.org/abs/astro-ph/0205436} {arXiv:astro-ph/0205436 [astro-ph]}
  \BibitemShut {NoStop}%
\bibitem [{\citenamefont {Lewis}(2019)}]{Lewis:2019xzd}%
  \BibitemOpen
  \bibfield  {author} {\bibinfo {author} {\bibfnamefont {A.}~\bibnamefont
  {Lewis}},\ }\href@noop {} {\  (\bibinfo {year} {2019})},\ \Eprint
  {http://arxiv.org/abs/1910.13970} {arXiv:1910.13970 [astro-ph.IM]}
  \BibitemShut {NoStop}%
\bibitem [{\citenamefont {Hotelling}(1936)}]{CCA}%
  \BibitemOpen
  \bibfield  {author} {\bibinfo {author} {\bibfnamefont {H.}~\bibnamefont
  {Hotelling}},\ }\href@noop {} {\bibfield  {journal} {\bibinfo  {journal}
  {Biometrika}\ }\textbf {\bibinfo {volume} {28}},\ \bibinfo {pages} {321}
  (\bibinfo {year} {1936})}\BibitemShut {NoStop}%
\bibitem [{\citenamefont {Wand}\ and\ \citenamefont
  {Jones}(1994)}]{wand1994kernel}%
  \BibitemOpen
  \bibfield  {author} {\bibinfo {author} {\bibfnamefont {M.}~\bibnamefont
  {Wand}}\ and\ \bibinfo {author} {\bibfnamefont {M.}~\bibnamefont {Jones}},\
  }\href {https://books.google.com/books?id=GTOOi5yE008C} {\emph {\bibinfo
  {title} {Kernel Smoothing}}},\ Chapman \& Hall/CRC Monographs on Statistics
  \& Applied Probability\ (\bibinfo  {publisher} {Taylor \& Francis},\ \bibinfo
  {year} {1994})\BibitemShut {NoStop}%
\bibitem [{\citenamefont {Aylor}\ \emph {et~al.}(2019)\citenamefont {Aylor},
  \citenamefont {Joy}, \citenamefont {Knox}, \citenamefont {Millea},
  \citenamefont {Raghunathan},\ and\ \citenamefont {Wu}}]{Aylor:2018drw}%
  \BibitemOpen
  \bibfield  {author} {\bibinfo {author} {\bibfnamefont {K.}~\bibnamefont
  {Aylor}}, \bibinfo {author} {\bibfnamefont {M.}~\bibnamefont {Joy}}, \bibinfo
  {author} {\bibfnamefont {L.}~\bibnamefont {Knox}}, \bibinfo {author}
  {\bibfnamefont {M.}~\bibnamefont {Millea}}, \bibinfo {author} {\bibfnamefont
  {S.}~\bibnamefont {Raghunathan}}, \ and\ \bibinfo {author} {\bibfnamefont
  {W.~L.~K.}\ \bibnamefont {Wu}},\ }\href {\doibase 10.3847/1538-4357/ab0898}
  {\bibfield  {journal} {\bibinfo  {journal} {Astrophys. J.}\ }\textbf
  {\bibinfo {volume} {874}},\ \bibinfo {pages} {4} (\bibinfo {year} {2019})},\
  \Eprint {http://arxiv.org/abs/1811.00537} {arXiv:1811.00537 [astro-ph.CO]}
  \BibitemShut {NoStop}%
\bibitem [{\citenamefont {Raveri}(2019)}]{Raveri:2019mxg}%
  \BibitemOpen
  \bibfield  {author} {\bibinfo {author} {\bibfnamefont {M.}~\bibnamefont
  {Raveri}},\ }\href@noop {} {\  (\bibinfo {year} {2019})},\ \Eprint
  {http://arxiv.org/abs/1902.01366} {arXiv:1902.01366 [astro-ph.CO]}
  \BibitemShut {NoStop}%
\bibitem [{\citenamefont {Knox}\ and\ \citenamefont
  {Millea}(2019)}]{Knox:2019rjx}%
  \BibitemOpen
  \bibfield  {author} {\bibinfo {author} {\bibfnamefont {L.}~\bibnamefont
  {Knox}}\ and\ \bibinfo {author} {\bibfnamefont {M.}~\bibnamefont {Millea}},\
  }\href@noop {} {\  (\bibinfo {year} {2019})},\ \Eprint
  {http://arxiv.org/abs/1908.03663} {arXiv:1908.03663 [astro-ph.CO]}
  \BibitemShut {NoStop}%
\bibitem [{\citenamefont {A.M.~Mathai}(1992)}]{Mathai:Quadraticforms}%
  \BibitemOpen
  \bibfield  {author} {\bibinfo {author} {\bibfnamefont {S.~B.~P.}\
  \bibnamefont {A.M.~Mathai}},\ }\href@noop {} {\emph {\bibinfo {title}
  {Quadratic Forms in Random Variables}}}\ (\bibinfo  {publisher} {Taylor and
  Francis},\ \bibinfo {year} {1992})\BibitemShut {NoStop}%
\bibitem [{\citenamefont {Patnaik}(1950)}]{10.2307/2332149}%
  \BibitemOpen
  \bibfield  {author} {\bibinfo {author} {\bibfnamefont {P.~B.}\ \bibnamefont
  {Patnaik}},\ }\href@noop {} {\bibfield  {journal} {\bibinfo  {journal}
  {Biometrika}\ }\textbf {\bibinfo {volume} {37}},\ \bibinfo {pages} {78}
  (\bibinfo {year} {1950})}\BibitemShut {NoStop}%
\end{thebibliography}%

\end{document}